\pdfoutput=1
\documentclass[10pt,twocolumn,letterpaper]{article}
\usepackage[pagenumbers]{cvpr}
\usepackage[utf8]{inputenc}
\usepackage[T1]{fontenc}
\usepackage{amsmath,amssymb,amsthm}

\usepackage{array,tabularx,booktabs,multirow}
\usepackage{xurl}
\usepackage{microtype}
\usepackage{colortbl}
\usepackage{graphicx}
\definecolor{linkblue}{rgb}{0.12,0.27,0.43}
\usepackage[breaklinks,colorlinks,allcolors=linkblue]{hyperref}
\providecolor{ridger}{RGB}{252,234,230}
\title{Seen but Not Connected:\\Recovering Vessel Trees Below the Threshold}
\author{Yihao Chen \quad Xuefeng Wang\\
College of Engineering, Peking University}
\date{}
\begin{document}
\addtocontents{toc}{\protect\setcounter{tocdepth}{-1}}
\makeatletter
\twocolumn[{\@maketitle\begin{minipage}{\textwidth}
\captionsetup{hypcap=false}
\centering
\includegraphics[width=\textwidth]{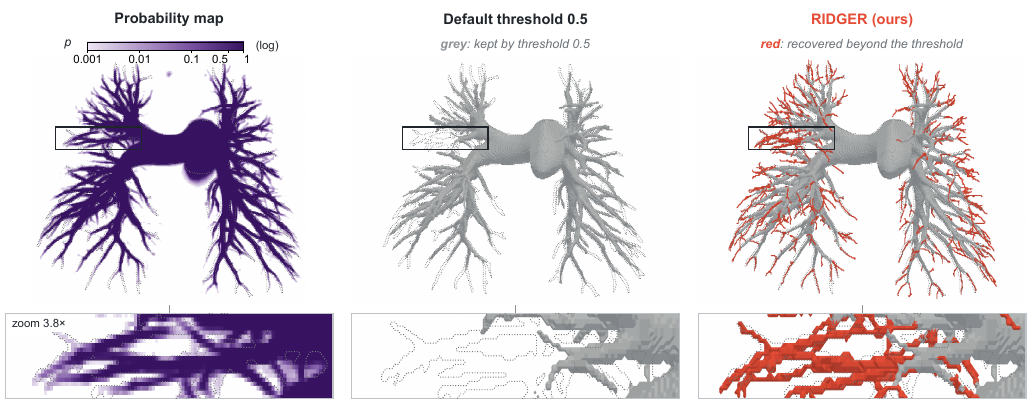}
\captionof{figure}{\textbf{The distal tree is in the map but not in the mask.} A pulmonary test patient with a large gain (HiPaS, nnU-Net; selection in Sec.~\ref{S-sec:supp-J-selection}). The reaching rate \(R\) (Sec.~\ref{sec:measurement}) is 57.8\% at 0.5 and 94.9\% with RIDGER, at off-reference volumes of 4.66\% and 8.48\%.}
\label{fig:teaser}
\end{minipage}\par\smallskip
}]
\makeatother
\begin{abstract}
Current segmentation networks reconstruct the distal branches of vessel trees incompletely. To improve completeness, existing methods train the network with topology-aware losses, or track and repair vessels after segmentation. Either way, the network output still passes a fixed threshold, usually 0.5, which ignores true vessels that the map may hold at low probability. A lower threshold keeps them but thickens existing vessels. We propose \textbf{RIDGER} (RIDGE Rooting), which extends the 0.5 mask along the trend of the probability map instead of lowering the threshold. RIDGER first finds vessel ridges in the map, which are the lines of locally highest probability, and keeps the long, confident ridges that join the mask as centrelines. RIDGER then estimates the vessel thickness adaptively: from the 0.5 mask for the proximal vessels and from the probability profile across the vessel for the distal ones. On the other hand, the commonly used metric, Dice, barely drops when the distal vessels are reconstructed incompletely. To indicate the performance of distal vessel reconstruction, we propose a new metric, the reaching rate (R). R is the ratio of the length that the mask covers and connects to the root to the full length of the last 20\,mm of each reference branch. When the proposed method, RIDGER, replaces nnU-Net's 0.5 threshold, the metric R rises from 77\% to 91\% (coronary) and from 80\% to 96\% (pulmonary), without retraining the network.

\end{abstract}
\section{Introduction}
\label{sec:related}

Deep networks segment the large vessels in CT well. We train a standard nnU-Net \cite{isensee2021nnunet} on each of two public datasets. ImageCAS-X annotates the coronary arteries in CT angiography (CTA) \cite{bransby2026imagecasx}, and HiPaS annotates the pulmonary arteries in chest CT \cite{chu2025hipas}. On their test patients, the two networks reach Dice scores of 89\% and 88\%. However, near the ends of the vessel tree, their masks often stop early or break off, and a high Dice hides this loss. We call the manual annotation the reference. In the last 20\,mm of each reference branch, the root-connected mask parts cover only 77\% (coronary) and 80\% (pulmonary) of the length. About a fifth of the distal tree is missing (Tab.~\ref{tab:main-coronary}). These distal branches matter clinically. For example, the small pulmonary vessels are related to the severity of emphysema and of thromboembolic pulmonary hypertension \cite{estepar2013vascular,rahaghi2016cteph}. This paper studies how to recover these distal branches.

Existing methods improve distal reconstruction in two ways. Topology-aware losses, such as clDice \cite{shit2021cldice}, Skeleton Recall \cite{kirchhoff2024skeleton} and cbDice \cite{shi2024cbdice}, change the training so that the network keeps thin vessels connected. Post-processing methods, such as vessel tracking \cite{wolterink2019centerline,liu2026topovst} and repair networks \cite{weng2023topology,drwiega2026mindthegap}, follow or reconnect the vessels after segmentation. In both cases, a fixed threshold (usually 0.5) turns the output probability map into a mask and discards all probabilities below it. Some readouts do use probabilities below 0.5, such as a lower threshold or hysteresis thresholding \cite{canny1986edge}. However, these readouts still judge each voxel by its own probability.

Below the threshold, the probability map still holds much of the distal tree. Along a distal branch that the 0.5 mask misses, the map usually keeps a weak response, while the background at the same distance from the mask does not. The networks have therefore seen these branches, but give them probabilities below 0.5. Our measurements support this. Of the distal reference centreline that the 0.5 mask misses, 59\% (coronary) and 58\% (pulmonary) have a voxel with a probability of at least 0.01 within 1\,mm, against only 7\% and 14\% of background points (Tab.~\ref{S-tab:supp-seen}). The training labels may be one reason, because manual references often stop before the vessel ends in the image \cite{garciauceda2018airway,bauer2015airway}. Networks retrained on extended labels support this (Sec.~\ref{sec:labels}).

A lower threshold recovers part of these branches, but inefficiently: most of the voxels it adds only thicken existing vessels. This happens because one threshold sets both how far a vessel extends and how thick it is. Lowering it to reach the faint tips also widens every vessel that the mask already contains. Most of the added volume therefore thickens existing vessels, while the gain in distal coverage comes from the small part that extends the tree. Our measurements show this split. When we lower the threshold from 0.5 to \(1/255\) for coronary networks trained with four losses, at most 16\% of the added volume extends the tree, yet this part gives at least 74\% of the gain in distal coverage (Sec.~\ref{S-sec:supp-S1-lengthen}).

\begin{figure*}[t]
\centering
\includegraphics[width=\textwidth]{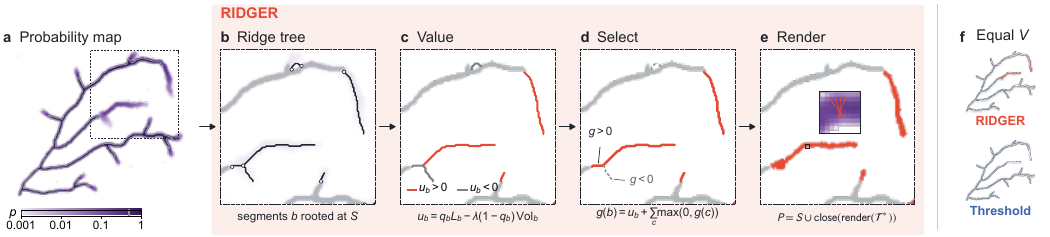}
\caption{\textbf{Overview of RIDGER} on a semi-synthetic 2D sketch (real coronary branches, synthetic probabilities). (a) Probability map \(p\). (b) Ridges beyond the default mask \(S\) (grey) form a tree rooted at \(S\). (c) Each segment gets a value \(u_b\) \eqref{eq:objective}. (d) One leaf-to-root pass \eqref{eq:dp} keeps the best rooted subtree. A faint segment stays because the branch beyond has positive value; a negative side branch is pruned. (e) Kept segments drawn at measured width. (f) At equal off-reference volume \(V\), RIDGER reaches reference ends (dotted) that the best threshold misses.}
\label{fig:algorithm}
\end{figure*}

We propose RIDGER (RIDGE Rooting), a training-free method that sets the length of the vessel tree and the width of the vessels separately (Fig.~\ref{fig:algorithm}). A thin distal vessel has a low probability, but its probability is still higher than that of the tissue around it. The vessel therefore forms a ridge in the probability map. A ridge is a thin line of locally highest probability. RIDGER starts from the 0.5 mask and follows these ridges outward. It decides branch by branch, not voxel by voxel. It adds a branch if the branch adds much length for little volume and stands out clearly from its surroundings. RIDGER can also cross a weak stretch of a ridge when the branches beyond it are long and confident enough. It thus reconnects breaks that otherwise need a separate repair step. Finally, RIDGER keeps the vessel width of the 0.5 mask and measures the width of each new branch from the probability profile across it. The width of a branch therefore does not depend on how far the tree extends.

Distal reconstruction also needs its own measure, because existing metrics such as Dice hardly see the distal tree. Dice measures the overlap between two masks in volume, so every voxel counts the same and the thick vessels dominate it. Distal branches are thin and hold few voxels, so losing them barely changes Dice, while a small change in the width of the thick vessels changes it a lot. A test on the coronary reference shows this. Cutting the last 10\,mm off every branch removes 43\% of the distal centreline but lowers Dice by only 4 points, whereas drawing every vessel one voxel thicker lowers it by 41 points (Sec.~\ref{S-sec:supp-C-quoted}). We therefore propose a new metric, the \emph{reaching rate} \(R\). \(R\) is the ratio of the length that the mask covers and connects to the root to the full length of the last 20\,mm of each reference branch. The reaching rate alone is not enough, because a mask can raise it just by adding volume, even when most of what it adds is not vessel. We therefore also measure the \emph{off-reference volume}, the mask volume outside the reference divided by the reference volume, and compare methods at equal off-reference volume (Fig.~\ref{fig:reach-volume}). At the same volume, a higher reaching rate means that the added volume extends vessels instead of thickening them or adding false ones. We compare RIDGER with two strong baselines on the same map. The first is the best threshold for each patient. The second is the \emph{float16 control}, which lowers the threshold below \(1/255\), the lowest nonzero 8-bit level, and then removes the voxels that only thicken existing vessels (Sec.~\ref{sec:measurement}). Even against this control, RIDGER keeps a clear lead at the same off-reference volume, so its gain does not come from added volume. RIDGER covers 91\% (coronary) and 96\% (pulmonary) of the distal reference length, against 85\% and 91\% for the control.

\textbf{From probability map to mask.} The usual readout is a fixed threshold. A lower threshold, hysteresis \cite{canny1986edge} and decision rules that are optimal for expected Dice or IoU \cite{nowozin2014optimal,dai2023rankseg} look below 0.5. They still judge single voxels by their probability. Pruning methods from neuron tracing judge branches by their length and coverage \cite{peng2011app,xiao2013app2}. They do not weigh the length of a branch against its volume. None of these readouts sets tree length and vessel width separately.

\textbf{Methods for distal reconstruction.} Classical methods use vesselness filters, ridge operators, Morse operators and minimal paths to extract curvilinear structures from the image intensity \cite{frangi1998multiscale,wang2018dimorsc,hu2021dmt,wang2020ddt,sirazitdinov2024centerline,lesage2009review}. Topology-aware losses \cite{shit2021cldice,kirchhoff2024skeleton,shi2024cbdice,stucki2024betti}, related architectures and test-time adaptation \cite{mosinska2018beyond,batra2019improved,hu2019topology,mao2026morvess,zong2026wprf,zhou2025topotta,zhang2023atm} keep thin vessels connected. Tracking methods follow the vessel from a seed \cite{aylward2002ridge,cohen1997minimal,wolterink2019centerline,naeem2025trexplorer,liu2026topovst,nadeem2021airway}. Repair networks reconnect broken trees \cite{weng2023topology,qiu2023corsegrec,oyer2025repair,drwiega2026mindthegap,goffart2026breakage}. Graph methods select connected subgraphs of the image \cite{turetken2016reconstructing,rempfler2016mccs,vicente2008connectivity,stuhmer2013tree}. These methods either take the mask from a fixed threshold or start from the image, and image evidence near the tips is weak. RIDGER uses sub-threshold probabilities and judges whole branches.

\textbf{Evaluation of distal reconstruction.} Tree length detected and branches detected \cite{lo2012exact,zhang2023atm} measure how much of the tree is found. They do not count false positives. Dice, clDice \cite{shit2021cldice} and the Betti errors measure overlap and topology. They count every branch outside the reference as an error \cite{reinke2024pitfalls,maierhein2024metrics}. However, manual references often end before the vessel ends in the image \cite{garciauceda2018airway,bauer2015airway}. Some works therefore complete the labels by self-learning \cite{wang2023airway,karimi2020noisy}, and others re-annotate the data \cite{bransby2026imagecasx,zeng2023imagecas}. Our evaluation measures the reaching rate, compares methods at equal off-reference volume, and asks blinded readers to judge the additions.

Our contributions are the following.
\begin{itemize}
\item \textbf{We find that networks already see the distal branches they miss.} These branches stay in the probability map below the threshold, and a lower threshold recovers them only by thickening every vessel.
\item \textbf{We propose RIDGER}, a training-free readout that extends the 0.5 mask along the ridges of the probability map and sets the width of new branches separately. It works on the map of any segmentation network and on other tree-like structures.
\item \textbf{We propose an evaluation of distal reconstruction} that compares the reaching rate (\(R\)) at equal off-reference volume. Under this evaluation, RIDGER reaches further than every threshold of the same map, and blinded readers judge nearly all of its additions to be vessels.
\end{itemize}

\section{Method: binarisation as a choice of tree}
\label{sec:method}
\label{sec:problem}

\textbf{Setting.} A network outputs a probability map \(p\). We write \(S\) for the root-connected default mask. \(S\) contains the components of the 0.5 mask that touch an anchor at the root of the tree. On pulmonary CT, the anchor is the largest predicted component. On coronary CTA, the anchor is the region around the aorta. The off-the-shelf model TotalSegmentator \cite{wasserthal2023totalsegmentator} segments the aorta. Neither anchor uses the reference. RIDGER returns a mask \(P\supseteq S\). We write \(\ell=\log\bigl(p/(1-p)\bigr)\) for the log odds of the map.

\subsection{A tree of ridge segments}
\label{sec:ridges}

\textbf{Candidates and linking.} To find the ridges, RIDGER first collects every voxel outside \(S\) with \(p\geq10^{-5}\). This candidate threshold was set on validation patients. RIDGER thins the candidates to a skeleton one voxel wide \cite{lee1994thinning}. It then links each skeleton voxel back to \(S\) along the cheapest path. A path \(\gamma\) has the cost
\begin{equation}
c(\gamma)=\int_\gamma\max\bigl(0,-\ell(x)\bigr)^2\,\mathrm dx,
\label{eq:cost}
\end{equation}
where \(x\) runs along the path and length is in millimetres. The cost is zero where \(p\geq0.5\) and grows quickly as the probability falls, so the cheapest paths stay on the ridges. These paths form a shortest-path tree rooted at \(S\).

\textbf{Segments and rooted subtrees.} We cut this tree at its bifurcations and where it leaves \(S\). The pieces are the segments \(b\). Each segment has one parent, the segment it attaches to on its way to \(S\), or \(S\) itself (Fig.~\ref{fig:algorithm}b). The children of a segment are the segments whose parent it is. A set of segments that contains the parent of each member is a \emph{rooted subtree}. Every connected extension of \(S\) along the ridges is a rooted subtree. Choosing a mask beyond \(S\) thus becomes choosing a rooted subtree. A threshold on the tree is one such choice. It keeps a segment when every probability on its path to \(S\) is at least a level \(k\). Sec.~\ref{sec:prune} gives each rooted subtree a score. A threshold is one of the rooted subtrees, so the best-scoring subtree scores no lower than any threshold on the tree (Sec.~\ref{S-sec:supp-B_proofs}).

\subsection{Subtree selection}
\label{sec:prune}

\textbf{Segment value and confidence.} A segment that runs alongside a vessel in \(S\) adds volume but does not raise the reaching rate. A segment that extends beyond a tip adds both. RIDGER therefore credits a segment \(b\) only with the part of its centreline that advances the tree. This length \(L_b\) covers the centreline points that lie more than 1.5\,mm from \(S\) or past an end point of the skeleton of \(S\). RIDGER weighs \(L_b\) against the volume \(\mathrm{Vol}_b\) that the segment adds when it is drawn at its measured width (Sec.~\ref{sec:width}). Both terms are weighted by a confidence. Probability is a poor confidence here, because the flank of a thick vessel can have a higher probability than the ridge of a thin vessel. RIDGER instead measures how much a segment stands out from its surroundings. We call the skeleton voxels of a segment its nodes. The ridge contrast \(\rho_v\) of a node \(v\) and the confidence \(q_b\) of a segment \(b\) are
\begin{equation}
\rho_v=\sigma\bigl(\ell_v-\bar\ell_v\bigr),\qquad q_b=\operatorname*{median}_{v\in b}\,\rho_v,
\label{eq:conf}
\end{equation}
where \(\sigma\) is the logistic function, \(\ell_v\) the log odds at \(v\), and \(\bar\ell_v\) the mean log odds over a hollow shell around \(v\).

\textbf{Objective.} The confidence weights both terms. The term \(q_bL_b\) is the advancing length that the segment is expected to add. The term \((1-q_b)\mathrm{Vol}_b\) is the volume that it may add wrongly. With a weight \(\lambda\) on the volume, RIDGER keeps the rooted subtree
\begin{equation}
\mathcal T^\ast(\lambda)=\arg\max_{\mathcal T\ \text{rooted}}\ \sum_{b\in\mathcal T}\underbrace{\bigl(q_b\,L_b-\lambda\,(1-q_b)\,\mathrm{Vol}_b\bigr)}_{u_b},
\label{eq:objective}
\end{equation}
where \(\mathcal T\) ranges over the rooted subtrees and \(u_b\) is the value of segment \(b\). The weight \(\lambda\) (mm\(^{-2}\)) converts volume (mm\(^3\)) into centreline length (mm).

\textbf{Solution.} A segment can be kept only together with its parent. One pass from the leaves to the root therefore computes, for each segment \(b\) with children \(c\),
\begin{equation}
g(b)=u_b+\sum_{c}\max\bigl(0,g(c)\bigr),
\label{eq:dp}
\end{equation}
where \(g(b)\) is the best value that the subtree below \(b\) can contribute. A child \(c\) is kept when its parent is kept and \(g(c)>0\). \emph{Proposition 1 (exactness).} The pass \eqref{eq:dp} returns a maximiser of the selection objective \eqref{eq:objective} in time linear in the number of segments (proof in Sec.~\ref{S-sec:supp-B_proofs}). The pass adapts strong pruning \cite{johnson2000prize} to the ridge tree.

\textbf{Proposition 2 (nested selections).} If \(\lambda\leq\lambda'\), then \(\mathcal T^\ast(\lambda')\subseteq\mathcal T^\ast(\lambda)\) (proof in Sec.~\ref{S-sec:supp-B-nested}). Raising \(\lambda\) therefore only removes segments, so one parameter moves RIDGER along a nested family of selections.

\subsection{Width and loop veto}
\label{sec:width}

\textbf{Width.} RIDGER measures the width at each node \(v\) of a kept segment. It casts eight rays in the plane normal to the segment. Along each ray, it finds the half-height edge. At this point, \(p\) has fallen halfway from its value \(p_v\) at \(v\) to the local background \(f_v\). The local background is the lowest probability within 3\,mm of \(v\). The median over the eight rays gives the radius at \(v\). RIDGER draws the segment as a chain of balls of this radius and leaves the width of \(S\) unchanged.

\textbf{Loop veto and closure.} A vessel tree has no loops. When RIDGER follows faint ridges, it may join two separate branches and create a loop. RIDGER rejects every drawn addition that would create a loop. Let \(\beta_1\) be the number of loops of the mask and \(\chi_{26}\) its Euler number under 26-connectivity. Let \(\Delta\) be the change that an addition causes in \(\beta_1\) or \(\chi_{26}\), and \(m\) the decrease in the number of connected components of the mask. An addition is accepted if it meets two conditions. It must leave the components of the background unchanged, so it encloses no cavity. It must also satisfy the loop condition
\begin{equation}
\Delta\beta_1=-m-\Delta\chi_{26}\leq 0
\label{eq:veto}
\end{equation}
(derivation in Sec.~\ref{S-sec:supp-B_proofs}). Two drawn tubes sometimes touch only along an edge or at a corner. The lumen is then not face-connected. A closure step therefore adds the fewest voxels that complete a face-connected path, one or two per touching pair. These voxels count towards the off-reference volume. In full, RIDGER returns
\begin{equation}
P=S\cup\operatorname{close}\Bigl(\operatorname{render}\bigl(\mathcal T^\ast(\lambda)\bigr)\Bigr),
\label{eq:ridger}
\end{equation}
where \(\operatorname{render}\) draws the selected segments under the loop veto and \(\operatorname{close}\) applies the closure. Because \(P\supseteq S\), RIDGER never lowers the reaching rate of the default mask. An addition that passes the loop condition \eqref{eq:veto} never adds a loop (Proposition 3, Sec.~\ref{S-sec:supp-B_proofs}). Proposition 3 does not cover the closure. Tab.~\ref{tab:ablation} gives the loop error before and after it.

\textbf{Settings.}\phantomsection\label{sec:impl} We set \(\lambda=300\,\text{mm}^{-2}\) for coronary CTA and \(\lambda=30\,\text{mm}^{-2}\) for pulmonary CT. We chose each value on validation patients; both networks of an anatomy share it. On a CPU thread, candidate extraction and width measurement take 4.4\,s per coronary scan and 4.5--5.9\,min per pulmonary scan. Sec.~\ref{S-sec:supp-A_method} gives constants, runtimes and code.

\section{Evaluating distal reconstruction}
\label{sec:measurement}

\textbf{Setting.} All readouts start from the same default mask \(S\) (Sec.~\ref{sec:method}). The root of the reference tree is read from the reference alone and is used only for scoring (Sec.~\ref{S-sec:supp-C-excl}). We fixed the scoring window, the tolerance and the reaching rate as the primary measure before designing RIDGER (Sec.~\ref{S-sec:supp-D-window}).

\textbf{Reaching rate and off-reference volume.} Let \(G\) be the reference mask. The distal window \(W\) is the reference centreline within 20\,mm upstream of each terminal. We sample it with arc-length weights \(w_i\). For a prediction \(P\), we set \(C_i(P)=1\) when sample \(i\) lies within 1\,mm of \(P\) and connects to the reference root through a path of predicted voxels that all lie within 1\,mm of \(G\). Otherwise \(C_i(P)=0\). A branch that is reached only through leakage outside the reference therefore does not count. We define
\begin{equation}
R(P)=100\,\tfrac{\sum_{i\in W}w_iC_i(P)}{\sum_{i\in W}w_i},\quad
V(P)=100\,\tfrac{\operatorname{vol}(P\setminus G)}{\operatorname{vol}(G)}.
\label{eq:rv}
\end{equation}
\(R\) is the \emph{reaching rate}. \(V\) is the \emph{off-reference volume}, the predicted volume outside the reference as a percentage of the reference volume. \(\operatorname{vol}(\cdot)\) is the volume of a voxel set.

\textbf{Equal-volume comparison.} Let \(p_8\) be the map with its probabilities rounded to 8 bits. For a level \(k\), the mask \(\Theta_S(k)\) keeps the 26-connected components of \(S\cup\{p_8\geq k\}\) that touch \(S\). The family of all thresholds of the map is \(\Theta_8=\{\Theta_S(k)\}_{k=1}^{255}\). The envelope
\begin{equation}
\mathcal E_8(d)=\max_{Q\in\Theta_8:\,V(Q)\leq d}R(Q)
\label{eq:envelope}
\end{equation}
is the highest \(R\) of any member \(Q\) of this family whose off-reference volume is at most \(d\). We call the maximising member the \emph{best 8-bit threshold}. For a readout \(P\), the \emph{equal-volume gain} is \(R(P)-\mathcal E_8(V(P))\). We compute the gain for each patient, with that patient's best threshold.

\textbf{Volume or length.} The \emph{off-reference length} is the length of predicted skeleton more than 1\,mm from \(G\). It penalises every added millimetre of branch but not thickening. We repeat the main comparisons at equal off-reference length, with the same envelope.

\textbf{Comparators.} The best 8-bit threshold cannot go below \(1/255\), but RIDGER reads the map in 16-bit floating point (float16), down to \(10^{-5}\). Part of its gain could therefore come only from these lower probabilities. The float16 control tests this. The control thresholds the map at 36 levels, 24 on the 8-bit grid and 12 from \(1/255\) down to \(p=10^{-4}\). It then trims each thresholded mask. Each added voxel is assigned to the nearest point of the skeleton of the mask. Voxels assigned to the skeleton of \(S\) are removed as thickening, and newly added branches are kept. The control also includes partly trimmed masks placed near RIDGER's per-patient volume. For each patient, the control takes the member with the highest \(R\) within that volume (Sec.~\ref{S-sec:supp-D-f16}). This choice favours the control. We also compare with two-level hysteresis \cite{canny1986edge}, freeze-and-grow \cite{nadeem2021airway} and three classical post-processors: minimal-path gap bridging \cite{cohen1997minimal}, ridge tracking \cite{aylward2002ridge} and discrete Morse extraction (DiMorSC) \cite{wang2018dimorsc}.

\textbf{Tuning.} Threshold-type comparators take each patient's best member, chosen with the reference. This favours the comparators. Classical post-processors and RIDGER use settings chosen on validation patients.

\textbf{Other scores.} We also report tree length detected (TD) and branches detected (BD) \cite{lo2012exact,zhang2023atm}. They are the shares of the reference centreline length and of the reference branches that the prediction covers and connects to the root. We further report Dice, clDice \cite{shit2021cldice} and the Betti errors \cite{schaap2009cat08,reinke2024pitfalls,maierhein2024metrics}. The Betti errors \(E_{\beta_0}\) and \(E_{\beta_1}\) are the absolute differences between prediction and reference in the numbers of connected components and of loops. Differences are in percentage points (points). Connectivity is 26-connectivity.

\section{Experiments}
\label{sec:experiments}

\providecolor{ridger}{RGB}{252,234,230}
\providecolor{subhead}{gray}{0.96}
\begin{table*}[t]
\caption{\textbf{Readouts of the same map within RIDGER's off-reference volume.} \(R\): reaching rate. Test-patient means (\%; $E_{\beta}$ absolute); TD: tree length detected, BD: branches detected (Sec.~\ref{sec:measurement}). Each comparator uses at most RIDGER's per-patient $V$ (values in grey); thresholds take each patient's best member. In the text, two values for an anatomy refer to the left and right network. Best per column and network in bold, second underlined. clDice counts every predicted centreline beyond the reference as false (Sec.~\ref{sec:readers}).}
\label{tab:main-coronary}\label{tab:main-lung}
\centering\scriptsize
\setlength{\tabcolsep}{2.75pt}
\begin{tabular}{@{}l*{8}{r}@{\hspace{9pt}}*{8}{r}@{}}
\toprule
Readout within RIDGER's $V$ & $V$ & $R\!\uparrow$ & TD$\!\uparrow$ & BD$\!\uparrow$ & Dice$\!\uparrow$ & clDice$\!\uparrow$ & $E_{\beta_0}\!\downarrow$ & $E_{\beta_1}\!\downarrow$ & $V$ & $R\!\uparrow$ & TD$\!\uparrow$ & BD$\!\uparrow$ & Dice$\!\uparrow$ & clDice$\!\uparrow$ & $E_{\beta_0}\!\downarrow$ & $E_{\beta_1}\!\downarrow$ \\
\midrule
\textbf{Coronary CTA} & \multicolumn{8}{c}{nnU-Net Dice+CE ($n=160$)} & \multicolumn{8}{c}{released nnU-Net ($n=160$)} \\
\cmidrule(lr){2-9}\cmidrule(l){10-17}
\rowcolor{subhead}\multicolumn{17}{@{}l}{\textit{Thresholds of the same map}} \\
\hspace{0.6em}Default threshold (0.5) & \textcolor{gray}{5.00} & 77.31 & 89.94 & 84.55 & 89.29 & 92.00 & 0.10 & 0.06 & \textcolor{gray}{5.27} & 75.13 & 88.95 & 83.24 & 89.32 & 91.58 & 0.09 & \textbf{0.10} \\
\hspace{0.6em}Best 8-bit threshold & \textcolor{gray}{6.54} & 78.99 & 90.60 & 85.67 & \underline{89.90} & 92.12 & 0.08 & \underline{0.05} & \textcolor{gray}{7.18} & 77.89 & 90.17 & 85.00 & \textbf{90.10} & 92.03 & 0.07 & 0.16 \\
\hspace{0.6em}Hysteresis \cite{canny1986edge} & \textcolor{gray}{5.55} & 78.80 & 90.39 & 85.31 & 89.41 & 92.11 & 0.15 & 0.06 & \textcolor{gray}{6.22} & 77.46 & 89.94 & 84.76 & 89.71 & \underline{92.07} & 0.11 & \underline{0.11} \\
\hspace{0.6em}Freeze-and-grow \cite{nadeem2021airway} & \textcolor{gray}{5.73} & 78.03 & 90.26 & 84.93 & 89.57 & 92.07 & 0.09 & \textbf{0.04} & \textcolor{gray}{6.14} & 76.51 & 89.62 & 84.09 & 89.68 & 91.87 & 0.08 & 0.13 \\
\addlinespace[1pt]
\rowcolor{subhead}\multicolumn{17}{@{}l}{\textit{Classical post-processing}} \\
\hspace{0.6em}Minimal-path bridging \cite{cohen1997minimal} & \textcolor{gray}{5.49} & 79.49 & 90.57 & 85.39 & 89.24 & 91.92 & 0.08 & 0.12 & \textcolor{gray}{5.87} & 77.32 & 89.83 & 84.36 & 89.23 & 91.51 & 0.07 & 0.21 \\
\hspace{0.6em}Ridge tracking \cite{aylward2002ridge} & \textcolor{gray}{5.19} & 78.11 & 90.27 & 85.17 & 89.29 & 92.06 & 0.09 & 0.12 & \textcolor{gray}{5.47} & 75.41 & 89.10 & 83.44 & 89.26 & 91.53 & 0.08 & 0.16 \\
\hspace{0.6em}DiMorSC \cite{wang2018dimorsc} & \textcolor{gray}{5.11} & 80.08 & 90.98 & 85.89 & 89.33 & \underline{92.20} & \textbf{0.05} & 0.11 & \textcolor{gray}{5.40} & 78.28 & 90.12 & 84.94 & 89.36 & 91.76 & \textbf{0.06} & 0.13 \\
\addlinespace[1pt]
\hspace{0.6em}Float16 control (ours) & \textcolor{gray}{5.83} & \underline{84.65} & \underline{92.39} & \underline{88.47} & 89.50 & \textbf{92.39} & \textbf{0.05} & 0.19 & \textcolor{gray}{6.36} & \underline{84.24} & \underline{92.57} & \underline{88.41} & 89.83 & \textbf{92.55} & \textbf{0.06} & 0.21 \\
\rowcolor{ridger} \hspace{0.6em}\textbf{RIDGER} & \textcolor{gray}{6.67} & \textbf{91.03} & \textbf{95.78} & \textbf{92.71} & \textbf{90.06} & 90.94 & 0.06 & 0.09 & \textcolor{gray}{7.24} & \textbf{91.25} & \textbf{95.79} & \textbf{92.66} & \underline{89.89} & 90.87 & \textbf{0.06} & 0.17 \\
\midrule
\textbf{Pulmonary CT} & \multicolumn{8}{c}{nnU-Net ($n=49$)} & \multicolumn{8}{c}{mask-prompted SR ($n=50$)} \\
\cmidrule(lr){2-9}\cmidrule(l){10-17}
\rowcolor{subhead}\multicolumn{17}{@{}l}{\textit{Thresholds of the same map}} \\
\hspace{0.6em}Default threshold (0.5) & \textcolor{gray}{8.75} & 79.86 & 87.63 & 78.49 & 87.72 & 86.83 & 5.65 & 4.51 & \textcolor{gray}{15.36} & 90.41 & 94.09 & 87.45 & 86.86 & \textbf{86.52} & 5.54 & 4.24 \\
\hspace{0.6em}Best 8-bit threshold & \textcolor{gray}{12.82} & 83.44 & 89.95 & 81.61 & \underline{88.28} & \underline{87.07} & 5.65 & \underline{4.29} & \textcolor{gray}{18.11} & 90.92 & 94.39 & 88.00 & 86.83 & 86.29 & 5.54 & 4.28 \\
\hspace{0.6em}Hysteresis \cite{canny1986edge} & \textcolor{gray}{12.01} & 83.02 & 89.69 & 81.25 & \textbf{88.36} & \textbf{87.08} & 5.88 & 4.31 & \textcolor{gray}{17.32} & 90.83 & 94.34 & 87.88 & \underline{86.95} & 86.37 & 5.56 & 4.18 \\
\hspace{0.6em}Freeze-and-grow \cite{nadeem2021airway} & \textcolor{gray}{10.02} & 81.43 & 88.68 & 79.80 & 88.26 & 86.84 & 5.65 & 4.43 & \textcolor{gray}{15.71} & 90.54 & 94.17 & 87.60 & \textbf{87.01} & \underline{86.50} & 5.54 & 4.18 \\
\addlinespace[1pt]
\rowcolor{subhead}\multicolumn{17}{@{}l}{\textit{Classical post-processing}} \\
\hspace{0.6em}Minimal-path bridging \cite{cohen1997minimal} & \textcolor{gray}{9.19} & 83.15 & 89.51 & 81.33 & 87.62 & 85.03 & 5.65 & 4.37 & \textcolor{gray}{15.89} & 92.06 & 95.00 & 89.17 & 86.69 & 83.80 & 5.54 & 4.18 \\
\hspace{0.6em}Ridge tracking \cite{aylward2002ridge} & \textcolor{gray}{9.65} & 85.24 & 90.68 & 83.05 & 87.47 & 83.47 & 5.65 & 5.37 & \textcolor{gray}{15.84} & 91.62 & 94.75 & 88.71 & 86.70 & 84.84 & 5.54 & 4.16 \\
\hspace{0.6em}DiMorSC \cite{wang2018dimorsc} & \textcolor{gray}{9.14} & 85.10 & 90.84 & 83.21 & 87.63 & 83.52 & 5.65 & 5.43 & \textcolor{gray}{15.42} & 90.89 & 94.37 & 87.85 & 86.84 & 85.85 & 5.54 & \underline{3.98} \\
\addlinespace[1pt]
\hspace{0.6em}Float16 control (ours) & \textcolor{gray}{12.09} & \underline{90.51} & \underline{94.17} & \underline{87.56} & 87.04 & 83.79 & 5.65 & \textbf{4.00} & \textcolor{gray}{17.61} & \underline{93.49} & \underline{95.88} & \underline{90.80} & 86.20 & 82.56 & 5.54 & \textbf{3.80} \\
\rowcolor{ridger} \hspace{0.6em}\textbf{RIDGER} & \textcolor{gray}{12.87} & \textbf{96.30} & \textbf{97.24} & \textbf{94.21} & 86.45 & 70.38 & 5.65 & 4.37 & \textcolor{gray}{18.14} & \textbf{96.35} & \textbf{97.46} & \textbf{94.22} & 85.87 & 70.78 & 5.54 & 4.32 \\
\bottomrule
\end{tabular}
\end{table*}

\textbf{Data and networks.} We evaluate on 160 coronary CTA test patients of ImageCAS-X \cite{bransby2026imagecasx} and on 60 pulmonary CT test patients of HiPaS \cite{chu2025hipas}. ImageCAS-X re-annotates the CTs of ImageCAS \cite{zeng2023imagecas}. The HiPaS reference is the pulmonary artery tree. All 40 labelled cases of ASOCA \cite{gharleghi2023asoca} form an external coronary cohort. The coronary maps come from eight networks: nnU-Net \cite{isensee2021nnunet} trained with the Dice+CE loss (three seeds), nnU-Net trained with the cbDice, clDice and Skeleton Recall (SR) losses, the released coronary nnU-Net, and CAS-Net \cite{dong2023casnet}. The pulmonary maps come from three networks: a residual-encoder nnU-Net \cite{isensee2024revisited}, an SR network, and an SR network fine-tuned to take the default mask as an extra input (mask-prompted SR). A pair is a network with its test set. The four primary pairs are nnU-Net Dice+CE (seed 1) and the released nnU-Net on coronary CTA, and nnU-Net and mask-prompted SR on pulmonary CT. When we give two values for an anatomy, they follow the order of its primary pairs. On pulmonary CT, 49 and 50 patients are scored, because ten references have no readable root and one more patient is excluded for the nnU-Net (Sec.~\ref{S-sec:supp-C-excl}). All settings are fixed on 80 coronary and 10 pulmonary validation patients. Intervals are 95\% patient bootstraps.

\begin{figure*}[t]
\centering
\includegraphics[width=\textwidth]{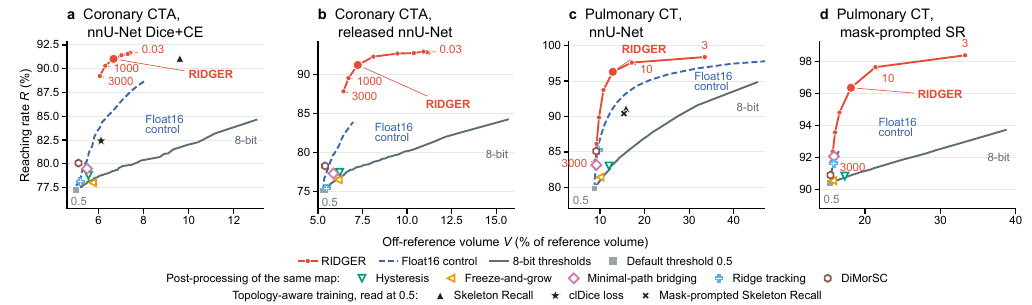}
\caption{\textbf{At the same off-reference volume, RIDGER reaches further than the other readouts of the same map.} Test-patient means of the four primary pairs. Red: RIDGER as \(\lambda\) varies (labels in mm\(^{-2}\)); large marker: the \(\lambda\) of this paper. Dashed blue: the float16 control, a lowered threshold with the thickening removed. Open markers: the post-processors of Tab.~\ref{tab:main-coronary}. Black markers: networks trained with topology-aware losses, read at 0.5 (cbDice, at \(R=66.7\), lies below panel a). A point higher up reaches further for the same added volume.}
\label{fig:reach-volume}
\end{figure*}

\subsection{Reaching rate at equal off-reference volume}
\label{sec:reach}

\textbf{Replacement for 0.5.} RIDGER extends the tree far at little cost in Dice. The reaching rate rises from 77\% to 91\% and from 75\% to 91\% on coronary CTA, and from 80\% to 96\% and from 90\% to 96\% on pulmonary CT (Tab.~\ref{tab:main-coronary}). Dice rises slightly on coronary CTA. On pulmonary CT it falls by about one point, because most additions lie outside the reference, and readers judged these additions to be vessels (Sec.~\ref{sec:readers}).

\begin{figure*}[t]
\centering
\includegraphics[width=\textwidth]{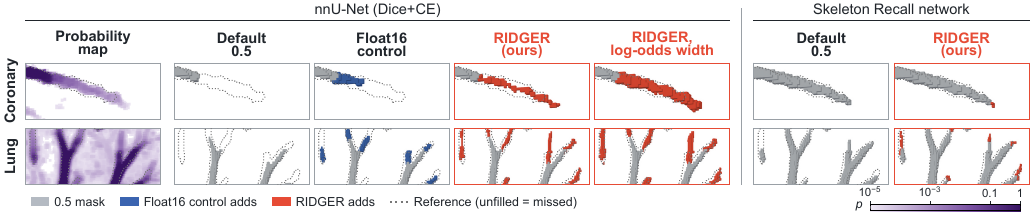}
\caption{\textbf{Readouts of a plain and a topology-aware network.} Grey: the 0.5 mask; colour: the voxels each readout adds. The float16 control is shown at RIDGER's off-reference volume for that patient. Log-odds width: the same selected tree with the width measured on the log odds (Sec.~\ref{sec:robust}). The map is shown down to RIDGER's candidate threshold \(10^{-5}\). The windows show the coronary patient with the median gain and one pulmonary patient. Secs.~\ref{S-sec:supp-J-selection} and~\ref{S-sec:supp-J-columns} give the full trees, the selection rules and per-column \(R\) and \(V\).}
\label{fig:qual}
\end{figure*}

\textbf{Equal off-reference volume.} At the same off-reference volume, RIDGER reaches much further than the best threshold of the same map, and on almost every patient. On coronary CTA it reaches 91\%, against 79\% and 78\% for the best threshold, and it reaches further on 144 and 149 of the 160 patients. On pulmonary CT it reaches 96\%, against 83\% and 91\% (Tab.~\ref{tab:currencies}, Fig.~\ref{fig:qual}). Our float16 control, a stronger baseline, closes part of this gap but stays behind, at 85\% and 84\% on coronary CTA (2.5 to 5.2 points behind on the four pairs at exactly RIDGER's volume, Sec.~\ref{S-sec:supp-D-matched}). The lead does not depend on the settings. It stays above 2.4 points when \(\lambda\) is tripled or divided by three, and above 5.5 coronary points when RIDGER stops at the control's \(10^{-4}\) (Sec.~\ref{S-sec:supp-F_ablation}).

\textbf{Tree scores.} The standard tree scores agree. At RIDGER's off-reference volume, no comparator detects more tree length or more branches than RIDGER on any of the four pairs (Tab.~\ref{tab:main-coronary}).

\subsection{Comparison with topology-aware training}
\label{sec:topology}

\textbf{Topology-aware losses.} A plain network read by RIDGER reaches as far as topology-aware networks read at 0.5. On coronary CTA, the Dice+CE nnU-Net read by RIDGER has an \(R\) of 91\%, the same as the Skeleton Recall network, and it adds less volume. On pulmonary CT, it reaches 96\%, against 91\% and 90\% for the two topology-aware networks (Tab.~\ref{S-tab:supp-E-topo}). RIDGER also extends every topology-aware network once it reads that network's map. TopoVST \cite{liu2026topovst}, a trained tracker that also reads the CT, reaches less far than the best threshold of its own input map. RIDGER reaches 8.0 points further than TopoVST with about a fifth of its off-reference volume (Sec.~\ref{S-sec:supp-E-topovst}).

\subsection{Known-truth repair and blinded reading}
\label{sec:repair}
\label{sec:readers}

\textbf{Truncated references.} When the missing part is known, RIDGER restores more than 92\% of it. We cut the last 10\,mm of every reference branch and start every readout from the cut reference. RIDGER restores 92--93\% of the removed tree. At the same off-reference volume, the thresholds, the float16 control and the classical post-processors restore at most 80\% (Tab.~\ref{S-tab:supp-repair-truncate}). The second test breaks branches inside the map. We lower the probability over a 2\,mm stretch of every first-generation coronary branch, so that the branch beyond the stretch is cut off from the root at any threshold above the faded value. Repair methods are built for this case. RIDGER rejoins 87--96\% of the broken branches, because it judges the branch beyond the stretch as a whole. The thresholds and the float16 control rejoin at most 1.4\%, because they judge each voxel alone and stop at the faded stretch (Sec.~\ref{S-sec:supp-repair-fade}).

\textbf{Blinded reading.} The truncation test checks additions whose truth is known. On real maps, the additions outside the reference have no truth, so readers judge them. Readers judged nearly all sampled additions to be vessels. A pre-registered blinded reading showed pulmonary additions of RIDGER and of the float16 control to six clinically trained readers, each with a clinical rotation in respiratory medicine. The readers did not know that two methods were compared. They judged 98\% of the additions of both methods to be vessels, but accepted only 17\% of the distant fragments that served as checks, so they did not simply accept everything they were shown (Sec.~\ref{S-sec:supp-I-design}). Two pulmonary-vascular physicians called no addition a non-vessel. They judged 83\% to be vessels and the rest indeterminate. On coronary CTA, two physicians judged 94\% of RIDGER's additions to be coronary arteries, against 98\% of the control's, and the interval of the difference includes zero. They called none of RIDGER's additions a non-vessel (Sec.~\ref{S-sec:supp-readers-coronary}).

\textbf{What the readings imply.} The readings suggest that most of the overlap loss comes from the incomplete reference, not from false additions. If the judged vessels count as correct, the pooled pulmonary Dice change becomes a gain of 0.9 to 2.1 points, and the skeleton-precision loss behind the clDice drop shrinks from 33.2 to 3.5 points (Tab.~\ref{S-tab:readers-scompletion}). RIDGER's additions also stay off the veins: 2.0\% of its added length lies within 1\,mm of the HiPaS pulmonary-vein label, against 5.0\% for the control (Sec.~\ref{S-sec:supp-readers-exploratory}). Both methods pass the reading, so they differ in the reaching rate (Tab.~\ref{tab:currencies}).

\subsection{Ablations}
\label{sec:explain}

\providecolor{ridger}{RGB}{252,234,230}
\providecolor{subhead}{gray}{0.96}
\begin{table}[t]
\caption{\textbf{Ablation of the selection.} Test-patient means for nnU-Net Dice+CE (coronary CTA) and nnU-Net (pulmonary CT). (a) Components added in turn; Gain: difference in reaching rate $R$ from the float16 control at the row's volume; a negative gain means that the control reaches further. (b) Other rules that select among RIDGER's segments. RIDGER's renderer draws the selected segments within RIDGER's volume. ``RIDGER leads by'': difference in $R$ (points, 95\% interval). Each rule takes its best setting per patient (APP2-style pruning: validation settings). Rules: Sec.~\ref{S-sec:supp-F_ablation}.}
\label{tab:ablation}
\centering\scriptsize
\setlength{\tabcolsep}{2pt}
\begin{tabular*}{\columnwidth}{@{\extracolsep{\fill}}l*{4}{r}*{4}{r}@{}}
\toprule
 & \multicolumn{4}{c}{Coronary CTA} & \multicolumn{4}{c}{Pulmonary CT} \\
\cmidrule(lr){2-5}\cmidrule(l){6-9}
 & $R\!\uparrow$ & $V$ & $E_{\beta_1}\!\downarrow$ & Gain$\uparrow$ & $R\!\uparrow$ & $V$ & $E_{\beta_1}\!\downarrow$ & Gain$\uparrow$ \\
\midrule
\rowcolor{subhead}\multicolumn{9}{@{}l}{\textit{(a) Components added in turn}} \\
Default threshold (0.5) & 77.31 & \textcolor{gray}{5.00} & 0.06 & $-$0.49 & 79.86 & \textcolor{gray}{8.75} & 4.51 & 0.00 \\
Rooted subtree & 90.93 & \textcolor{gray}{6.56} & 0.51 & 6.51 & 96.17 & \textcolor{gray}{12.20} & 4.80 & 6.47 \\
+ loop veto & 90.93 & \textcolor{gray}{6.56} & 0.06 & 6.51 & 96.17 & \textcolor{gray}{12.19} & 4.51 & 6.47 \\
\rowcolor{ridger}+ closure: \textbf{RIDGER} & 91.03 & \textcolor{gray}{6.67} & 0.09 & 6.38 & 96.30 & \textcolor{gray}{12.87} & 4.37 & 5.79 \\
\bottomrule
\end{tabular*}

\vspace{5pt}
\begin{tabular*}{\columnwidth}{@{\extracolsep{\fill}}lrrrr@{}}
\toprule
 & \multicolumn{2}{c}{Coronary CTA} & \multicolumn{2}{c}{Pulmonary CT} \\
\cmidrule(lr){2-3}\cmidrule(l){4-5}
 & $R\!\uparrow$ & RIDGER leads by & $R\!\uparrow$ & RIDGER leads by \\
\midrule
\rowcolor{subhead}\multicolumn{5}{@{}l}{\textit{(b) Other selection rules within RIDGER's $V$}} \\
Nearest segments first & 86.56 & 4.47 [3.29, 5.69] & 81.18 & 15.12 [12.21, 18.13] \\
Per-segment gate & 89.66 & 1.37 [0.71, 2.10] & 92.82 & 3.48 [2.67, 4.31] \\
Budget-greedy & 88.16 & 2.87 [1.86, 3.91] & 90.96 & 5.34 [4.33, 6.39] \\
Threshold on the tree & 89.52 & 1.51 [0.68, 2.56] & 88.11 & 8.20 [6.23, 10.24] \\
Union greedy & 87.66 & 3.37 [2.39, 4.40] & 92.00 & 4.30 [3.44, 5.20] \\
APP2-style pruning & 89.30 & 1.73 [1.05, 2.54] & 88.51 & 7.79 [5.94, 9.80] \\
\bottomrule
\end{tabular*}
\end{table}

\textbf{Selection rule.} On the same ridge tree, the subtree selection beats every other selection rule. With the ridge tree, the renderer and the off-reference volume fixed, every other rule reaches less far than RIDGER, even with its setting chosen for each patient (Tab.~\ref{tab:ablation}b). A threshold on the same ridge tree (Sec.~\ref{sec:ridges}) falls 1.5 points behind on coronary CTA and 8.2 points behind on pulmonary CT. APP2-style pruning from neuron tracing \cite{xiao2013app2}, with settings chosen on validation patients, falls 1.7 to 7.8 points behind (Sec.~\ref{S-sec:supp-F-app2}).

\textbf{Components.} The rooted subtree alone gains 3.1 to 7.5 points over the float16 control on the four pairs, and the loop veto brings the loop error back to that of the default mask (Tab.~\ref{tab:ablation}a; Sec.~\ref{S-sec:supp-F_ablation}). The ridge contrast adds up to 2.1 points over a constant confidence or the raw probability (Sec.~\ref{S-sec:supp-F-confidence}).

\subsection{Robustness}
\label{sec:robust}

\providecolor{ridger}{RGB}{252,234,230}
\providecolor{subhead}{gray}{0.96}
\begin{table}[t]
\caption{\textbf{Reaching rate at equal off-reference volume and at equal off-reference length.} Test-patient means of the reaching rate \(R\) (\%). Each comparator uses at most RIDGER's per-patient off-reference volume (top) or off-reference length (bottom). Paired differences with 95\% intervals are in Sec.~\ref{S-sec:supp-D-matched} and Tab.~\ref{S-tab:supp-E-length}.}
\label{tab:currencies}
\centering\scriptsize
\setlength{\tabcolsep}{2.6pt}
\begin{tabular}{@{}l*{2}{r}@{\hspace{6pt}}*{2}{r}@{\hspace{6pt}}r@{}}
\toprule
 & \multicolumn{2}{c}{Coronary CTA} & \multicolumn{2}{c}{Pulmonary CT} & Airway \\
\cmidrule(lr){2-3}\cmidrule(lr){4-5}\cmidrule(l){6-6}
Readout & Dice+CE & released & nnU-Net & m.-p.\ SR & nnU-Net \\
\midrule
\rowcolor{subhead}\multicolumn{6}{@{}l}{\textit{At equal off-reference volume}} \\
\hspace{0.6em}Best 8-bit threshold & 78.99 & 77.89 & 83.44 & 90.92 & 75.66 \\
\hspace{0.6em}Float16 control & 84.65 & 84.24 & 90.51 & 93.49 & 88.28 \\
\rowcolor{ridger}\hspace{0.6em}\textbf{RIDGER} & \textbf{91.03} & \textbf{91.25} & \textbf{96.30} & \textbf{96.35} & \textbf{95.25} \\
\addlinespace[2pt]
\rowcolor{subhead}\multicolumn{6}{@{}l}{\textit{At equal off-reference length}} \\
\hspace{0.6em}Best 8-bit threshold & 84.64 & 84.25 & 94.82 & 93.71 & 87.83 \\
\hspace{0.6em}Float16 control & 88.55 & 88.67 & \textbf{96.70} & 96.25 & 93.63 \\
\rowcolor{ridger}\hspace{0.6em}\textbf{RIDGER} & \textbf{91.03} & \textbf{91.25} & 96.30 & \textbf{96.35} & \textbf{95.25} \\
\bottomrule
\end{tabular}
\end{table}

\textbf{Off-reference length and width.} The lead over the float16 control does not come from the width rule alone. At equal off-reference length, RIDGER still leads the best threshold on both anatomies and the control on coronary CTA (Tab.~\ref{tab:currencies}). With the width measured on the log odds, RIDGER leads the control by 2.8 to 5.1 points (Sec.~\ref{S-sec:supp-F-width}). When both methods use RIDGER's width rule, it leads by 1.1 to 3.2 points, with every interval above zero (Tab.~\ref{S-tab:supp-selection-contrasts}).

\textbf{Other networks and scoring windows.} RIDGER gains 3.4 to 17.4 points over the best threshold for every network on the ImageCAS-X and HiPaS references, and 2.0 to 6.9 points on the original ImageCAS labels (Tab.~\ref{S-tab:supp-E-all}). The gain stays positive for distal windows from 10\,mm to the whole tree, for coverage tolerances from 0.5 to 2\,mm, and under map noise (Sec.~\ref{S-sec:supp-F-window}; Tab.~\ref{S-tab:supp-mapchanges}). CAS-Net gives a small probability to large parts of the image. For these maps, RIDGER first sharpens the map with a rule that reads only the map (map normalisation). It then gains 15.6 points over the best threshold on the test patients, and 15.0 points on 80 validation patients with the rule unchanged (Sec.~\ref{S-sec:supp-J-casnet}).

\providecolor{ridger}{RGB}{252,234,230}
\providecolor{subhead}{gray}{0.96}
\begin{table}[t]
\caption{\textbf{Other tree-like structures.} Test means (\%); \(R\): reaching rate. \(\lambda\) is set for each domain on validation data, and each comparator uses at most RIDGER's volume (area in 2D).}
\label{tab:domains}
\centering\scriptsize
\setlength{\tabcolsep}{2.6pt}
\begin{tabular}{@{}l*{2}{r}@{\hspace{6pt}}*{2}{r}@{\hspace{6pt}}*{2}{r}@{}}
\toprule
 & \multicolumn{2}{c}{Airway CT (18)} & \multicolumn{2}{c}{Retina (150)} & \multicolumn{2}{c}{Roads (49)} \\
\cmidrule(lr){2-3}\cmidrule(lr){4-5}\cmidrule(l){6-7}
Readout & $R\!\uparrow$ & clDice$\!\uparrow$ & $R\!\uparrow$ & clDice$\!\uparrow$ & $R\!\uparrow$ & clDice$\!\uparrow$ \\
\midrule
\rowcolor{subhead}\multicolumn{7}{@{}l}{\textit{Thresholds of the same map}} \\
\hspace{0.6em}Default threshold (0.5) & 70.74 & 84.10 & 83.33 & 87.92 & 80.43 & 88.80 \\
\hspace{0.6em}Best 8-bit threshold & 75.66 & \textbf{85.53} & 85.98 & 89.00 & 81.15 & \textbf{88.85} \\
\rowcolor{ridger}\hspace{0.6em}\textbf{RIDGER} & \textbf{95.25} & 78.36 & \textbf{88.36} & \textbf{89.09} & \textbf{85.21} & 86.71 \\
\bottomrule
\end{tabular}
\end{table}

\textbf{External cohort and other domains.} Without retuning, RIDGER gains 6.2 and 6.5 points over the best threshold on the external ASOCA cohort, and 4.1 and 3.7 points over the float16 control. The interval of the second network includes zero. With \(\lambda\) set for each domain, it also gains 19.6 points on airway CT \cite{stoverud2024aeropath}, 2.4 points on retinal images \cite{jin2022fives} and 4.1 points on road maps \cite{mnih2013machine} (Tab.~\ref{tab:domains}; Sec.~\ref{S-sec:supp-H_domains}).

\subsection{Retraining on extended labels}
\label{sec:labels}

Networks retrained on labels extended below the threshold reach further at 0.5. We added RIDGER's additions to the training labels and retrained. The pulmonary nnU-Nets then reach 11 to 12 points further at 0.5 than networks of the same seeds trained on the original labels. The coronary nnU-Nets reach 4 to 5 points further. Labels extended by the float16 control give a similar gain (Sec.~\ref{S-sec:supp-L_distillation}).

\section{Limitations and conclusion}
\label{sec:limits}
\label{sec:costs}

\textbf{Selection objective.} The selection objective \eqref{eq:objective} penalises added volume but not added length. At equal off-reference length, RIDGER still leads the float16 control on coronary CTA and is level with it on pulmonary CT (Tab.~\ref{tab:currencies}). A length penalty in the objective does not change this (Sec.~\ref{S-sec:supp-F-length-price}).

\textbf{Rendered width.} RIDGER draws its distal coronary branches about 0.2\,mm thinner than the reference, because the half-height edge of a faint ridge lies inside the lumen. Measuring the width on the log odds halves this error (Fig.~\ref{fig:qual}), and RIDGER still leads the float16 control (Sec.~\ref{sec:robust}).

\textbf{Overlap scores.} Overlap scores count every addition outside the reference as an error. With RIDGER, the clDice of the pulmonary nnU-Net falls from 87\% to 70\%, and the coronary one from 92\% to 91\%. Tab.~\ref{tab:domains} shows the same clDice cost on airways and roads, and on airways RIDGER detects fewer branches than the float16 control (Sec.~\ref{S-sec:supp-H_domains}).

\textbf{Input.} RIDGER grows the tree from the root-connected 0.5 mask. It cannot rejoin a vessel part that the 0.5 mask separates from the root. This happens in one ASOCA patient (Sec.~\ref{S-sec:supp-J-split}). RIDGER also first sharpens maps with a high background probability (map normalisation). We tested this step only on CAS-Net (Sec.~\ref{S-sec:supp-J-casnet}).

\label{sec:conclusion}
\textbf{Conclusion.} We proposed RIDGER and an evaluation of distal reconstruction. RIDGER needs no training and selects the best rooted subtree of the ridges of a probability map. The evaluation compares the reaching rate at equal off-reference volume. First, the distal tree that the 0.5 mask misses is largely present in the probability map. A lower threshold recovers it only by also thickening vessels, and networks retrained on labels extended below the threshold reach further at 0.5. Second, RIDGER recovers this distal tree. In place of the 0.5 threshold, it raises the reaching rate from 77\% to 91\% on coronary CTA and from 80\% to 96\% on pulmonary CT. At equal off-reference volume, it reaches further than the best threshold of the same map for every network we tested, and further than the float16 control on the four primary pairs. Third, overlap scores count the additions outside the reference as errors, but blinded readers judged 98\% of the sampled pulmonary additions to be vessels and 94\% of the coronary additions to be coronary arteries.

\section*{LLM Usage Statement}
Anthropic Claude and OpenAI Codex helped write experiment, analysis and figure-plotting code, run and monitor experiments, survey related work, discuss ideas, and draft and revise the manuscript and its appendix. DeepSeek and Xiaomi MiMo helped monitor experiments, check their outputs and give feedback on drafts. All figures are plotted or rendered from result files by scripts; no figure content was produced by a generative model. The authors designed the study, reviewed all generated code and text, verified every reported number against the result files, and take full responsibility for the content of this paper.

{\small
\bibliographystyle{ieeenat_fullname}
\bibliography{refs}
}
\clearpage
\makeatletter
\makeatother
\renewcommand{\thelemma}{S\arabic{lemma}}
\setcounter{section}{0}\setcounter{table}{0}\setcounter{figure}{0}\setcounter{equation}{0}\setcounter{lemma}{0}
\renewcommand{\thesection}{S\arabic{section}}
\renewcommand{\thetable}{S\arabic{table}}
\renewcommand{\thefigure}{S\arabic{figure}}
\renewcommand{\theequation}{S\arabic{equation}}
\renewcommand{\theHsection}{S\arabic{section}}
\renewcommand{\theHtable}{S\arabic{table}}
\renewcommand{\theHfigure}{S\arabic{figure}}
\renewcommand{\theHequation}{S\arabic{equation}}
\def\theHlemma{S\arabic{lemma}}
\twocolumn[{\centering\Large\textbf{Appendix}\par\vspace{1.5em}}]
\addtocontents{toc}{\protect\setcounter{tocdepth}{1}}

\input{supp_v2/0_map}
\section{What scores and maps see}
\label{S-sec:supp-S1_scores}

This section gives the measurements behind three statements of the introduction: how Dice and the reaching rate \(R\) respond when the reference is edited (Sec.~\ref{S-sec:supp-C-quoted}), how much of the missed distal tree the probability map still sees (Sec.~\ref{S-sec:supp-C-seen}), and what a lower threshold adds (Sec.~\ref{S-sec:supp-S1-lengthen}).

\subsection{Scores under edits of the reference}
\label{S-sec:supp-C-quoted}

The introduction says that Dice hardly sees the distal tree (Sec.~\ref{sec:related}). We test this on the reference itself, where the answer is known. We edit each reference in one way and score the edited copy against the original, so any change in a score is the effect of that edit.

\textbf{Dice responds to thickness.} When the reference of each of the 160 coronary test patients is dilated by one voxel, the tree keeps its full length, yet Dice falls by 41.2 points. \(R\), the tree length detected (TD) and the branches detected (BD) do not move, because no branch became shorter.

\textbf{\(R\) responds to length.} When the last 10\,mm of every terminal is removed, the tree loses its distal part, and \(R\) falls by 42.6 points. Dice falls by only 3.9 points, about a tenth of what one voxel of thickness cost, because the removed branches are thin and hold few voxels. TD and BD fall by 10.9 and 26.4 points, less than \(R\), so \(R\) is the most sensitive of the three to a shortened tree. This is why \(R\) is the primary measure (Sec.~\ref{sec:measurement}).

\textbf{clDice responds to centreline outside the reference.} We add thin false tubes worth 2.0\% of the reference volume to each of the 49 scored pulmonary CT test patients. clDice falls by 22.3 points, because its skeleton precision \(T_\text{prec}\), the share of the predicted skeleton that lies inside the reference, falls by 36.3 points. RIDGER's pulmonary additions also carry a long skeleton outside the reference: its length is 60.7\% (nnU-Net) and 45.2\% (mask-prompted SR) of the length of the reference centreline. clDice counts this skeleton as false whether or not it is a vessel, so the blinded reading judges it instead (Sec.~\ref{sec:readers}).

\begin{details}
\textbf{False branches.} Each false tube has a radius of 0.5\,mm and a length of 15\,mm. The tubes of one patient together hold 2.0\% of the volume of that patient's pulmonary reference. The score of a patient is the average over five random placements of the tubes.

\textbf{Pulmonary additions.} We take the part of the skeleton of RIDGER's additions that lies more than 1\,mm from the reference. Its length, pooled over patients, is divided by the length of the reference centreline.
\end{details}

\subsection{What the map still sees beyond the threshold}
\label{S-sec:supp-C-seen}
RIDGER reads the map below 0.5, which only helps if the map still responds along the branches that the 0.5 mask misses (Sec.~\ref{sec:related}). We measure this directly. A distal centreline sample is missed when the root-connected 0.5 mask does not reach it, and a missed sample is seen when the map holds \(p\geq0.01\) within 1\,mm of it. The same test on background points at the same distance from the mask shows how often the map responds where there is no vessel. On the nnU-Net maps, the map sees 59.4\% (coronary CTA, Dice+CE, seed 1) and 57.8\% (pulmonary CT) of the missed centreline, against 6.7\% and 13.7\% of the background points. The map therefore responds along more than half of the missed distal tree and rarely elsewhere, and this response is what RIDGER follows (Sec.~\ref{sec:method}). Tab.~\ref{S-tab:supp-seen} gives the missed share, the seen share and the background share for every network and reference.

\begin{details}
\textbf{Definition.} The window is that of the reaching rate \(R\): the reference centreline within 20\,mm upstream of each terminal. A sample is missed when the network's root-connected 0.5 mask does not reach it under \(R\). To test whether a missed sample is seen, we read the 8-bit map with linear interpolation and take the maximum over a ball of 1\,mm around the sample. The sample is seen when this maximum is at least 0.01. For background, we apply the same test to points at the same distance from the default mask and at least 3\,mm from the reference.

\textbf{Other networks and references.} The map sees more of the missed distal tree than of background for every network and every reference. On ImageCAS-X the seen share is 42.1--98.9\% over the 12 coronary networks, and background is at most 12.0\%. CAS-Net has the highest seen share and also the highest background (Sec.~\ref{S-sec:supp-J-casnet}). The original ImageCAS labels draw a different tree. Against them the seen share is lower and still above background for every network. The six released coronary networks are the models published with the ImageCAS-X benchmark \cite{bransby2026imagecasx}.
\end{details}

\begin{table*}[t]
\caption{\textbf{Seen share of the missed distal tree for every network and reference.} Rows are networks, grouped by reference. Columns give the patient mean (\%) with 95\% patient-bootstrap interval. Missed: share of the distal reference centreline (20\,mm window) not reached at 0.5; seen: share of the missed centreline with \(p\geq0.01\) within 1\,mm; background: the same test at distance-matched points away from the reference. \(n\): patients with a missed sample.}
\label{S-tab:supp-seen}
\centering\scriptsize
\setlength{\tabcolsep}{3pt}
\begin{tabular}{@{}lcccc@{}}
\toprule
Network & \(n\) & Missed & Seen (\(p\geq0.01\)) & Background (\(p\geq0.01\)) \\
\midrule
\rowcolor{subhead}\multicolumn{5}{@{}l}{\textit{Coronary CTA, ImageCAS-X labels (160 test patients)}} \\
\hspace{0.5em}nnU-Net Dice+CE, seed 1 & 155 & 22.7 [20.3, 25.2] & 59.4 [54.9, 63.7] & 6.7 [5.1, 8.6] \\
\hspace{0.5em}nnU-Net Dice+CE, seed 2 & 151 & 23.0 [20.6, 25.4] & 61.1 [57.0, 65.2] & 6.1 [4.5, 7.9] \\
\hspace{0.5em}nnU-Net Dice+CE, seed 3 & 151 & 22.3 [19.9, 24.7] & 59.3 [54.9, 63.7] & 6.0 [4.5, 7.6] \\
\hspace{0.5em}cbDice & 158 & 33.3 [30.6, 36.1] & 73.1 [69.9, 76.2] & 5.3 [4.1, 6.7] \\
\hspace{0.5em}clDice & 145 & 17.5 [15.4, 19.8] & 77.0 [72.7, 81.0] & 7.7 [5.9, 9.6] \\
\hspace{0.5em}Skeleton Recall & 116 & 9.0 [7.5, 10.6] & 59.1 [52.8, 65.4] & 8.8 [6.4, 11.8] \\
\hspace{0.5em}released nnU-Net & 155 & 24.9 [22.4, 27.4] & 64.3 [59.9, 68.5] & 5.3 [4.0, 6.7] \\
\hspace{0.5em}CAS-Net & 153 & 27.8 [25.2, 30.5] & 98.9 [98.0, 99.6] & 12.0 [10.3, 14.0] \\
\hspace{0.5em}3D-FFR-UNet & 158 & 42.0 [38.6, 45.5] & 56.7 [52.8, 60.7] & 4.3 [3.3, 5.5] \\
\hspace{0.5em}Swin-UNETR & 157 & 32.4 [29.6, 35.1] & 65.3 [61.8, 69.0] & 5.2 [3.8, 6.8] \\
\hspace{0.5em}ImageCAS coarse-to-fine & 156 & 34.7 [32.1, 37.4] & 79.7 [76.5, 82.7] & 6.2 [4.9, 7.8] \\
\hspace{0.5em}ADE-HTL & 156 & 28.1 [25.5, 30.8] & 42.1 [37.7, 46.3] & 7.5 [5.8, 9.4] \\
\midrule
\rowcolor{subhead}\multicolumn{5}{@{}l}{\textit{Coronary CTA, original ImageCAS labels (same CTs)}} \\
\hspace{0.5em}nnU-Net Dice+CE, seed 1 & 144 & 40.0 [36.8, 43.3] & 23.0 [19.2, 27.0] & 5.3 [3.8, 7.0] \\
\hspace{0.5em}nnU-Net Dice+CE, seed 2 & 144 & 39.9 [36.6, 43.2] & 22.8 [19.0, 26.8] & 5.1 [3.6, 6.9] \\
\hspace{0.5em}nnU-Net Dice+CE, seed 3 & 145 & 39.3 [36.1, 42.5] & 23.4 [19.5, 27.5] & 5.2 [3.7, 7.0] \\
\hspace{0.5em}cbDice & 145 & 45.4 [42.2, 48.4] & 33.3 [29.2, 37.5] & 4.6 [3.4, 6.2] \\
\hspace{0.5em}clDice & 143 & 36.9 [33.9, 40.1] & 27.5 [23.4, 31.9] & 5.7 [4.1, 7.6] \\
\hspace{0.5em}Skeleton Recall & 139 & 32.5 [29.5, 35.6] & 17.4 [14.0, 21.0] & 6.1 [4.2, 8.3] \\
\hspace{0.5em}released nnU-Net & 143 & 41.8 [38.5, 45.1] & 27.0 [23.3, 31.1] & 4.7 [3.3, 6.3] \\
\hspace{0.5em}CAS-Net & 147 & 44.0 [40.9, 47.2] & 59.7 [54.9, 64.6] & 10.8 [8.9, 12.9] \\
\hspace{0.5em}3D-FFR-UNet & 146 & 51.7 [48.0, 55.5] & 33.9 [29.8, 38.1] & 6.3 [4.5, 8.4] \\
\hspace{0.5em}Swin-UNETR & 145 & 46.2 [42.7, 49.8] & 32.4 [28.5, 36.5] & 6.4 [4.6, 8.5] \\
\hspace{0.5em}ImageCAS coarse-to-fine & 147 & 48.3 [45.3, 51.3] & 41.2 [37.1, 45.5] & 5.5 [4.3, 6.9] \\
\hspace{0.5em}ADE-HTL & 145 & 44.0 [40.9, 47.1] & 21.3 [17.5, 25.3] & 5.7 [4.1, 7.6] \\
\midrule
\rowcolor{subhead}\multicolumn{5}{@{}l}{\textit{Coronary CTA, ASOCA (40 cases)}} \\
\hspace{0.5em}nnU-Net Dice+CE, seed 1 & 40 & 21.4 [16.7, 26.6] & 36.6 [27.2, 46.4] & 6.9 [4.4, 9.8] \\
\hspace{0.5em}nnU-Net Dice+CE, seed 2 & 38 & 22.6 [17.6, 27.9] & 40.7 [31.2, 50.5] & 6.5 [4.3, 9.1] \\
\hspace{0.5em}nnU-Net Dice+CE, seed 3 & 40 & 25.9 [20.3, 32.2] & 41.5 [32.1, 51.2] & 6.7 [4.4, 9.4] \\
\hspace{0.5em}cbDice & 40 & 27.3 [21.7, 33.3] & 47.5 [38.0, 57.2] & 6.4 [4.3, 8.8] \\
\hspace{0.5em}clDice & 38 & 20.9 [16.3, 25.9] & 39.1 [29.6, 49.1] & 9.9 [5.8, 15.0] \\
\hspace{0.5em}Skeleton Recall & 39 & 18.4 [13.4, 24.1] & 35.8 [25.2, 46.7] & 9.2 [4.6, 15.4] \\
\hspace{0.5em}released nnU-Net & 39 & 26.4 [20.4, 33.0] & 45.2 [35.2, 55.0] & 7.8 [4.8, 11.5] \\
\midrule
\rowcolor{subhead}\multicolumn{5}{@{}l}{\textit{Pulmonary CT, HiPaS (60 test patients)}} \\
\hspace{0.5em}nnU-Net & 49 & 20.2 [16.7, 24.1] & 57.8 [53.9, 61.9] & 13.7 [12.1, 15.5] \\
\hspace{0.5em}Skeleton Recall & 49 & 9.2 [7.1, 11.5] & 46.8 [42.3, 51.3] & 17.5 [15.3, 20.1] \\
\hspace{0.5em}mask-prompted SR & 49 & 9.6 [7.4, 12.0] & 35.9 [32.4, 39.5] & 15.4 [13.5, 17.9] \\
\bottomrule
\end{tabular}
\end{table*}

\subsection{What a lower threshold adds}
\label{S-sec:supp-S1-lengthen}

The introduction says that a lower threshold recovers branches inefficiently, because one threshold sets both the length and the width of the tree (Sec.~\ref{sec:related}). We measure how the volume that a lower threshold adds divides between the two. On a fixed subset of 16 coronary validation patients, we lower the rooted 0.5 mask of each network to the lowest 8-bit level, 1/255, and split the added voxels into two kinds. A voxel lengthens the tree when it lies within 1\,mm of the reference and its nearest reference sample becomes newly connected to the root. Every other added voxel thickens the tree. We average per patient over four losses: Dice+CE and Skeleton Recall have three seeds each, and clDice and cbDice have one. The lengthening voxels are a small part of what the threshold adds, 3.6\% (Skeleton Recall) to 16.2\% (cbDice) of the added volume, yet they carry 74.1\% to 99.8\% of the distal gain in \(R\). Most of the added volume therefore only thickens the tree. RIDGER avoids this by measuring the width of each new branch separately (Sec.~\ref{sec:width}).

\section{Method details and constants}
\label{S-sec:supp-A_method}

This section specifies RIDGER step by step, lists its constants and where each comes from, and gives its runtime and code.

RIDGER reads the probability map alone. It has three inputs: the network's probability field \(p\), the voxel spacing, and the root-connected default mask \(S\). The field \(p\) is stored as float16 log odds \(\ell\). It reads no image intensity and no reference. The two anatomies run the same code and differ only in \(\lambda\), the anchor and the voxel spacing (Sec.~\ref{S-sec:supp-A-anatomy}).

\subsection{Step-by-step specification}
\label{S-sec:supp-A-steps}

RIDGER runs the anchor, the map normalisation and steps 1 to 8 in this order. The Details block after the list gives the thresholds, windows and tie rules of every step under the same labels.

\textbf{Anchor.} \(S\) is the 0.5 mask restricted to the components that touch the inference anchor: the largest predicted component on pulmonary CT, the predicted aortic neighbourhood in coronary CTA. \(S\) is never changed, so \(P\supseteq S\) always. No reference label enters.

\textbf{Map normalisation.} This step divides the log odds by a temperature \(T\) so that the share of voxels outside \(S\) above the candidate floor stays within a bound. Every later step reads \(\ell/T\). \(S\) is formed before this step and does not depend on \(T\).

\textbf{1. Candidates.} The candidate set is every voxel outside \(S\) with \(p\geq10^{-5}\), together with \(S\). The floor \(10^{-5}\) is the lowest ridge that RIDGER reads beyond \(S\).

\textbf{2. Skeleton and shortest-path tree.} The candidate set is thinned to its Lee skeleton \cite{lee1994thinning}, and a shortest-path tree rooted at \(S\) is built on all candidate voxels. Each skeleton voxel keeps its path to \(S\), so each skeleton branch is joined to \(S\) along the ridge and never through a straight chord.

\textbf{3. Segments.} The tree is cut into segments \(b\) at its bifurcations and where a path leaves \(S\). Every segment has one parent segment, or \(S\), so the segments form the tree on which step 5 selects.

\textbf{4. Segment statistics.} Each segment receives the three quantities of the selection objective. The gain \(L_b\) is the centreline length, in millimetres, by which the segment advances beyond \(S\). The confidence \(q_b\) is the median ridge contrast of its voxels, and the volume \(\mathrm{Vol}_b\) is what the segment renders outside \(S\).

\textbf{5. Exact selection.} The utility of segment \(b\) is \(u_b=q_bL_b-\lambda(1-q_b)\mathrm{Vol}_b\). One pass from the leaves to the root and one pass back find the best rooted subtree (Eqs.~\eqref{eq:objective} and~\eqref{eq:dp}). The two passes touch each segment twice, are exact, and have no tuning parameter besides \(\lambda\).

\textbf{6. Width and rendering.} The radius at each tree voxel is read from the map at half height, as the median over eight rays in the plane normal to the tree. A kept segment is rendered as the union of the balls of these radii around its tree voxels. The union of the kept segments with \(S\) is reduced to the part 26-connected to \(S\).

\textbf{7. Loop veto.} The additions of the rendered union are applied to \(S\) in a fixed order, each tested on the voxel mask before it is accepted. An addition is accepted when it does not raise the loop count \(\beta_1\) of the mask (Eq.~\eqref{eq:veto}).

\textbf{8. Face-to-face closure.} Two foreground voxels may touch through an edge or a corner without a face path between them. When at least one of the two was added by RIDGER, the fewest voxels that complete a face path bridge them. \(S\) is never changed. Every closure voxel is counted in \(V\). The loop count is not tested again after closure.

\begin{details}
\textbf{Anchor.} In coronary CTA the anchor keeps the 0.5 components that come within 5\,mm of an aorta mask predicted from the CT by TotalSegmentator~\cite{wasserthal2023totalsegmentator}, and falls back to the whole 0.5 mask when none does.

\textbf{Map normalisation.} For each temperature \(T\) of the grid \(\{1,\tfrac12,\tfrac14,\tfrac18\}\), \(H_T\) is computed on the full stored grid. It is the share of the voxels outside \(S\) with \(\ell\geq T\ln\frac{10^{-5}}{1-10^{-5}}\). This is the condition \(\sigma(\ell/T)\geq10^{-5}\). The cutoff follows the float16 convention of step 1. RIDGER takes the largest \(T\) of the grid with \(H_T\leq h_0=28.88\%\). When no \(T\) of the grid qualifies, RIDGER takes \(T=\tfrac18\) and reports the patient. This did not occur on any map of this paper. \(\ell/T\) is computed in float32 and stored as float16. \(S\) is formed from \(\ell>0\). For \(T=1\) the map is unchanged.

\textbf{1. Candidates.} The cutoff is applied to the float16 log odds. It is the smallest float16 value not below \(\ln\frac{10^{-5}}{1-10^{-5}}\approx-11.51\), so no stored value is rounded across it. All work runs in the bounding box of the candidates with a margin of 3.1\,mm, and at least five voxels, on each side.

\textbf{2. Skeleton and shortest-path tree.} The shortest-path tree takes the 26 voxel neighbours as arcs. The cost of an arc is its Euclidean length in millimetres times the mean of the two endpoint costs \(\max(0,-\ell)^2\), and a voxel adjacent to \(S\) starts a path with the cost of the step from \(S\). Ties are broken by voxel index. Every voxel on the path from a skeleton voxel to \(S\) is marked as part of the tree, and the tree is the set of marked voxels. Spurs are not trimmed separately: the selection of step 5 removes any spur that does not pay for itself.

\textbf{3. Segments.} A bifurcation belongs to its incoming segment, and each of its outgoing branches starts a new segment. A path that leaves \(S\) also starts a new segment.

\textbf{4. Segment statistics.} Each tree voxel \(v\) has a path edge to its ancestor, of physical length \(e_v\) (for a root voxel, the distance to the nearest surface voxel of \(S\)).
\begin{itemize}[leftmargin=*,itemsep=1pt,topsep=2pt]
\item \emph{Gain \(L_b\).} A point \emph{advances} in two cases. In the first, it lies more than 1.5\,mm from the surface of \(S\). In the second, its nearest skeleton voxel of \(S\) is a skeleton tip, and the point lies ahead of the tip, on the outward side of the tip's direction. The direction of a tip is taken from the 1.5\,mm of the skeleton of \(S\) behind it. An edge counts half its length for each of its two ends that advances. \(L_b\) is the sum of these lengths over the segment's edges, in millimetres.
\item \emph{Confidence \(q_b\).} For a tree voxel \(v\), the ridge contrast is \(\rho_v=\sigma(\ell_v-\bar\ell_v)\), where \(\bar\ell_v\) is the mean log odds over the hollow shell around \(v\). The shell is the cube of \(9\times9\times9\) voxels centred on \(v\) without its central \(3\times3\times3\) cube, 702 voxels, with the map reflected at its border. Because the shell is a voxel neighbourhood, its physical extent follows the spacing. \(q_b\) is the median of \(\rho_v\) over the voxels of the segment.
\item \emph{Volume \(\mathrm{Vol}_b\).} The volume the segment renders (step 6) outside \(S\), in \(\text{mm}^3\). The selection uses this measured-width volume and excludes closure voxels.
\end{itemize}

\textbf{5. Exact selection.} The best rooted subtree is found as follows.
\begin{enumerate}[leftmargin=*,itemsep=1pt,topsep=2pt]
\item Set \(g(b)\gets u_b\) for every segment.
\item Visit segments from the leaves towards the root (children before parents). For each segment \(c\) with parent segment \(b\), if \(g(c)>0\), add \(g(c)\) to \(g(b)\).
\item Visit segments from the root outward. Keep a segment if \(g(b)>0\) and its parent segment is kept (or its parent is \(S\)).
\item The kept set is a rooted subtree whose total \(u\) equals the sum of \(\max(0,g)\) over the segments whose parent is \(S\). We check this equality on every patient.
\end{enumerate}

\textbf{6. Width and rendering.} At each tree voxel \(v\), the tangent is the direction between the points 1.5\,mm before and after \(v\) along the tree. Eight rays leave \(v\) in the plane normal to the tangent, at \(45^\circ\) steps, and sample \(p\) every 0.1\,mm out to 3\,mm by trilinear interpolation. Per ray, the background \(f\) is the lowest sampled value, and the edge is the first sample below \(f+(p_v-f)/2\), located by linear interpolation between its two neighbours. A ray with no crossing within 3\,mm takes the radius 3\,mm. The radius of \(v\) is the median over the eight rays. The voxel set of \(v\) is the ball of that radius around \(v\), excluding \(S\). The ball is clipped to voxels whose probability is at least the half-height level in their direction. This level uses the per-ray backgrounds, interpolated linearly around the circle. The voxels of a kept segment are the union over its tree voxels. The union of the kept segments with \(S\) is reduced to the part 26-connected to \(S\).

\textbf{7. Loop veto.} The test of one addition comes first, and the order of the additions follows. For an addition, take the mask in a box that extends one voxel beyond it (the collar). The addition must leave the 6-connected partition of the background in the collar unchanged, which preserves the global second Betti number. Let \(m\) be the reduction in the number of 26-connected foreground components, counted on the labels of the whole current mask, and \(\Delta\chi_{26}\) the change of the 26-connected Euler number in the collar. The addition is accepted if \(\Delta\beta_1=-m-\Delta\chi_{26}\leq0\) (Eq.~\eqref{eq:veto}). An addition with no voxel connected to the current mask is rejected. Each connected component of the rendered additions is tried whole first, in order of decreasing highest log odds. A rejected component is split into single voxels, tried in order of decreasing log odds. The voxels that remain are retried in rounds, again by decreasing log odds, until a round accepts none.

\textbf{8. Face-to-face closure.} Let \(M\) be the mask after step 7 and \(A=M\setminus S\). Consider every pair of foreground voxels that meets three conditions. The two voxels are 26-adjacent but not face-adjacent (adjacent through an edge or a corner). At least one of them lies in \(A\). They are not face-connected inside the \(2\times2\) or \(2\times2\times2\) box they span. Each such pair is bridged by the fewest voxels that complete a face path: one voxel for an edge pair, one or two for a corner pair. Among equally short bridges, the one whose weakest voxel has the highest probability wins, then the one with the highest summed probability, then the lowest voxel index. Pairs are processed in order of voxel index and neighbour offset, and a pair already joined by an earlier bridge is skipped. The pass repeats until no pair remains, and we check that the result contains none. Every closure voxel is added to \(A\).

\textbf{Differences between the two anatomies.}\phantomsection\label{S-sec:supp-A-anatomy} The two anatomies share one code path, the candidate threshold \(10^{-5}\) and the same advance test, shell, ray and veto settings. Coronary CTA takes \(\lambda=300\,\text{mm}^{-2}\), which is not at either end of its grid. Pulmonary CT takes \(\lambda=30\,\text{mm}^{-2}\). Each value is shared by the two networks of its anatomy and is applied unchanged to ASOCA.

\textbf{Choice of \(\lambda\) and the renderer.} \(\lambda\) and the renderer were chosen once, on validation patients, by one criterion that we call the balanced gain. Validation uses 80 coronary patients per network and the pulmonary patients of Sec.~\ref{S-sec:supp-C-splits}. For one network, we take the patient-mean gain over the best 8-bit threshold and the patient-mean gain over the float16 control, both at equal off-reference volume, and average the two. The balanced gain is the mean of this average over the two networks of the anatomy. For each renderer (the measured width of step 6, or a thin rendering of fixed radius) and each anatomy, we took the grid value of \(\lambda\) with the largest balanced gain. Ties go to the smaller mean \(V\), then to the smaller \(\lambda\). The measured width has the larger mean over the four networks of the two anatomies (7.66 against 7.49 points), so RIDGER renders with it. The coronary grid is \(\{0.03,0.1,0.3,1,3,10,30,100,300,1000,3000\}\)\,mm\(^{-2}\). The pulmonary grid is its upper part, \(\{3,10,30,100,300,1000,3000\}\)\,mm\(^{-2}\).
\end{details}

\subsection{Constants and their origin}
\label{S-sec:supp-A-constants}

Table~\ref{S-tab:supp-A-constants} lists every constant with its origin. \(\lambda\), the renderer and the candidate threshold are chosen on validation patients, and the normalisation bound \(h_0\) is computed from validation maps. Every other constant is a design value fixed before validation and never tuned, or a measurement read from the map by the stated rule. Section~\ref{S-sec:supp-F_ablation} varies \(\lambda\) and the floor. When \(\lambda\) is tripled or divided by three, the lead over the float16 control stays above 2.4 points (Sec.~\ref{sec:reach}).
\begin{table*}[t]
\caption{\textbf{Constants of RIDGER.} Coronary and pulmonary values are equal unless two are given. ``Validation'' denotes selection on validation patients, ``default'' a fixed design value, ``measurement'' a value read from the map.}
\label{S-tab:supp-A-constants}
\centering\scriptsize
\setlength{\tabcolsep}{3pt}
\begin{tabular}{@{}llp{5.9cm}l@{}}
\toprule
Constant & Value (coronary / pulmonary) & Role & Origin \\
\midrule
Anchor distance & 5\,mm to the predicted aorta (coronary) & which 0.5 components form \(S\) & default \\
Normalisation grid & \(T\in\{1,\tfrac12,\tfrac14,\tfrac18\}\) & temperatures of the map normalisation, largest first & default \\
Normalisation bound & \(h_0=28.88\%\) & largest admitted share \(H_T\) of voxels outside \(S\) above the floor; twice the largest validation \(H_1\) (14.44\%) & validation \\
Candidate floor & \(p\geq10^{-5}\), float16 log odds \(\geq-11.51\) & lowest ridge read beyond \(S\) & validation \\
Arc set and cost & 26 neighbours; \(\max(0,-\ell)^2\) per mm & shortest-path tree & default \\
Skeleton & Lee thinning & candidate branches & default \\
Advance distance & 1.5\,mm from \(S\); ahead of a skeleton tip & which centreline length counts as advancing in the selection (\(L_b\)) & default \\
Tip direction & 1.5\,mm of skeleton behind the tip & ``ahead'' test & default \\
Shell & \(9^3\) voxels minus central \(3^3\) (702), mean & background of the ridge contrast & default \\
Confidence \(q_b\) & median of \(\sigma(\ell_v-\bar\ell_v)\), no calibration & weights \(L_b\) and the volume term of the selection objective & default \\
Weight \(\lambda\) & 300 / 30\,mm\(^{-2}\) & rendered volume to centreline length & validation \\
Tangent & points 1.5\,mm either side along the tree & plane of the width rays & default \\
Width rays & 8 rays, 0.1\,mm steps, to 3\,mm, trilinear & radius by half height & measurement \\
Background & lowest value within 3\,mm per ray & half-height level & measurement \\
Radius bound & 3\,mm if no crossing; median over rays & safety bound & default \\
Renderer & measured width (step 6) & final mask & validation \\
Loop veto & collar of one voxel; 6-connected background; \(\Delta\chi_{26}\) & no addition raises \(\beta_1\) & default \\
Closure & face paths of 1 or 2 voxels; no free value & 26- to face-connected & default \\
Candidate box margin & 3.1\,mm, at least 5 voxels & crop only & default \\
\bottomrule
\end{tabular}
\end{table*}

\subsection{Runtime and memory}
\label{S-sec:supp-A-runtime}

RIDGER processes the map of a coronary CT in seconds and that of a pulmonary CT in minutes, on one CPU thread. Table~\ref{S-tab:supp-A-runtime} gives the time of each stage and the peak memory, as means over the test patients of each network; scoring is not included.
\begin{table*}[t]
\caption{\textbf{Runtime and peak memory.} Means over test patients, one CPU thread, scoring excluded. Candidates include the tree and the measured widths.}
\label{S-tab:supp-A-runtime}
\centering\scriptsize
\setlength{\tabcolsep}{3pt}
\begin{tabular}{@{}lrrrrr@{}}
\toprule
Network & \(n\) & Candidates (s) & Selection (s) & Closure (s) & Peak (GB) \\
\midrule
Coronary, both networks & 320 & 4.4 & 0.0004 & 2.0 & 1.6 \\
Pulmonary, nnU-Net & 59 & 351 & 0.14 & 10.3 & 5.3 \\
Pulmonary, mask-prompted SR & 60 & 271 & 0.11 & 6.2 & 5.9 \\
\bottomrule
\end{tabular}
\end{table*}

\begin{details}
Candidates with their widths take most of the time: 4.5 to 5.9\,min per pulmonary CT. The exact selection takes at most 0.31\,s on any pulmonary patient, so the exactness of Proposition 1 adds no measurable time.
\end{details}

\subsection{Code}
\label{S-sec:supp-A-code}

The code holds the readout, the scoring of \(R\) and the off-reference volume \(V\), the 8-bit envelope, the script that selects \(\lambda\) on validation patients from the shipped validation table, a synthetic demo, and tests that reproduce the archived test masks and scores voxel for voxel when pointed at the public data. It ships no network weights or images, and it will be made publicly available.

\section{Proofs}
\label{S-sec:supp-B_proofs}

This section proves the remark and the three propositions of Sec.~\ref{sec:method} and reports how closely the final masks follow the selections that the propositions cover.

We use the notation of Sec.~\ref{sec:method} and Sec.~\ref{S-sec:supp-A_method} without change: $S$ is the root-connected default mask, $p$ the probability field with log odds $\ell$, segments $b$ form a tree whose roots have parent $S$, $L_b$ the advancing centreline length of segment $b$, $\mathrm{Vol}_b$ its rendered volume, $q_b$ its confidence, $u_b=q_bL_b-\lambda(1-q_b)\mathrm{Vol}_b$ its utility, and $\mathcal T^\ast(\lambda)$ the rooted subtree selected by \eqref{eq:objective}.

\subsection{Thresholds as rooted subtrees}

\textbf{Remark (thresholds on the tree are rooted subtrees).} For any probability level $t$, keeping every segment whose path to $S$ stays at $p\geq t$ yields a rooted subtree.

\begin{proof}
Let $b$ be a segment whose path to $S$ stays at $p\geq t$. The path from $b$ to $S$ runs through its parent segment $\mathrm{par}(b)$: by construction of the segment tree (Sec.~\ref{sec:ridges}; step~3 of Sec.~\ref{S-sec:supp-A-steps}), each segment has exactly one parent, and the path to $S$ visits the parent before reaching $S$. Every voxel on the path from $b$ to $S$ lies on the path from $\mathrm{par}(b)$ to $S$ or on $b$ itself. Since the path from $b$ stays at $p\geq t$, so does the sub-path from $\mathrm{par}(b)$. Thus $\mathrm{par}(b)$ meets the same condition. A set that contains the parent of every segment it contains is a rooted subtree (Sec.~\ref{sec:ridges}), so the threshold rule returns a rooted subtree.
\end{proof}

The threshold rule on the segment tree is therefore one point of RIDGER's search space. RIDGER's optimum of \eqref{eq:objective} is at least the value of \eqref{eq:objective} at the threshold rule. Tab.~\ref{tab:ablation}b measures the same comparison in \(R\): the threshold on the same ridge tree falls 1.5 points behind on coronary CTA and 8.2 points on pulmonary CT.

\subsection{Exactness and linear time}

\textbf{Proposition 1 (exactness).} The pass \eqref{eq:dp} returns a maximiser of \eqref{eq:objective} in time linear in the number of segments.

\begin{proof}
We show by induction from the leaves that $g(b)$ equals the maximum total utility of a rooted subtree hanging from $b$.

\emph{Base.} If $b$ is a leaf, $g(b)=u_b$ by \eqref{eq:dp} with no children. The only rooted subtree hanging from $b$ is $\{b\}$, whose utility is $u_b$.

\emph{Induction step.} Let $b$ have children $c_1,\dots,c_r$. By the induction hypothesis, $g(c_j)$ is the best value of a rooted subtree hanging from $c_j$. A rooted subtree hanging from $b$ includes $b$ and, for each child $c_j$, either no segment of the branch below $c_j$ or one rooted subtree hanging from $c_j$. Because subtrees hanging from distinct children are disjoint, the total utility is $u_b+\sum_j v_j$ where $v_j$ is either $0$ or the value of a rooted subtree hanging from $c_j$. The best such value is $\max(0,g(c_j))$, because joining is worthwhile exactly when $g(c_j)>0$. Therefore $g(b)=u_b+\sum_j\max(0,g(c_j))$, which is \eqref{eq:dp}.

\emph{Maximiser.} The second pass (step~5 of Sec.~\ref{S-sec:supp-A-steps}) keeps a segment when $g(b)>0$ and its parent is kept or is $S$. The kept set is exactly the set of segments with $g>0$ along their path to $S$. Its total utility equals $\sum_{b:\,\mathrm{par}(b)=S}\max(0,g(b))$, the maximum of \eqref{eq:objective}.

\emph{Linear time.} The pass touches each segment twice (leaf-to-root, root-to-leaf). Each touch does constant work per child. With $n$ segments the total work is $O(n)$.
\end{proof}

\subsection{Nestedness in the weight}
\label{S-sec:supp-B-nested}

\textbf{Proposition 2 (nested selections).} If $\lambda\leq\lambda'$, then $\mathcal T^\ast(\lambda')\subseteq\mathcal T^\ast(\lambda)$.

\begin{proof}
Each $u_b=q_bL_b-\lambda(1-q_b)\mathrm{Vol}_b$ is non-increasing in $\lambda$ because $(1-q_b)\mathrm{Vol}_b\geq0$. By \eqref{eq:dp}, $g(b)$ is built from $u$ values and $\max(0,\cdot)$ along the subtree hanging from $b$; since each $u$ is non-increasing and $\max(0,\cdot)$ is monotone, every $g(b)$ is non-increasing in $\lambda$. A segment is kept when $g>0$ along its path to $S$. As $\lambda$ grows, the set of segments with $g(b)>0$ can only shrink. Thus $\mathcal T^\ast(\lambda')\subseteq\mathcal T^\ast(\lambda)$ when $\lambda\leq\lambda'$.
\end{proof}

Like thresholds, RIDGER's selections form a nested family indexed by one number.

\subsection{From selections to masks}
\label{S-sec:supp-B-masks}

Propositions 1 and 2 cover the selected segments, and the mask is drawn from them. Three later steps change the mask: rendered tubes of neighbouring segments overlap, the loop veto may refuse voxels, and the closure adds voxels. We measure their effect on the volume and on nestedness. The selection objective counts more volume than the mask adds on coronary maps, and about as much on pulmonary CT. The selections are nested, and the masks are nearly nested.

The table below has one row per network. Its volume columns are test-patient means, in voxels, of the volume added outside \(S\). Its first column, \(\Sigma\,\mathrm{Vol}_b\), is the summed volume of the selected segments, which is the volume that the selection objective counts. Overlap is the part of this sum that neighbouring tubes share, Veto is the volume that the loop veto refuses, Closure is the volume that the closure adds, and \(|P\setminus S|\) is the volume that the final mask adds. The last column, Gap, is the mean over patients of \((\Sigma\,\mathrm{Vol}_b-|P\setminus S|)/|P\setminus S|\), with 95\% patient-bootstrap intervals.

The objective's volume differs from the mask's, but every comparison of the main text holds the mask's \(V\) fixed.

\begin{center}
\scriptsize
\setlength{\tabcolsep}{2.5pt}
\begin{tabular}{lrrrrrr}
\toprule
Network & \(\Sigma\,\mathrm{Vol}_b\) & Overlap & Veto & Closure & \(|P\setminus S|\) & Gap (\%) \\
\midrule
Coronary, Dice+CE & 3065 & 1928 & 0.9 & 49 & 1185 & 66 [56, 77] \\
Coronary, released & 3177 & 1882 & 1.4 & 48 & 1341 & 78 [68, 88] \\
Pulmonary, nnU-Net & 20344 & 5332 & 10.3 & 2685 & 17686 & 16 [9, 24] \\
Pulmonary, m.-p.\ SR & 10162 & 1710 & 3.9 & 2267 & 10715 & \(-4\) [\(-10\), 3] \\
\bottomrule
\end{tabular}
\end{center}

\begin{details}
Lists of values follow the order coronary Dice+CE, released coronary nnU-Net, pulmonary nnU-Net, mask-prompted SR.

\textbf{Volume.} On coronary maps, overlap makes up 59 to 63\% of the summed segment volume. Averaged over patients, the selection objective therefore counts 66 and 78\% more voxels than the mask adds, and both intervals exclude zero. On pulmonary CT, overlap (17 to 26\%) and the closure (15 to 21\% of the final addition) nearly cancel. The gap is +16\% and \(-4\)\%, and the second interval includes zero. The loop veto removes about 10 voxels per patient at most (means of 0.9 to 10.3), far less than any other column. \(\lambda\) is chosen on validation patients with this objective in place.

\textbf{Nestedness.} We compare every pair of consecutive values on RIDGER's \(\lambda\) grid, on validation patients. The selections are nested at all 800, 800, 42 and 42 such pairs, as Proposition 2 states. A mask departs from nestedness when it holds voxels at the larger \(\lambda\) that are missing at the smaller \(\lambda\). On coronary maps, 25.5\% and 38.4\% of the pairs have such voxels. The median is 2 voxels (0.2\% of the added volume, at most 2.4\%). On pulmonary CT, every pair has such voxels, with a median of 112 and 55 (1.7 and 1.4\%, at most 3.9\%). The departures are frequent but small.
\end{details}

\subsection{Reaching rate and loops}

\textbf{Proposition 3 (the reaching rate does not fall; additions that pass the veto add no loop).} $R(P)\geq R(S)$, and no rendered addition that passes \eqref{eq:veto} increases the loop count of the mask.

\begin{proof}
\emph{Reaching rate.} $P\supseteq S$ by construction (\eqref{eq:ridger}). Adding voxels never disconnects a covered sample from the root, so $C_i(P)\geq C_i(S)$ for every sample $i$, and $R(P)\geq R(S)$.

\emph{Loops.} Let $a$ be an addition accepted by \eqref{eq:veto}, and $K$ its collar, the box one voxel beyond it. (i) The Euler number is a sum of local cell counts, and every cell that changes lies in $K$, so $\Delta\chi_{26}$ measured in $K$ is the global change. (ii) A background component can only be split, created or removed where background voxels are removed, that is inside $K$. If the 6-connected partition of the background in $K$ is unchanged, no global background component changes, and $\Delta\beta_2=0$. The test is sufficient, not necessary: it may reject an addition that was safe. (iii) $m$ is counted on the component labels of the whole mask, so $\Delta\beta_0=-m$ globally. The Euler--Poincar\'e relation $\Delta\chi=\Delta\beta_0-\Delta\beta_1+\Delta\beta_2$ then gives $\Delta\beta_1=-m-\Delta\chi_{26}$, and the veto $\Delta\beta_1\leq0$ says the loop count does not rise.
\end{proof}

The closure of Sec.~\ref{sec:width} runs after the veto and is not retested, so the guarantee stops at the veto. The closure raises the coronary loop error from 0.06 to 0.09 (Dice+CE) and from 0.10 to 0.17 (released nnU-Net). Tab.~\ref{tab:ablation}a reports the loop error with and without the closure.

\section{Measures and comparators}
\label{S-sec:supp-D_comparators}

This section defines the scoring of Sec.~\ref{sec:measurement} and every comparator of Tab.~\ref{tab:main-coronary} exactly as it was run.

Every comparator reads the same saved map as RIDGER and is held to \emph{RIDGER's volume} on each patient. A comparator that produces a family of masks gives each patient its best admissible member. Where a comparator has a setting to choose, the setting is chosen on validation patients under the same per-patient volume limit. The four primary pairs of anatomy and network are coronary CTA with nnU-Net Dice+CE (\(n=160\)), coronary CTA with the released nnU-Net (\(n=160\)), pulmonary CT with nnU-Net (\(n=49\)) and pulmonary CT with the mask-prompted SR (\(n=50\)). Lists of four values follow this order.

The volume limit is applied per patient: for patient \(i\), a member is admissible only if its integer off-reference voxel count does not exceed that of RIDGER on the same patient. Throughout, \(S\) is the default mask \(p\geq0.5\) restricted to its root-connected part and \(G\) is the reference. The off-reference volume \(V\) is reported in \% of the reference volume as in the main text, and it is compared on the voxel counts of each patient.

\subsection{Scoring window and tolerance}
\label{S-sec:supp-D-window}

The reaching rate \(R\) is scored on the distal part of the reference, with a fixed tolerance. The window is the reference centreline within 20\,mm upstream of each terminal. A centreline sample of the window counts as connected only through predicted voxels within 1\,mm of the reference. Both values, and the choice of the windowed \(R\) as the primary endpoint, were fixed in the evaluation code before RIDGER was designed. The gain stays positive for windows from 10\,mm to the whole tree and for tolerances from 0.5 to 2\,mm (Sec.~\ref{S-sec:supp-F-window}).

\subsection{Composition of the off-reference volume}
\label{S-sec:supp-F-V}
The off-reference volume \(V\) counts every predicted voxel outside the reference once. It therefore counts three different things the same way: the default mask's own disagreement with the reference, widening next to the default mask, and added branches outside the reference, whether true or false. When a reference stops early, an added branch adds to \(V\) even if it is a vessel. On pulmonary CT, most of RIDGER's \(V\) is inherited from the default mask, and its added branches are almost all vessel. The split pools the voxels of the 49 scored pulmonary test patients (nnU-Net). Of RIDGER's \(V\), 69.1\% is the default mask's own, so RIDGER's changes account for less than a third. Widening next to the default mask is 3.7\%, and added branches outside the reference are 27.2\%. Readers judge 97.9\% of those branches to be vessels (Sec.~\ref{S-sec:supp-I_readers}). Reading \(V\) as error is therefore conservative for RIDGER, because the additions that readers judge to be vessels still count in it.

\subsection{The 8-bit threshold family and its envelope}
\label{S-sec:supp-D-u8}

The 8-bit family contains every mask that a single global threshold of the stored map produces, one member per stored level and 255 per patient. Each patient takes the member with the largest \(R\) among those that fit RIDGER's volume on that patient. This best member is therefore the best threshold that the patient can be given at that volume. The envelope of Eq.~\eqref{eq:envelope} is this selection evaluated at every volume. The equal-volume gain is \(R(P)-\mathcal E_8(V(P))\) per patient, averaged over patients.

\begin{details}
\textbf{Members.} The network stores its probability map as 8-bit codes \(p_8=\operatorname{rint}(255\,p)\). For each level \(k=1,\dots,255\), the voxels with \(p_8\geq k\) are exactly those with \(p\geq(k-\tfrac12)/255\). Level 1 therefore keeps \(p\geq 0.00196\), level 128 keeps \(p\geq 0.50\) and level 255 keeps \(p\geq 0.998\). The member \(\Theta_S(k)\) is the union of \(S\) with the 26-connected components of \(S\cup\{p_8\geq k\}\) that touch \(S\). Components that do not reach \(S\) are dropped. \(S\) is never shrunk, even when \(k\) is above the level of \(S\). Levels whose masks coincide remain separate members with the same volume.

\textbf{Selection.} Let \(B_i\) be RIDGER's integer off-reference voxel count on patient \(i\). Members are grouped by their attained integer \(V\), and within each group the largest \(R\) is kept. The \emph{admissible} member is the one with the largest \(R\) among groups with \(V\leq B_i\). Ties in \(R\) go to the smaller attained \(V\), then to the member name. We do not interpolate between members or resize any member, so the selected member can have a smaller volume than \(B_i\) (Tab.~\ref{tab:main-coronary} lists each actual \(V\)).

\textbf{Length.} The length-matched comparison of Sec.~\ref{S-sec:supp-D-length} applies the same selector with the off-reference centreline length in place of \(V\).
\end{details}

\subsection{The float16 control}
\label{S-sec:supp-D-f16}

RIDGER reads the map down to \(10^{-5}\), below the lowest 8-bit level, so part of its gain could come from these lower probabilities alone (Sec.~\ref{sec:measurement}). The float16 control tests this, and RIDGER leads it by 2.48 to 5.15 points of \(R\) on the four pairs at exactly RIDGER's volume (Tab.~\ref{S-tab:supp-D-interp}).

The control is stronger than the 8-bit family in three ways. First, it reads the map down to \(p=10^{-4}\), below the lowest 8-bit level. Second, a radial trim removes the voxels by which a threshold thickens the vessels of \(S\) and keeps the newly added branches. Third, partly trimmed masks add members near RIDGER's per-patient volume. Each patient then takes its best admissible member, a choice that favours the control. The control has 36 thresholded members plus the pruned variants of the lowest 8-bit member. Every member starts from a thresholded mask and is scored with the same root rule as RIDGER.

\textbf{Reading at exactly RIDGER's volume.}
\label{S-sec:supp-D-matched}
Members are discrete, so a patient's best admissible member usually has a smaller off-reference volume than RIDGER. On coronary Dice+CE, the control's mean off-reference volume is 5.83 against RIDGER's 6.67. We therefore read each patient's \(R\) at exactly RIDGER's off-reference voxel count, by linear interpolation along the upper concave envelope of all members' (\(V\), \(R\)) points. The interpolated point is not a mask that any member produces. RIDGER's volume lies inside the members' range on every patient. Interpolation lifts the float16 control by 1.88 and 1.85 coronary points and 0.65 and 0.38 pulmonary points.

Tab.~\ref{S-tab:supp-D-interp} gives RIDGER's gain over the best admissible member and over the interpolated envelope, for the control and for the 8-bit family. Read at exactly RIDGER's volume, every interval of the gain over the control excludes zero, and RIDGER is ahead on 110 and 117 of the 160 coronary patients. The airway comparison is in Sec.~\ref{S-sec:supp-H_domains}.

\begin{details}
\textbf{Thresholds.} The 36 thresholded members are 24 levels on the 8-bit grid and 12 levels below it on the float16 map. Each is followed by the radial trim.
\begin{itemize}[leftmargin=*,itemsep=1pt,topsep=2pt]
\item \emph{24 levels on the 8-bit grid.} The levels are 24 distinct integers on a logarithmic grid from \(k=1\) to \(k=255\). Where rounding gives the same integer twice, the level moves to the next unused integer. The result is \(k=1,2,\dots,11\) followed by \(14, 18, 23, 29, 37, 47, 60, 76, 97, 124, 157, 200, 255\). The three levels above \(k=128\) return \(S\) itself.
\item \emph{12 float16 levels.} The map is also stored as float16 log odds. The twelve probabilities are log-spaced from \(1/255\) down to \(10^{-4}\) (Tab.~\ref{S-tab:supp-D-f16}). For each probability, the cutoff is the smallest float16 log odds that is not below its exact log odds. The member is the union of \(S\) with the components of the float16 map above the cutoff that touch \(S\). The lowest level, \(10^{-4}\), is the floor stated in Sec.~\ref{sec:measurement}.
\end{itemize}

\begin{table*}[t]
\caption{\textbf{The twelve float16 levels of the float16 control.} Probability \(p\) and its log odds \(\ln\frac{p}{1-p}\), log-spaced from \(1/255\) to \(10^{-4}\). Level 1 is \(1/255\).}
\label{S-tab:supp-D-f16}
\centering\scriptsize
\begin{tabular}{rcc@{\hspace{16pt}}rcc}
\toprule
Level & \(p\) & Log odds & Level & \(p\) & Log odds \\
\midrule
1 & \(3.92\times10^{-3}\) & \(-5.54\) & 7 & \(5.30\times10^{-4}\) & \(-7.54\) \\
2 & \(2.81\times10^{-3}\) & \(-5.87\) & 8 & \(3.80\times10^{-4}\) & \(-7.88\) \\
3 & \(2.01\times10^{-3}\) & \(-6.21\) & 9 & \(2.72\times10^{-4}\) & \(-8.21\) \\
4 & \(1.44\times10^{-3}\) & \(-6.54\) & 10 & \(1.95\times10^{-4}\) & \(-8.54\) \\
5 & \(1.03\times10^{-3}\) & \(-6.87\) & 11 & \(1.40\times10^{-4}\) & \(-8.88\) \\
6 & \(7.40\times10^{-4}\) & \(-7.21\) & 12 & \(1.00\times10^{-4}\) & \(-9.21\) \\
\bottomrule
\end{tabular}
\end{table*}

\textbf{Radial trim.} The trim deletes the shell that a lower threshold adds around the vessels of \(S\) and keeps the new branches. It has no geometric parameter, and every threshold is followed by the same operator. Let \(F\) be the thresholded mask. The part \(F\setminus S\) is divided into territories, each owned by one skeleton site of \(F\) (a geodesic Voronoi partition of \(F\setminus S\) along the skeleton). A territory is protected when its skeleton site lies outside \(S\). A territory of an old site of \(S\) is also protected when that old site lies within one local radius, along the skeleton, of a site outside \(S\). The local radius is the distance to the boundary of \(F\) at that site. Each remaining territory is deleted whole, in the fixed order of its owner's voxel index. A deletion is kept only if it leaves the local topology unchanged. The test covers the one-voxel neighbourhood: the foreground and background partitions and the Euler number must be unchanged under both 6- and 26-connectivity. A deletion that fails this test is restored. As a result, no bridge is cut, and no cavity or handle is opened or closed. The result contains \(S\) and lies in \(F\), and \(S\) is never altered. If a patient exceeds the default memory limit, the identical operator runs with a larger limit (5\,M foreground voxels, 130\,M directed edges, 48\,GiB) and the same deadline, local limits, order and tests.

\textbf{Pruning prefixes (12 to 20 members).} The prefixes are partly trimmed masks of the lowest 8-bit level, and they give the control members close to RIDGER's volume. The trim is run with \(F\) the pool \(p_8\geq1\). The territories that the trim accepts for deletion are recorded in processing order, \(1,\dots,N\), where \(N\) depends on the patient. Prefix \(j\) is \(F\) with the territories \(1,\dots,j\) removed, so prefix 0 is \(F\) itself and prefix \(N\) is the full trim. Twelve prefixes sit at \(j=\lfloor hN/11\rfloor\) for \(h=0,\dots,11\). Up to eight more are added by bisection between neighbouring sampled prefixes whose off-reference volumes straddle a target. The target is a per-patient voxel count close to RIDGER's off-reference volume. The bisection adds members near that volume and removes none. Each step takes the mid-index and uses \(V\) only, and the search stops once a member lies within 0.25\% below the target. Table~\ref{S-tab:supp-D-counts} gives the resulting counts per patient.

\begin{table*}[t]
\caption{\textbf{Member counts per patient.} Minimum and maximum over test patients (a single number means that all patients are equal). ``Control'' is trimmed levels plus float16 levels plus prefixes. The same members enter every selection of Sec.~\ref{sec:measurement}.}
\label{S-tab:supp-D-counts}
\centering\scriptsize
\setlength{\tabcolsep}{3pt}
\begin{tabular}{lrrrrrrr}
\toprule
Network & \(n\) & 8-bit & Trimmed & Float16 & Prefix & Control & Hysteresis \\
\midrule
Coronary, nnU-Net Dice+CE & 160 & 255 & 24 & 12 & 12--20 & 48--56 & 32 \\
Coronary, released nnU-Net & 160 & 255 & 24 & 12 & 12--19 & 48--55 & 32 \\
Pulmonary, nnU-Net & 49 & 255 & 24 & 12 & 13--18 & 49--54 & 32 \\
Pulmonary, mask-prompted SR & 50 & 255 & 24 & 12 & 13--18 & 49--54 & 32 \\
\bottomrule
\end{tabular}
\end{table*}

\textbf{Selection of the control.} Each patient takes the best admissible member of the pooled trimmed levels, float16 levels and prefixes. The admissible-member rule and the tie rule are those of Sec.~\ref{S-sec:supp-D-u8}.

\begin{table}[t]
\caption{\textbf{The control at RIDGER's exact volume.} Gain of RIDGER in $R$ (points) over each patient's best member and over the interpolated envelope, with 95\% patient-bootstrap intervals. B/T/W counts the patients better, tied and worse in the interpolated comparison.}
\label{S-tab:supp-D-interp}
\centering
\scriptsize
\setlength{\tabcolsep}{1.3pt}
\begin{tabular}{llrrr}
\toprule
Network & Family & Best member & Interpolated & B/T/W \\
\midrule
Coronary, Dice+CE & float16 & 6.38 [4.85, 8.07] & 4.50 [3.40, 5.69] & 110/30/20 \\
 & 8-bit & 12.04 [10.27, 13.87] & 10.75 [9.15, 12.40] & 144/12/4 \\
Coronary, released & float16 & 7.01 [5.69, 8.41] & 5.15 [4.04, 6.36] & 117/25/18 \\
 & 8-bit & 13.36 [11.67, 15.12] & 11.94 [10.39, 13.54] & 148/9/3 \\
Pulmonary, nnU-Net & float16 & 5.79 [4.18, 7.54] & 5.15 [3.63, 6.79] & 46/1/2 \\
 & 8-bit & 12.86 [10.14, 15.66] & 12.59 [9.90, 15.37] & 48/1/0 \\
Pulmonary, m.-p.\ SR & float16 & 2.86 [2.16, 3.63] & 2.48 [1.81, 3.22] & 46/1/3 \\
 & 8-bit & 5.42 [4.18, 6.78] & 5.32 [4.09, 6.67] & 49/1/0 \\
\bottomrule
\end{tabular}
\end{table}

\textbf{Extending the control to \(10^{-5}\).} RIDGER's lead stays when the control reads as low as RIDGER does. RIDGER reads down to \(10^{-5}\) and the control stops at \(10^{-4}\), so we added seven levels to the control on the same spacing down to \(10^{-5}\). On coronary Dice+CE, 28 of the new members fit RIDGER's volume, on 11 patients. On the released nnU-Net, 3 do, on 3 patients. They add none on pulmonary CT, where every such member exceeds RIDGER's off-reference volume. With these levels RIDGER's gain over each patient's best admissible member is 6.34 [4.82, 8.02], 7.00 [5.68, 8.41], 5.79 [4.18, 7.54] and 2.86 [2.16, 3.63] points of \(R\) on the four pairs. It differs from the best-member column of Tab.~\ref{S-tab:supp-D-interp} by at most 0.04 points. Read at exactly RIDGER's volume, the gain is 3.97 [2.97, 5.08], 4.84 [3.78, 5.95], 5.15 [3.63, 6.75] and 2.43 [1.76, 3.16]. The extra levels take about half a point off the coronary lead at most, and every interval still excludes zero.
\end{details}

\subsection{Hysteresis and the classical post-processors}
\label{S-sec:supp-D-classical}

The other comparators are two further threshold-type families and three classical post-processors, all held to RIDGER's volume. Hysteresis and freeze-and-grow each produce a family of masks, and each patient takes its best admissible member. Minimal-path bridging, ridge tracking and discrete Morse extraction each run with one setting. That setting is chosen on validation patients with the volume cap in force, from a grid that is extended once when the choice falls on its permissive edge.

\textbf{Hysteresis.} Two-level hysteresis keeps weak voxels that connect to strong ones. A voxel is kept when its 26-connected component of the float16 map above the low probability contains a voxel above the high probability and touches \(S\). The family covers a grid of 32 pairs of low and high probability. Each patient takes the best admissible of the 32 members, so no pair is fixed in advance.

\textbf{Freeze-and-grow.} Freeze-and-grow lowers the threshold step by step and freezes the growth wherever it leaks. We adapt the airway method of Nadeem et al.~\cite{nadeem2021airway} to the probability map alone, without its CT wall tests and its learned terminal classifier. Growth starts at \(S\) and lowers the probability from 0.5 to \(10^{-4}\). The accepted mask at every step is a member, and each patient takes the best admissible one. The three parameters of the growth are chosen once, on all 20 pulmonary validation patients, by the mean equal-volume gain over the 8-bit family.

\textbf{Minimal-path bridging, ridge tracking, DiMorSC.} All three start from \(S\), read only the saved map and add centreline that is rendered as tubes. Minimal-path bridging runs shortest paths from each skeleton terminal of \(S\) to forward ridge maxima of the map. Ridge tracking follows a ridge graph of the log probability forward from the terminals of \(S\). DiMorSC is the discrete Morse extraction of \cite{wang2018dimorsc}, run on the map within 6\,mm of \(S\).

Each of the three runs with one setting per method and anatomy, shared by the two networks. The setting is tuned under the same volume cap that applies on the test patients. On each validation patient the output is admissible only if its integer off-reference voxel count does not exceed RIDGER's, and otherwise the patient falls back to \(S\). The chosen setting has the largest patient-mean \(R\) after this fallback. Five of the six final settings come from the extended grid (Tab.~\ref{S-tab:supp-D-final}). The final settings alone are run on the test patients, once.

\begin{details}
\textbf{Hysteresis grid.} The grid has four high probabilities \(\{0.3,0.5,0.7,0.9\}\). Each has eight low probabilities log-spaced from \(10^{-3}\) up to the high value, which is the last of the eight. \(S\) itself is not added to the mask.

\textbf{Freeze-and-grow rule.} The probability is lowered in steps of a fixed number of octaves. At each step the 26-connected region grows from the accepted mask. A cube of half-width 2\,mm around each accepted skeleton site counts old and proposed voxels. The local frontier is frozen when two conditions hold: the proposed volume exceeds the leak ratio times the old volume, and at least a minimum number of new voxels is proposed. All 419 patient results of the four primary pairs respect RIDGER's integer off-reference voxel count. No selected member therefore exceeds RIDGER's volume.

\textbf{Freeze-and-grow grid.} The grid has 12 settings: step \(\{1,2\}\) octaves, leak ratio \(\{2,3,5\}\), minimum new volume \(\{32,128\}\) voxels. The setting is chosen on all 20 pulmonary validation patients, whereas RIDGER's own rules use 10 of the 20 (Sec.~\ref{S-sec:supp-C-splits}). Ties in the mean equal-volume gain go to the smaller mean \(V\). The chosen setting is a step of two octaves, ratio 2 and a minimum of 128 voxels.

\textbf{Operators of the three post-processors.} Each renders added centreline as a tube with a measured half-amplitude radius clipped to 0.35 to 0.60\,mm, and the output is \(S\) united with the tubes.
\begin{itemize}[leftmargin=*,itemsep=1pt,topsep=2pt]
\item \emph{Minimal-path bridging.} The cost of a path is the physical integral of the clipped \(-\ln p\) along it. The cost has the unit of a length (mm), because the integrand has none. A path is accepted when its length is at most a maximum length \(L\) and its cost is at most the stopping cost.
\item \emph{Ridge tracking.} We build the ridge graph of the log probability (Hessian at a Gaussian scale of 0.6\,mm) with weak and strong hysteresis at \(p\geq10^{-6}\) and \(p\geq10^{-3}\). The graph is tracked forward from the terminals of \(S\) up to a maximum rooted tracked length \(L\). Tracking uses an attachment radius of 1.6\,mm and a tangent cosine of at least 0.7. Tracking can optionally be gated by a cross-section contrast ratio, which has no unit.
\item \emph{DiMorSC.} The extraction runs on the native Freudenthal complex of the map within 6\,mm of \(S\), at a Gaussian scale of 0.6\,mm. Arcs with persistence above the threshold in log units are kept. The components attached to \(S\) (inside it, or within 1.6\,mm of its surface) are retained with all their edges, cycles included.
\end{itemize}

\textbf{Grids.} Stage one of the validation grid is
\begin{itemize}[leftmargin=*,itemsep=1pt,topsep=2pt]
\item minimal path: float16 map, \(L\in\{2,4,8\}\)\,mm, stopping cost \(\in\{8,32,128,\text{none}\}\) (12 settings);
\item ridge tracking: map \(\in\{\)8-bit, float16\(\}\), \(L\in\{2,4,8\}\)\,mm, contrast gate \(\in\{\text{ratio }1,\text{off}\}\) (12);
\item DiMorSC: map \(\in\{\)8-bit, float16\(\}\), persistence \(\in\{4,2,1,0.5,0.25,0.125\}\) log units, 6\,mm domain (12 on coronary; on pulmonary the float16 map was infeasible in memory, so 6 on the 8-bit map).
\end{itemize}
Stage two extends the grid when a stage-one choice lies on the permissive edge of an extendable axis (\(L=8\)\,mm, or persistence 0.125). The axis is then extended once on the validation patients, to \(L\in\{12,16\}\)\,mm or persistence \(\in\{0.0625,0.03125\}\). The choice is made again on the union of the two stages and is final. Values at the end of an axis (no stopping cost, gate off) need no extension.

\textbf{Tuning rule.} Validation uses 80 coronary patients per network and the 7 rooted pulmonary validation patients of Sec.~\ref{S-sec:supp-C-splits} per network. The shared setting maximises the mean of the two networks' patient-mean \(R\), with equal weight per network. Each patient's \(R\) is taken after the fallback to \(S\). Ties go to the smaller \(V\), then to the less permissive setting. Table~\ref{S-tab:supp-D-final} lists the final settings with their validation \(R\) and \(V\).

\begin{table*}[t]
\caption{\textbf{Final validation settings of the classical post-processors.} \(L\) is the maximum path or tracking length. The stopping cost and contrast gate are as in the text. ``Grid'' is the number of validation settings evaluated, stage two included. Validation patient means of \(R\) and \(V\) (\%) after the volume cap.}
\label{S-tab:supp-D-final}
\centering\scriptsize
\setlength{\tabcolsep}{3pt}
\begin{tabular}{llllrrr}
\toprule
Anatomy & Method & Setting & Extended & Grid & \(R\!\uparrow\) & \(V\) \\
\midrule
Coronary & Minimal path & float16, \(L=4\)\,mm, stop 8 & no & 12 & 80.14 & \textcolor{gray}{6.00} \\
Coronary & Ridge tracking & float16, \(L=12\)\,mm, ratio 1 & yes & 20 & 78.51 & \textcolor{gray}{5.75} \\
Coronary & DiMorSC & 8-bit, persistence 0.03125 & yes & 16 & 80.47 & \textcolor{gray}{5.55} \\
Pulmonary & Minimal path & float16, \(L=12\)\,mm, stop 8 & yes & 20 & 88.75 & \textcolor{gray}{16.23} \\
Pulmonary & Ridge tracking & 8-bit, \(L=16\)\,mm, ratio 1 & yes & 20 & 89.67 & \textcolor{gray}{16.43} \\
Pulmonary & DiMorSC & 8-bit, persistence 0.03125 & yes & 8 & 89.17 & \textcolor{gray}{15.94} \\
\bottomrule
\end{tabular}
\end{table*}
\end{details}

\subsection{Matching by length instead of volume}
\label{S-sec:supp-D-length}

The length-matched comparison of Sec.~\ref{sec:measurement} holds the off-reference length fixed in place of \(V\), with the same members and the same selection. The off-reference length is the physical length of the predicted skeleton whose voxels lie more than 1\,mm from the reference foreground. Each 26-neighbour skeleton edge gives half its Euclidean length to each end. The members (8-bit family, float16 control with its prefixes, hysteresis) enter unchanged. For each patient the admissible member maximises \(R\) among those with length at most RIDGER's. Ties go to the shorter length, then to member name. Patient means and 95\% intervals use 20{,}000 patient bootstrap draws. Under this matching, RIDGER leads the best 8-bit threshold on both anatomies and the float16 control on coronary CTA, and it is level with the control on pulmonary CT (Tab.~\ref{tab:currencies}).

\section{Data, splits and exclusions}
\label{S-sec:supp-C_data}

This section lists the datasets and splits, and the test patients that could not be scored.

\subsection{Datasets and splits}
\label{S-sec:supp-C-splits}

No network saw any of our test patients. Our networks are trained on the training patients of their own dataset, and Tab.~\ref{S-tab:supp-data} lists the splits and the test patients scored for each primary network. The released coronary nnU-Net and CAS-Net were trained by the ImageCAS-X authors on the official split, and we run inference with them as released. \(\lambda\) and every rule are fixed on validation patients, and all test results use the readout of the main text. Pulmonary rules are fixed on the first 10 of the 20 HiPaS validation patients by case number, 7 of them with a readable root. ASOCA is an external test set. Its 40 cases are scored with inference and readout unchanged. HiPaS shares scans with the PARSE 2022 challenge \cite{luo2023parse}, and no shared test scan was used for training.

\begin{details}
\textbf{HiPaS and PARSE 2022.} Of the 170 HiPaS training scans, 114 are PARSE scans that HiPaS labelled (111 from the PARSE training set and 3 from its validation set). Three test scans, hipas\_221, 223 and 234, are PARSE validation scans and were never used for training. Only 234 is scored, because 221 and 223 have no readable root.
\end{details}

\begin{table*}[t]
\caption{\textbf{Datasets, splits and test patients scored per primary network.} Numbers in parentheses are the pulmonary validation patients used to fix the rules. Pulmonary patients without a readable reference root are not scored for \(R\) (Sec.~\ref{S-sec:supp-C-excl}); ASOCA is not used for training or validation.}
\label{S-tab:supp-data}
\centering
\scriptsize
\setlength{\tabcolsep}{3pt}
\begin{tabular}{@{}llllrrrr@{}}
\toprule
Dataset & Anatomy & Reference & Primary network & Train & Val & Test & Scored \\
\midrule
ImageCAS-X & coronary CTA & ImageCAS-X & nnU-Net Dice+CE & 559 & 80 & 160 & 160 \\
ImageCAS-X & coronary CTA & ImageCAS-X & released nnU-Net & 559 & 80 & 160 & 160 \\
HiPaS & pulmonary CT & HiPaS & nnU-Net & 170 & 20 (10) & 60 & 49 \\
HiPaS & pulmonary CT & HiPaS & mask-prompted SR & 170 & 20 (10) & 60 & 50 \\
ASOCA & coronary CTA & ASOCA & both coronary networks & n/a & n/a & 40 & 40 \\
\bottomrule
\end{tabular}
\end{table*}

\subsection{Excluded test patients}
\label{S-sec:supp-C-excl}

\textbf{Rule.} A patient is scored for the reaching rate \(R\) only if the reference has a readable root. A patient without one is left out of every comparison of \(R\), and no score is imputed. The root is read from the reference alone.

\textbf{Counts.} The rule excludes no coronary patient and ten of the 60 pulmonary test patients. All 160 ImageCAS-X test patients have a readable root, and no anchor failed. The same ten pulmonary patients lack a root under both networks, which leaves 50 for the mask-prompted SR. The nnU-Net pulmonary roster also lacks hipas\_111 and has 49. On this patient the comparator pipeline exceeded its foreground-voxel cap before any readout was scored. The Skeleton Recall pulmonary roster lacks hipas\_111 and hipas\_149 for the same reason and has 48 patients (Tab.~\ref{S-tab:supp-E-topo}). Every exclusion applies to all readouts of a network alike.

\begin{details}
\textbf{Excluded patients.} The ten pulmonary patients without a readable root are hipas\_127, 159, 175, 181, 212, 216, 221, 223, 237 and 238.

\textbf{Why the root is unreadable.} The pulmonary trunk is present in all ten patients. The root is the reference skeleton voxel with the largest inscribed radius, in the pulmonary trunk. In these ten patients, the root fails a distance test on the skeleton of this wide trunk. The scorer resamples the reference to 0.5\,mm, re-skeletonises it, and keeps the root only if that skeleton passes within 3\,mm. In the ten excluded patients the root lies inside the trunk (inscribed radius 10.5 to 17.5\,mm), but the resampled skeleton passes 3.5 to 31.6\,mm away. In every scored patient the distance is 0.06 to 2.91\,mm (median 0.57), so the 3\,mm test separates the two groups without overlap.

\textbf{Slice spacing.} A label-only screen covers 59 of the 60 pulmonary test patients, all but hipas\_111, and flags seven whose header slice spacing may not be physical: hipas\_103, 143, 149, 158, 162, 178 and 193. The screen flags a patient when the craniocaudal extent of the artery label exceeds 1.3 times its left-right extent. The seven patients stay in every result. Dropping them changes the four pulmonary mean gains as follows. For nnU-Net, the gain over the best 8-bit threshold changes by +0.03 and the gain over the float16 control by +0.10. For mask-prompted SR, the same two gains change by +0.06 and \(-\)0.04 points.
\end{details}

\providecolor{ridger}{RGB}{252,234,230}
\section{Full results}
\label{S-sec:supp-E_results}

This section gives the results of Secs.~\ref{sec:reach} and~\ref{sec:topology} for every network and reference, the comparison with a trained tracker, and how the patients of Figs.~\ref{fig:teaser} and~\ref{fig:qual} were chosen.

\textbf{Conventions.} The reaching rate \(R\) and the off-reference volume \(V\) are given in percent of the reference. Gains are means over patients in points of \(R\), with 95\% patient-bootstrap intervals. Where three counts are given, they are the patients better, tied and worse. A patient is tied when the gain is exactly zero, which happens when both readouts select the same mask or reach equally far.

\subsection{All networks and references}
\label{S-sec:supp-E-all}

\begin{figure*}[t]
\centering
\includegraphics[width=\textwidth]{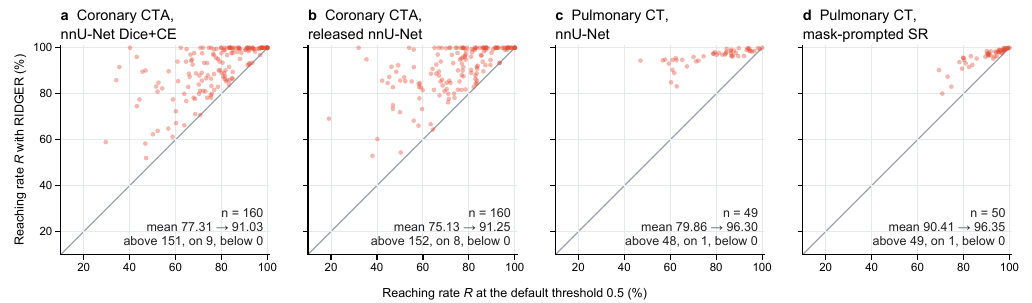}
\caption{\textbf{Reaching rate \(R\) of each patient at the default threshold and with RIDGER.} Each dot is one test patient of a primary pair. Dots above the diagonal gain \(R\), and dots on the diagonal are unchanged. No patient loses \(R\), because RIDGER keeps the default mask.}
\label{S-fig:supp-per-patient}
\end{figure*}

\begin{figure*}[t]
\centering
\includegraphics[width=\textwidth]{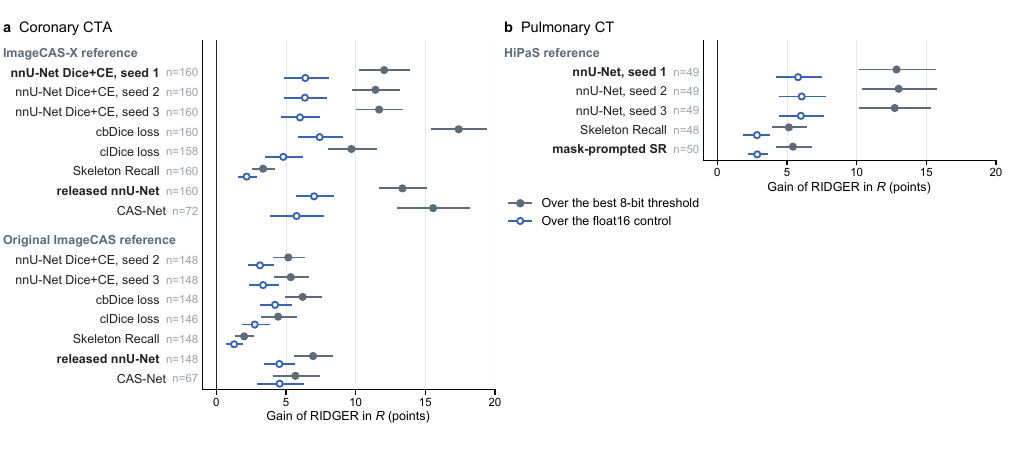}
\caption{\textbf{Equal-volume gain of RIDGER for every network and reference.} Gain in \(R\) (points) over the best 8-bit threshold (filled) and over the float16 control (open) at RIDGER's per-patient off-reference volume, with 95\% patient-bootstrap intervals (values in Tab.~\ref{S-tab:supp-E-all}). The names of the four primary networks are in bold.}
\label{S-fig:supp-gain-forest}
\end{figure*}

\textbf{Every network and reference (Tab.~\ref{S-tab:supp-E-all}).} The comparison of the four primary pairs extends to every network and to both coronary reference conventions. The table lists each map twice, read at 0.5 and by RIDGER. The two gain columns compare RIDGER with the best 8-bit threshold of the same map and with the float16 control, both held at RIDGER's per-patient $V$. On the ImageCAS-X and HiPaS references, all eight coronary networks and all five pulmonary networks gain over both comparators, with every interval above zero. On the original ImageCAS labels the gains are smaller and stay above zero. The gains over the best 8-bit threshold run from 1.99 to 6.94 points for the seven coronary networks scored on these labels.

\textbf{Each patient (Fig.~\ref{S-fig:supp-per-patient}).} The figure plots \(R\) of every test patient of the four primary pairs, at the default threshold against RIDGER. No dot lies below the diagonal, because RIDGER keeps the default mask.

\textbf{Gains by network (Fig.~\ref{S-fig:supp-gain-forest}).} The figure draws the two gain columns of the table with their intervals. RIDGER adds most where a network reaches least at 0.5. On the ImageCAS-X and HiPaS references, the smallest gain over the best 8-bit threshold is that of Skeleton Recall on coronary CTA (\(R\) = 91.08 at 0.5). The largest is that of the cbDice loss (\(R\) = 66.70 at 0.5). The gain changes little between training seeds. Three Dice+CE coronary seeds gain 6.38, 6.35 and 6.00 points over the float16 control, and three pulmonary nnU-Net seeds gain 5.95 $\pm$ 0.14.

\begin{details}
\textbf{Original ImageCAS labels.} The seven coronary networks scored on these labels also gain over the float16 control. The lowest lower bound of these gains is 0.72.

\textbf{Patients scored.} CAS-Net is scored on 72 of the 160 test patients. Its comparators exceeded the foreground-voxel cap on 48 patients, and 40 patients have no stored map. On the original ImageCAS labels, 12 test patients have no readable root, so every network is scored on 148 (146 for the clDice loss, 67 for CAS-Net). The primary Dice+CE network (seed 1) is not scored on these labels.

\textbf{CAS-Net.} Both CAS-Net rows are read after the map normalisation (\(T=\tfrac12\) for all 72 patients, Sec.~\ref{S-sec:supp-J-casnet}). This normalisation sharpens no other map of the table.
\end{details}

\begin{table*}[t]
\caption{\textbf{All maps at 0.5 and read by RIDGER.} Test-patient means (\%). Gains are in points of $R$ at RIDGER's per-patient off-reference volume (grey), with 95\% patient-bootstrap intervals, over each patient's best 8-bit threshold and over the float16 control. The last column counts patients better, tied and worse than the control. Seeds are independent training runs. The pulmonary nnU-Net seed 1 is the network of the main text. The CAS-Net rows cover the 72 evaluable patients; on the original ImageCAS labels a patient without a defined \(R\) is not counted. $^\dagger$Read after the map normalisation (\(T=\tfrac12\)), with the best 8-bit threshold taken on the map that RIDGER reads.}
\label{S-tab:supp-E-all}
\centering\scriptsize
\setlength{\tabcolsep}{3pt}
\begin{tabular*}{\textwidth}{@{\extracolsep{\fill}}lrrr@{\hspace{6pt}}rrrrr@{}}
\toprule
 & & \multicolumn{2}{c}{0.5} & \multicolumn{2}{c}{RIDGER} & \multicolumn{3}{c}{Gain of RIDGER over} \\
\cmidrule(lr){3-4}\cmidrule(lr){5-6}\cmidrule(l){7-9}
Network & $n$ & $R\!\uparrow$ & $V$ & $R\!\uparrow$ & $V$ & best 8-bit threshold & float16 control & B/T/W \\
\midrule
\multicolumn{9}{@{}l}{\textbf{Coronary CTA, ImageCAS-X reference}} \\
\hspace{0.5em}nnU-Net Dice+CE, seed 1 & 160 & 77.31 & \textcolor{gray}{5.00} & \cellcolor{ridger}91.03 & \cellcolor{ridger}\textcolor{gray}{6.67} & 12.04 [10.27, 13.87] & 6.38 [4.85, 8.07] & 111/37/12 \\
\hspace{0.5em}nnU-Net Dice+CE, seed 2 & 160 & 77.04 & \textcolor{gray}{4.62} & \cellcolor{ridger}90.57 & \cellcolor{ridger}\textcolor{gray}{6.51} & 11.42 [9.74, 13.17] & 6.35 [4.88, 7.91] & 110/37/13 \\
\hspace{0.5em}nnU-Net Dice+CE, seed 3 & 160 & 77.70 & \textcolor{gray}{5.46} & \cellcolor{ridger}91.17 & \cellcolor{ridger}\textcolor{gray}{7.30} & 11.68 [10.03, 13.43] & 6.00 [4.67, 7.45] & 111/37/12 \\
\hspace{0.5em}cbDice loss & 160 & 66.70 & \textcolor{gray}{4.44} & \cellcolor{ridger}91.79 & \cellcolor{ridger}\textcolor{gray}{8.05} & 17.39 [15.43, 19.41] & 7.41 [5.85, 9.11] & 111/40/9 \\
\hspace{0.5em}clDice loss & 158 & 82.42 & \textcolor{gray}{6.11} & \cellcolor{ridger}95.09 & \cellcolor{ridger}\textcolor{gray}{8.97} & 9.69 [7.98, 11.50] & 4.80 [3.51, 6.24] & 83/69/6 \\
\hspace{0.5em}Skeleton Recall & 160 & 91.08 & \textcolor{gray}{9.61} & \cellcolor{ridger}94.75 & \cellcolor{ridger}\textcolor{gray}{10.56} & 3.35 [2.57, 4.18] & 2.17 [1.52, 2.88] & 67/82/11 \\
\hspace{0.5em}released nnU-Net & 160 & 75.13 & \textcolor{gray}{5.27} & \cellcolor{ridger}91.25 & \cellcolor{ridger}\textcolor{gray}{7.24} & 13.36 [11.67, 15.12] & 7.01 [5.69, 8.41] & 123/29/8 \\
\hspace{0.5em}CAS-Net$^\dagger$ & 72 & 73.62 & \textcolor{gray}{5.52} & \cellcolor{ridger}93.35 & \cellcolor{ridger}\textcolor{gray}{12.65} & 15.56 [12.96, 18.24] & 5.75 [3.88, 7.70] & 48/14/10 \\
\addlinespace[2pt]
\multicolumn{9}{@{}l}{\textbf{Coronary CTA, original ImageCAS reference}} \\
\hspace{0.5em}nnU-Net Dice+CE, seed 2 & 148 & 60.11 & \textcolor{gray}{1.62} & \cellcolor{ridger}67.54 & \cellcolor{ridger}\textcolor{gray}{2.17} & 5.16 [4.08, 6.36] & 3.13 [2.25, 4.11] & 80/51/17 \\
\hspace{0.5em}nnU-Net Dice+CE, seed 3 & 148 & 60.68 & \textcolor{gray}{1.73} & \cellcolor{ridger}67.97 & \cellcolor{ridger}\textcolor{gray}{2.25} & 5.34 [4.12, 6.68] & 3.35 [2.35, 4.48] & 73/56/19 \\
\hspace{0.5em}cbDice loss & 148 & 54.64 & \textcolor{gray}{1.35} & \cellcolor{ridger}68.71 & \cellcolor{ridger}\textcolor{gray}{2.42} & 6.19 [4.94, 7.55] & 4.21 [3.10, 5.43] & 79/50/19 \\
\hspace{0.5em}clDice loss & 146 & 63.26 & \textcolor{gray}{1.93} & \cellcolor{ridger}70.41 & \cellcolor{ridger}\textcolor{gray}{2.72} & 4.43 [3.22, 5.81] & 2.75 [1.81, 3.85] & 62/64/20 \\
\hspace{0.5em}Skeleton Recall & 148 & 67.47 & \textcolor{gray}{2.74} & \cellcolor{ridger}69.76 & \cellcolor{ridger}\textcolor{gray}{3.01} & 1.99 [1.34, 2.72] & 1.27 [0.72, 1.89] & 39/98/11 \\
\hspace{0.5em}released nnU-Net & 148 & 58.22 & \textcolor{gray}{1.61} & \cellcolor{ridger}68.61 & \cellcolor{ridger}\textcolor{gray}{2.21} & 6.94 [5.57, 8.38] & 4.52 [3.45, 5.65] & 88/47/13 \\
\hspace{0.5em}CAS-Net$^\dagger$ & 67 & 54.24 & \textcolor{gray}{1.58} & \cellcolor{ridger}66.83 & \cellcolor{ridger}\textcolor{gray}{3.55} & 5.67 [4.04, 7.45] & 4.54 [2.94, 6.26] & 36/21/10 \\
\addlinespace[2pt]
\multicolumn{9}{@{}l}{\textbf{Pulmonary CT, HiPaS reference}} \\
\hspace{0.5em}nnU-Net, seed 1 & 49 & 79.86 & \textcolor{gray}{8.75} & \cellcolor{ridger}96.30 & \cellcolor{ridger}\textcolor{gray}{12.87} & 12.86 [10.14, 15.66] & 5.79 [4.18, 7.54] & 47/1/1 \\
\hspace{0.5em}nnU-Net, seed 2 & 49 & 79.51 & \textcolor{gray}{8.77} & \cellcolor{ridger}95.08 & \cellcolor{ridger}\textcolor{gray}{12.16} & 13.01 [10.35, 15.75] & 6.05 [4.44, 7.76] & 46/1/2 \\
\hspace{0.5em}nnU-Net, seed 3 & 49 & 79.78 & \textcolor{gray}{8.92} & \cellcolor{ridger}95.34 & \cellcolor{ridger}\textcolor{gray}{12.51} & 12.73 [10.18, 15.36] & 5.99 [4.43, 7.65] & 46/2/1 \\
\hspace{0.5em}Skeleton Recall & 48 & 91.11 & \textcolor{gray}{15.81} & \cellcolor{ridger}97.06 & \cellcolor{ridger}\textcolor{gray}{18.98} & 5.13 [3.92, 6.43] & 2.84 [1.87, 3.81] & 45/1/2 \\
\hspace{0.5em}mask-prompted SR & 50 & 90.41 & \textcolor{gray}{15.36} & \cellcolor{ridger}96.35 & \cellcolor{ridger}\textcolor{gray}{18.14} & 5.42 [4.18, 6.78] & 2.86 [2.16, 3.63] & 49/1/0 \\
\bottomrule
\end{tabular*}
\end{table*}

\subsection{Threshold curves}
\label{S-sec:supp-E-frontier}

The tables of the main text hold every readout at RIDGER's volume. Fig.~\ref{fig:reach-volume} of the main text plots the patient-mean reaching rate \(R\) against the off-reference volume \(V\) on the four primary pairs, so it shows the ranking at other volumes too. RIDGER lies above the curve that the 8-bit thresholds trace as the threshold is lowered. At RIDGER's mean \(V\) this curve sits 11.8 points below RIDGER's mean \(R\) on coronary CTA and 13.1 points below it on pulmonary CT. The classical post-processors lie on or near the curve. The figure also draws the curve that RIDGER traces when \(\lambda\) varies over the validation grid. On each pair, RIDGER's curve lies above both the sweep of the 8-bit thresholds and the sweep of the float16 control. RIDGER's place above both sweeps is therefore not specific to the \(\lambda\) used in this paper.

\subsection{Topology-aware losses, all metrics}
\label{S-sec:supp-E-topo}

A plain network read by RIDGER reaches as far as the topology-aware networks read at 0.5. Table~\ref{S-tab:supp-E-topo} gives every score of the networks compared in Sec.~\ref{sec:topology}, each at 0.5 and read by RIDGER; \(V\) is not matched across its rows. On coronary CTA, the Dice+CE nnU-Net read by RIDGER has \(R\) = 91.03 at a \(V\) of 6.67, against 91.08 at 9.61 for Skeleton Recall at 0.5. On pulmonary CT, the nnU-Net read by RIDGER has \(R\) = 96.30, against 91.11 and 90.41 for the two topology-aware networks at 0.5.

RIDGER also extends the topology-aware networks themselves. From 0.5 to RIDGER, \(R\) rises for every network of the table, by 3.67 to 25.09 points. Dice moves by at most 0.89 points on coronary CTA. clDice falls for every network: it changes by $-$5.37 to $-$0.71 points on coronary CTA and by $-$16.44 to $-$15.74 on pulmonary CT. The fall in clDice follows from the skeleton that RIDGER adds outside the reference (Sec.~\ref{S-sec:supp-C-quoted}).

\begin{details}
\textbf{Scores.} The cbDice scores come from the official implementation (commit a970c1d4). It uses its default morphological skeleton of ten iterations and distances in voxels. The scores are evaluated on the foreground bounding box of prediction and reference. All foreground labels are merged. cbDice at 0.5 is scored for the four primary networks. $E_{\beta_1}$ is the absolute error in the number of loops under native 26-connectivity.
\end{details}

\begin{table*}[t]
\caption{\textbf{All metrics for the topology-aware losses.} Test-patient means (\%; $E_{\beta_1}$ absolute) on ImageCAS-X (coronary) and HiPaS (pulmonary) references, at 0.5 and read by RIDGER. clDice loss: two patients failed an input check. $V$ differs across rows. A dash marks a score that was not stored.}
\label{S-tab:supp-E-topo}
\centering\scriptsize
\setlength{\tabcolsep}{2.6pt}
\begin{tabular}{@{}lr*{6}{r}@{\hspace{8pt}}*{6}{r}@{}}
\toprule
 & & \multicolumn{6}{c}{0.5} & \multicolumn{6}{c}{RIDGER} \\
\cmidrule(lr){3-8}\cmidrule(l){9-14}
Network & $n$ & $R\!\uparrow$ & $V$ & Dice$\!\uparrow$ & clDice$\!\uparrow$ & cbDice$\!\uparrow$ & $E_{\beta_1}\!\downarrow$ & $R\!\uparrow$ & $V$ & Dice$\!\uparrow$ & clDice$\!\uparrow$ & cbDice$\!\uparrow$ & $E_{\beta_1}\!\downarrow$ \\
\midrule
\rowcolor{subhead}\multicolumn{14}{@{}l}{\textit{Coronary CTA}} \\
\hspace{0.5em}clDice loss & 158 & 82.42 & \textcolor{gray}{6.11} & 90.02 & 92.65 & -- & 0.11 & \cellcolor{ridger}95.09 & \cellcolor{ridger}\textcolor{gray}{8.97} & \cellcolor{ridger}89.90 & \cellcolor{ridger}87.28 & \cellcolor{ridger}85.74 & \cellcolor{ridger}0.21 \\
\hspace{0.5em}cbDice loss & 160 & 66.70 & \textcolor{gray}{4.44} & 87.93 & 90.45 & -- & 0.07 & \cellcolor{ridger}91.79 & \cellcolor{ridger}\textcolor{gray}{8.05} & \cellcolor{ridger}88.81 & \cellcolor{ridger}88.99 & \cellcolor{ridger}86.34 & \cellcolor{ridger}0.26 \\
\hspace{0.5em}Skeleton Recall & 160 & 91.08 & \textcolor{gray}{9.61} & 90.38 & 92.56 & -- & 0.16 & \cellcolor{ridger}94.75 & \cellcolor{ridger}\textcolor{gray}{10.56} & \cellcolor{ridger}90.19 & \cellcolor{ridger}90.57 & \cellcolor{ridger}88.54 & \cellcolor{ridger}0.16 \\
\hspace{0.5em}nnU-Net Dice+CE & 160 & 77.31 & \textcolor{gray}{5.00} & 89.29 & 92.00 & 88.79 & 0.06 & \cellcolor{ridger}91.03 & \cellcolor{ridger}\textcolor{gray}{6.67} & \cellcolor{ridger}90.06 & \cellcolor{ridger}90.94 & \cellcolor{ridger}88.46 & \cellcolor{ridger}0.09 \\
\hspace{0.5em}released nnU-Net & 160 & 75.13 & \textcolor{gray}{5.27} & 89.32 & 91.58 & 88.45 & 0.10 & \cellcolor{ridger}91.25 & \cellcolor{ridger}\textcolor{gray}{7.24} & \cellcolor{ridger}89.89 & \cellcolor{ridger}90.87 & \cellcolor{ridger}88.22 & \cellcolor{ridger}0.17 \\
\rowcolor{subhead}\multicolumn{14}{@{}l}{\textit{Pulmonary CT}} \\
\hspace{0.5em}nnU-Net & 49 & 79.86 & \textcolor{gray}{8.75} & 87.72 & 86.83 & 82.26 & 4.51 & \cellcolor{ridger}96.30 & \cellcolor{ridger}\textcolor{gray}{12.87} & \cellcolor{ridger}86.45 & \cellcolor{ridger}70.38 & \cellcolor{ridger}71.91 & \cellcolor{ridger}4.37 \\
\hspace{0.5em}Skeleton Recall & 48 & 91.11 & \textcolor{gray}{15.81} & 87.07 & 85.87 & -- & 4.08 & \cellcolor{ridger}97.06 & \cellcolor{ridger}\textcolor{gray}{18.98} & \cellcolor{ridger}85.95 & \cellcolor{ridger}70.00 & \cellcolor{ridger}72.56 & \cellcolor{ridger}4.19 \\
\hspace{0.5em}mask-prompted SR & 50 & 90.41 & \textcolor{gray}{15.36} & 86.86 & 86.52 & 81.88 & 4.24 & \cellcolor{ridger}96.35 & \cellcolor{ridger}\textcolor{gray}{18.14} & \cellcolor{ridger}85.87 & \cellcolor{ridger}70.78 & \cellcolor{ridger}73.59 & \cellcolor{ridger}4.32 \\
\bottomrule
\end{tabular}
\end{table*}

\subsection{A trained tracker: TopoVST}
\label{S-sec:supp-E-topovst}
\label{S-sec:supp-D-topovst}

TopoVST \cite{liu2026topovst} is a trained tracker that reads the CT, and it reaches less far than the best 8-bit threshold of the map from which it starts. It tracks a vessel skeleton from seeds. At each node, a graph network reads the CT around it and predicts the radius and the directions to follow. We retrained it on the ImageCAS-X training patients and seeded it from the skeleton of the default mask \(S\) of the Dice+CE map, so the tracker starts from the same anchor as RIDGER. Its main readout is \(S\) united with tubes of the predicted radius along the tracked skeleton. This readout is scored against the comparator of RIDGER, the best 8-bit threshold of the same map at equal off-reference volume.

\textbf{Result.} On the test patients, TopoVST has a reaching rate \(R\) of 83.07 at an off-reference volume \(V\) of 35.80. This is 1.42 points below the best 8-bit threshold of the same map at that volume, and the interval excludes zero (Tab.~\ref{S-tab:supp-E-topovst}). On ASOCA the two are level, with an interval that includes zero. RIDGER's \(R\) is higher than TopoVST's by 7.96 [6.30, 9.72] points on the test patients and by 3.26 [0.59, 6.03] on ASOCA, at less off-reference volume (6.67 against 35.80 on the test patients).

\begin{table*}[t]
\caption{\textbf{TopoVST against the same comparators as RIDGER.} Gains are \(R\) minus the best 8-bit threshold of the same Dice+CE map at equal off-reference volume and at equal off-reference length, with 95\% patient-bootstrap intervals. On 2 of 160 test patients the tracked tree is shorter than every threshold, and its equal-length gain is undefined.}
\label{S-tab:supp-E-topovst}
\centering\scriptsize
\setlength{\tabcolsep}{3pt}
\begin{tabular}{llrrrrrr}
\toprule
Set & Readout & \(R\!\uparrow\) & \(V\) & Dice$\!\uparrow$ & clDice$\!\uparrow$ & Gain at equal \(V\) & Gain at equal length \\
\midrule
Test & \(S\) & 77.31 & \textcolor{gray}{5.00} & 89.29 & 92.00 & -- & -- \\
Test & TopoVST, \(S\cup\) tubes & 83.07 & \textcolor{gray}{35.80} & 79.99 & 90.99 & \(-\)1.42 [\(-\)2.42, \(-\)0.46] & 0.05 [\(-\)0.91, 1.02] \\
Test & TopoVST, tubes & 82.75 & \textcolor{gray}{34.15} & 73.81 & 90.52 & \(-\)1.66 [\(-\)2.66, \(-\)0.71] & \(-\)0.23 [\(-\)1.19, 0.74] \\
\rowcolor{ridger}Test & \textbf{RIDGER} & 91.03 & \textcolor{gray}{6.67} & 90.06 & 90.94 & 12.04 [10.27, 13.87] & 6.39 [5.05, 7.88] \\
\midrule
ASOCA & \(S\) & 78.61 & \textcolor{gray}{13.03} & 80.86 & 80.34 & -- & -- \\
ASOCA & TopoVST, \(S\cup\) tubes & 82.91 & \textcolor{gray}{36.63} & 76.36 & 76.92 & 0.40 [\(-\)2.06, 3.04] & 0.73 [\(-\)1.58, 3.24] \\
ASOCA & TopoVST, tubes & 82.29 & \textcolor{gray}{31.70} & 69.50 & 75.88 & 0.02 [\(-\)2.32, 2.61] & 0.11 [\(-\)2.23, 2.68] \\
\rowcolor{ridger}ASOCA & \textbf{RIDGER} & 86.17 & \textcolor{gray}{16.23} & 81.20 & 75.33 & 6.15 [3.85, 8.72] & 3.55 [1.69, 5.71] \\
\bottomrule
\end{tabular}
\end{table*}

\begin{details}
\textbf{Training.} TopoVST was retrained with the reference centrelines and CT of the ImageCAS-X training patients. The published settings were kept: the sample mix (30\% off the centreline, 10\% outside the lumen), scales, loss, Adam at \(5\times10^{-4}\) and batch 8. Training ran for 20{,}000 steps, saving every 2{,}000. The released sample generator needed two corrections. Its grid normalisation placed radius and lumen labels up to one voxel off. Its tangent labels assumed centreline point ids that run along each line. ImageCAS-X ids do not, and 54\% of on-centreline samples lost their direction. The checkpoint was chosen by centreline overlap OV \cite{schaap2009cat08} on 16 validation patients (step 18{,}000, OV 0.892). Test and ASOCA patients were tracked once with it.

\textbf{Tracking.} The method's wavefront cap, sphere-response threshold and step size were kept. Three caps of the released Circle-of-Willis configuration were removed: the limit of 20 seed rounds, the limit of 600 iterations and loop closing at 30\,mm. Tracking takes a median of 665\,s per test patient on one GPU. The tubes of the main readout are kept where they touch \(S\). The second readout of Tab.~\ref{S-tab:supp-E-topovst} is the tubes alone.

\textbf{Length and width.} The off-reference volume of TopoVST comes from the width of its tubes. On test patients, 93\% of tracked nodes lie inside \(S\), and the tracked skeleton runs off the reference for less length than RIDGER's (49.9 against 83.0\,mm per patient). The tubes are wider than the distal reference vessels. No test or ASOCA patient has TopoVST at or below RIDGER's off-reference volume, so the two are not compared at equal volume. At equal off-reference length, which does not count width, TopoVST ties the best 8-bit threshold (Tab.~\ref{S-tab:supp-E-topovst}).

\textbf{Centreline overlap.} RIDGER has the highest centreline recall, the lowest centreline precision and the lowest OV of the three readouts. OV is scored on the 160 test patients, with each mask's skeleton resampled like the tracked graph. It is 0.904 for \(S\) (recall 0.888, precision 0.923), 0.890 for TopoVST (recall 0.914, precision 0.876) and 0.881 for RIDGER (recall 0.950, precision 0.843). OV counts every centreline beyond the reference as an error. Sec.~\ref{S-sec:supp-I_readers} shows what such additions are.
\end{details}

\subsection{Selection rules of Figs.~\ref{fig:teaser} and~\ref{fig:qual}}
\label{S-sec:supp-J-selection}
\label{S-sec:supp-J_qualitative}

\begin{figure*}[t]
\centering
\includegraphics[width=\textwidth]{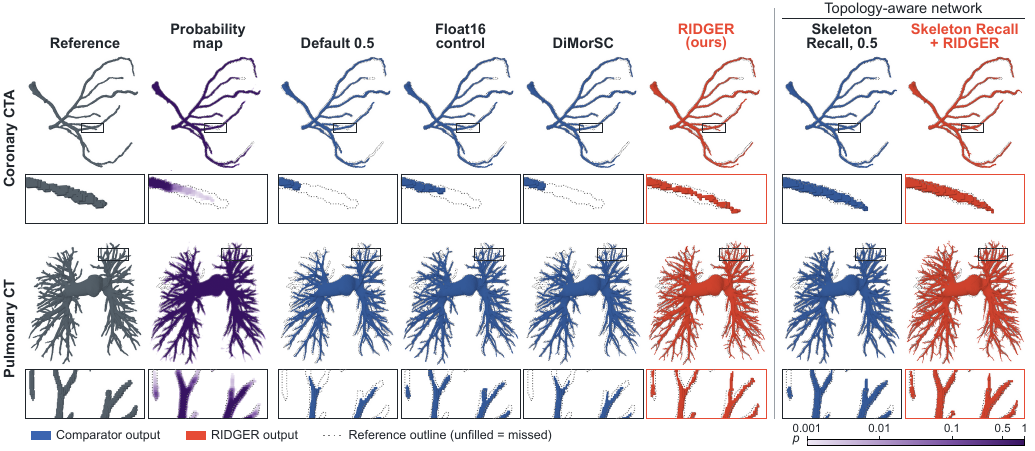}
\caption{\textbf{Full trees of the patients of Fig.~\ref{fig:qual}.} Columns after Default 0.5 read the nnU-Net map (maximum projection, log scale) at RIDGER's per-patient off-reference volume; the right block reads a Skeleton Recall network on the same CT. Each column shows its whole output in one colour (comparators blue, RIDGER vermilion). The dotted outline is the reference, unfilled where missed. Boxes mark the windows magnified in the strips and in Fig.~\ref{fig:qual}.}
\label{S-fig:supp-qual-full}
\end{figure*}

The patients and the windows of the two figures were chosen as follows. The coronary patient of Fig.~\ref{fig:qual} is the test patient whose gain over the float16 control is closest to the median. The pulmonary patient of Fig.~\ref{fig:qual} is an illustrative patient from the middle half of the pulmonary patients ranked by this gain. The pulmonary patient of Fig.~\ref{fig:teaser} was selected for a large gain in reaching rate over 0.5, the third largest of the 49 scored patients. Fig.~\ref{S-fig:supp-qual-full} shows the whole trees of both patients of Fig.~\ref{fig:qual}, with a DiMorSC column and two columns for a Skeleton Recall network. The boxes mark the windows that Fig.~\ref{fig:qual} magnifies. Sec.~\ref{S-sec:supp-Z_gallery} adds ten further patients chosen by a fixed rule across the range of the gain.

\begin{details}
\textbf{Fig.~\ref{fig:teaser}.} The pulmonary patient of Fig.~\ref{fig:teaser} is read on the nnU-Net map. Its \(R\) is 57.8\% at 0.5 and 94.9\% with RIDGER. This gain of 37.1 points is the third largest of the 49 scored patients (median 13.8). Its off-reference volume is 4.66\% at 0.5 and 8.48\% with RIDGER.

\textbf{Coronary patient.} The top row shows the coronary nnU-Net Dice+CE test patient whose gain over the float16 control is closest to the median. The median is taken among patients with a nonzero gain (123 of 160), and ties are broken by case identifier. The selected patient has a gain of 4.48 points over the float16 control (median 4.48).

\textbf{Pulmonary patient.} The bottom row shows an illustrative pulmonary nnU-Net patient whose gain over the float16 control lies in the middle half of the ranked gain distribution. Patients flagged by the slice-spacing screen (Sec.~\ref{S-sec:supp-C-excl}) and patients whose scans also appear in the PARSE dataset are excluded before ranking. The selected patient has a gain of 4.77 points over the float16 control, and the interquartile range of this gain is 1.28 to 11.06.

\textbf{Windows.} Each window magnifies a part of the tree that only RIDGER connects to the root. The coronary inset shows the largest 26-connected component of the voxels that RIDGER and the reference share and that the best 8-bit threshold lacks. The view is cropped to this component with a margin of 6\,mm and projected through its full depth. The pulmonary inset is a window of fixed size. Candidate windows are centred on a grid with 2\,mm spacing, and a window must meet three conditions. (a) It holds reference centreline that RIDGER root-connects. The default mask, the best float16 control and DiMorSC all leave this centreline unconnected to the root. (b) Every RIDGER 26-component in the window holds reference that RIDGER root-connects. (c) RIDGER's off-reference skeleton in the window is nonzero. Each piece of it leaves the reference within 3\,mm of a reference terminal. The length of this skeleton does not exceed the length of condition (a). Windows are ranked by the length of condition (a). The window shown is one of the first six non-overlapping windows that meet all three conditions. The projected slab is 28.7\,mm deep for the pulmonary window and 40\,mm deep for the coronary inset, centred on the chosen component.
\end{details}

\subsection{Per-column measurements}
\label{S-sec:supp-J-columns}

On the nnU-Net map of both patients, RIDGER reaches further than the best 8-bit threshold, the float16 control and DiMorSC, and no comparator uses more off-reference volume \(V\) than RIDGER. Its \(R\) is 97.59 against 93.11 for the float16 control on the coronary patient, and 94.40 against 89.63 on the pulmonary patient. The details list \(R\) and \(V\) of every column of Fig.~\ref{S-fig:supp-qual-full}.

\begin{details}
\textbf{Coronary patient.} \(R\) is 79.95 at 0.5, 80.33 for the best 8-bit threshold, 93.11 for the float16 control and 97.59 for RIDGER. \(V\) is 3.54, 4.07, 3.93 and 4.08 in the same order. DiMorSC reaches 82.55 at 3.70. Only Fig.~\ref{fig:qual} of the main text shows a column of RIDGER drawn with the log-odds width. This column reaches 97.98 at 5.50.

\textbf{Pulmonary patient.} \(R\) is 72.28 at 0.5, 77.62 for the best 8-bit threshold, 89.63 for the float16 control and 94.40 for RIDGER. \(V\) is 1.65, 3.56, 2.93 and 3.59 in the same order. DiMorSC reaches 81.57 at 1.81. RIDGER drawn with the log-odds width reaches 94.45 at 4.69.

\textbf{Skeleton Recall network.} The right block reads a Skeleton Recall network on the same CT. On the coronary patient, its 0.5 mask reaches 99.19 at a \(V\) of 7.62, and RIDGER leaves \(R\) at 99.19 with a \(V\) of 7.82. On the pulmonary patient, its 0.5 mask reaches 91.13 at 4.77, and RIDGER raises \(R\) to 96.70 at 7.13.
\end{details}

\section{Known-truth repair}
\label{S-sec:supp-G_repair}

This section gives the three known-truth tests behind Sec.~\ref{sec:repair}: tips cut from the reference, first-generation branches severed, and the map faded inside a branch.

\textbf{Three tests.} Two tests cut the reference on the test patients. Every readout then starts from the damaged tree in place of the default mask \(S\), and we call this starting mask the damaged core. The first cut removes the last 10\,mm of every terminal (the cut-tip experiment). The second severs every first-generation coronary branch with a 2\,mm gap. The third test leaves the reference whole and fades the map inside these branches (Sec.~\ref{S-sec:supp-repair-fade}). In all three, each readout uses at most RIDGER's off-reference volume \(V\) on each patient, and the reaching rate \(R\) is scored against the intact reference.

\textbf{Conventions.} Pairs of values are in the order Dice+CE, released nnU-Net. Lists of four are in the order of Tab.~\ref{S-tab:supp-repair-truncate}. APP2-style pruning is a selection rule applied to RIDGER's own ridge tree (lower part of Tab.~\ref{tab:ablation}; Sec.~\ref{S-sec:supp-F-app2}). It is not one of the classical post-processors and is reported apart from them.

\subsection{Cut tips}
\begin{figure*}[t]
\centering
\includegraphics[width=\textwidth]{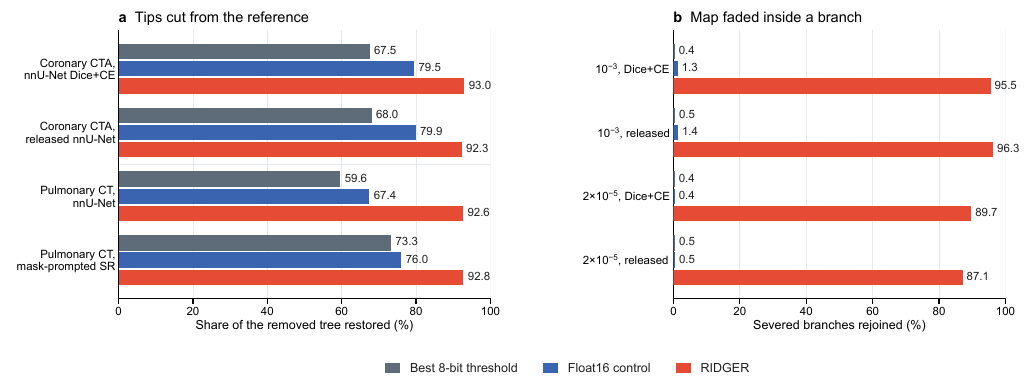}
\caption{\textbf{Known-truth repair.} (a) Share of the removed tree that each readout restores after the last 10\,mm of every terminal is cut from the reference. (b) Share of the severed branches that each readout rejoins after the map is faded inside every first-generation coronary branch, at two fade depths (Sec.~\ref{S-sec:supp-repair-fade}). Every readout stays within RIDGER's off-reference volume.}
\label{S-fig:supp-repair-bars}
\end{figure*}

This test removes the last 10\,mm of every terminal from each reference of the four primary pairs. The restored share is the part of the removed tree that a readout recovers. Because the removed tree is known, this share shows whether a readout puts its volume on the missing vessels.

\begin{figure}[t]
\centering
\includegraphics[width=\linewidth]{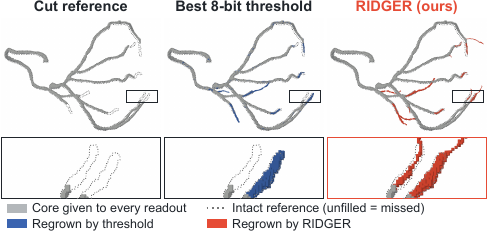}
\caption{\textbf{Repair of cut tips.} Coronary Dice+CE map. For both readouts, the damaged core (grey) is the reference minus the last 10\,mm of every terminal, and each readout stays within RIDGER's off-reference volume. Dotted: intact reference. Recovered: 79.4\% (8-bit) and 99.5\% (RIDGER) on this median-gain patient.}
\label{S-fig:repair}
\end{figure}

RIDGER restores 92.3 to 93.0\% of the missing tree on all four primary pairs (Tab.~\ref{S-tab:supp-repair-truncate}; Fig.~\ref{S-fig:supp-repair-bars}a; Fig.~\ref{S-fig:repair} shows the median-gain coronary patient). The best 8-bit threshold restores 59.6 to 73.3\%, and the float16 control at most 79.9\%. RIDGER thus puts its volume on the removed vessels. A threshold or the control at the same volume leaves at least a fifth of these vessels missing. The classical post-processors and APP2-style pruning are in the details.

\begin{details}
\textbf{Float16 control.} The control restores 67.4 to 79.9\% of the missing tree, more than the best 8-bit threshold on every pair. RIDGER leads the control on every pair, scored on every patient, and every interval is above zero. It also restores 13.5 [11.5, 15.7], 12.4 [10.5, 14.4], 25.2 [19.3, 31.8] and 16.7 [11.8, 22.5] points more of the missing tree. Its \(R\) is higher by 5.09 [4.24, 6.00], 4.69 [3.92, 5.54], 10.46 [8.00, 13.25] and 6.82 [4.78, 9.33] points. The control stops below RIDGER's volume, at a \(V\) of 2.81, 3.11, 5.87 and 9.34.

\textbf{Classical post-processors.} Each of the five is read at RIDGER's per-patient volume on the same damaged coronary trees (160 patients per map), where RIDGER restores 93.0 and 92.3\%. Discrete Morse extraction restores 61.1 and 61.2\%, ridge tracking 44.1 and 22.0\%, minimal-path bridging 34.6 and 35.7\%, two-level hysteresis 28.5 and 36.8\%, and freeze-and-grow 1.0 and 1.1\%. RIDGER's \(R\) exceeds theirs by 13.7 to 38.7 points, every interval above zero.

\textbf{APP2-style pruning.} This selection rule prunes RIDGER's ridge tree with a background threshold and a length cut, and it comes closest to RIDGER. It restores 90.2 and 85.8\% with the setting \(p_\mathrm{bg}=10^{-4.5}\), \(l_{\min}=4\)\,mm of the explanatory analysis in Sec.~\ref{S-sec:supp-F-app2}. RIDGER's \(R\) is higher than that of this rule by 1.82 [1.52, 2.13] and 3.22 [1.92, 4.68] points, and both intervals are above zero.
\end{details}

\begin{table*}[t]
\caption{\textbf{Cut-tip repair on the four primary pairs.} The last 10\,mm of every terminal is removed from each reference; every readout starts from the damaged core. $R$ is the reaching rate, $V$ the off-reference volume in \% of the reference. Restored is the share of the missing tree recovered (\%, patient means). The gain over the best 8-bit threshold is in $R$ points (all-case paired means). The float16 control restores 79.5, 79.9, 67.4 and 76.0\% and the classical post-processors at most 61.2\% (text of this section).}
\label{S-tab:supp-repair-truncate}
\centering\scriptsize
\setlength{\tabcolsep}{3pt}
\begin{tabular}{@{}llrrrrr@{}}
\toprule
Network & Readout & $n$ & $R\!\uparrow$ & $V$ & Restored (\%) & Gain over best 8-bit threshold \\
\midrule
\multirow{3}{*}{Coronary, nnU-Net Dice+CE} & Damaged core & 160 & 57.43 & \textcolor{gray}{0.00} & 0.00 & \\
 & Best 8-bit threshold & 160 & 86.82 & \textcolor{gray}{3.33} & 67.49 & \\
 & \cellcolor{ridger}\textbf{RIDGER} & \cellcolor{ridger}160 & \cellcolor{ridger}96.57 & \cellcolor{ridger}\textcolor{gray}{3.47} & \cellcolor{ridger}92.98 & \cellcolor{ridger}+9.75  \\
\midrule
\multirow{3}{*}{Coronary, released nnU-Net} & Damaged core & 160 & 57.43 & \textcolor{gray}{0.00} & 0.00 & \\
 & Best 8-bit threshold & 160 & 87.12 & \textcolor{gray}{3.74} & 68.03 & \\
 & \cellcolor{ridger}\textbf{RIDGER} & \cellcolor{ridger}160 & \cellcolor{ridger}96.36 & \cellcolor{ridger}\textcolor{gray}{3.87} & \cellcolor{ridger}92.28 & \cellcolor{ridger}+9.24  \\
\midrule
\multirow{3}{*}{Pulmonary, nnU-Net} & Damaged core & 49 & 58.90 & \textcolor{gray}{0.00} & 0.00 & \\
 & Best 8-bit threshold & 49 & 83.68 & \textcolor{gray}{6.36} & 59.63 & \\
 & \cellcolor{ridger}\textbf{RIDGER} & \cellcolor{ridger}49 & \cellcolor{ridger}97.41 & \cellcolor{ridger}\textcolor{gray}{6.60} & \cellcolor{ridger}92.64 & \cellcolor{ridger}+13.73  \\
\midrule
\multirow{3}{*}{Pulmonary, mask-prompted SR} & Damaged core & 50 & 58.89 & \textcolor{gray}{0.00} & 0.00 & \\
 & Best 8-bit threshold & 50 & 89.31 & \textcolor{gray}{9.52} & 73.28 & \\
 & \cellcolor{ridger}\textbf{RIDGER} & \cellcolor{ridger}50 & \cellcolor{ridger}97.32 & \cellcolor{ridger}\textcolor{gray}{10.13} & \cellcolor{ridger}92.75 & \cellcolor{ridger}+8.01  \\
\bottomrule
\end{tabular}
\end{table*}

\subsection{Severed first-generation branches}

This test breaks the starting mask and leaves the map intact. It asks whether RIDGER regrows a branch that the mask has lost but the map still holds. The test cuts a 2\,mm gap in every first-generation coronary branch of the reference. Here the damaged core is the root-connected part of the damaged reference. A branch is severed when the cut leaves its distal part disconnected from the root. This occurs in 141 of 160 patients. On pulmonary CT, the same cut severs no branch, so Tab.~\ref{S-tab:supp-repair-gap} holds the two coronary maps.

RIDGER rejoins 91.7\% and 90.6\% of the severed branches on the Dice+CE and released nnU-Net maps, and the best 8-bit threshold rejoins 64.9\% and 67.3\% (Tab.~\ref{S-tab:supp-repair-gap}). RIDGER also leads the float16 control, with every interval above zero. It rejoins 19.1 [14.1, 24.1] and 16.0 [11.1, 21.0] points more of the severed branches, and its \(R\) is higher by 3.70 [2.70, 4.87] and 3.59 [2.62, 4.63] points.

\begin{details}
APP2-style pruning of RIDGER's ridge tree comes closest to RIDGER and rejoins 87.1 and 85.3\%. Hysteresis, minimal-path bridging, ridge tracking and DiMorSC rejoin 0.2 to 41.6\% at RIDGER's per-patient volume.
\end{details}

\begin{table*}[t]
\caption{\textbf{Severed first-generation coronary branches.} A 2\,mm gap is cut in every first-generation branch. The damaged core is the root-connected part of the damaged reference. Restored is the share of severed branches rejoined (\%, patient means among the 141 of 160 cases with a severed branch). The gain over the best 8-bit threshold is in $R$ points (all-case paired means).}
\label{S-tab:supp-repair-gap}
\centering\scriptsize
\setlength{\tabcolsep}{3pt}
\begin{tabular}{@{}llrrrrr@{}}
\toprule
Network & Readout & $n$ & $R\!\uparrow$ & $V$ & Restored (\%) & Gain over best 8-bit threshold \\
\midrule
\multirow{3}{*}{Coronary, nnU-Net Dice+CE} & Damaged core & 160 & 65.71 & \textcolor{gray}{0.00} & 0.00 & \\
 & Best 8-bit threshold & 160 & 90.84 & \textcolor{gray}{3.50} & 64.88 & \\
 & \cellcolor{ridger}\textbf{RIDGER} & \cellcolor{ridger}160 & \cellcolor{ridger}97.77 & \cellcolor{ridger}\textcolor{gray}{3.61} & \cellcolor{ridger}91.73 & \cellcolor{ridger}+6.92  \\
\midrule
\multirow{3}{*}{Coronary, released nnU-Net} & Damaged core & 160 & 65.71 & \textcolor{gray}{0.00} & 0.00 & \\
 & Best 8-bit threshold & 160 & 91.62 & \textcolor{gray}{3.92} & 67.30 & \\
 & \cellcolor{ridger}\textbf{RIDGER} & \cellcolor{ridger}160 & \cellcolor{ridger}97.73 & \cellcolor{ridger}\textcolor{gray}{4.04} & \cellcolor{ridger}90.60 & \cellcolor{ridger}+6.12  \\
\bottomrule
\end{tabular}
\end{table*}

\subsection{Faded map}
\label{S-sec:supp-repair-fade}
This test breaks branches inside the map and leaves the reference whole. Over a 2\,mm interval of every first-generation coronary branch, the map is lowered to a fade depth \(p_{\mathrm{dip}}\), and the default mask then breaks at the fade. No readout starts from a damaged reference in this test. The default mask, the candidates and RIDGER at the same \(\lambda\) are recomputed from the faded map. Every readout is read at RIDGER's off-reference volume on this map. A branch is severed when its distal part is root-connected in the original default mask but not in the faded one. The rejoin rate is the share of severed branches whose distal part the readout connects to the root again.

RIDGER rejoins 87.1 to 96.3\% of the severed branches at the two primary fade depths with one \(\lambda\), and up to 97.8\% across all four depths. The thresholds and the float16 control rejoin at most 1.4\% (Fig.~\ref{S-fig:supp-repair-bars}b). They judge each voxel alone and stop at the fade, whereas RIDGER judges the branch beyond the fade as a whole. At \(p_{\mathrm{dip}}=10^{-3}\), RIDGER rejoins 95.5 and 96.3\% of severed branches, the best 8-bit threshold 0.4 and 0.5\%, and the float16 control 1.3 and 1.4\%. At \(2\times10^{-5}\), RIDGER rejoins 89.7 and 87.1\%, the thresholds and the float16 control 0.4 and 0.5\%. RIDGER's \(R\) exceeds theirs by about 67 to 70 points. The classical post-processors and APP2-style pruning are in the details.

\begin{details}
\textbf{Protocol.} The map is lowered within the branch's local radius plus one voxel. The two primary depths are \(p_{\mathrm{dip}}=10^{-3}\) and \(2\times10^{-5}\). Both values lie above RIDGER's candidate threshold \(10^{-5}\). With the map unfaded, RIDGER reproduces its result on every patient. On the Dice+CE and released nnU-Net maps, the fade severs at least one branch in 122 and 118 patients (224 and 217 branches).

\textbf{APP2-style pruning.} This selection rule on RIDGER's ridge tree rejoins branches while the fade stays above its background threshold and rejoins none below it. The setting of the cut-tip test (\(p_\mathrm{bg}=10^{-4.5}\), \(l_{\min}=4\)\,mm) rejoins 91.5 and 88.5\% at \(10^{-3}\), where RIDGER leads it by 2.96 [1.93, 4.16] and 5.20 [2.70, 8.14] points of \(R\). At \(2\times10^{-5}\), the fade lies below this threshold and the setting rejoins none.

Two further depths, \(10^{-2}\) and \(10^{-4}\), give the same picture. RIDGER rejoins 97.8 and 97.7\% of the severed branches at \(10^{-2}\), and 91.5 and 91.7\% at \(10^{-4}\).

\textbf{Classical post-processors.} The classical post-processors rejoin few branches at RIDGER's volume. At \(10^{-3}\) the best of them, ridge tracking, rejoins 11.2 and 8.3\%. At \(2\times10^{-5}\), hysteresis, minimal-path bridging and freeze-and-grow rejoin 0.4 and 0.5\%, ridge tracking rejoins 10.3 and 6.9\%, and discrete Morse extraction rejoins none. At this depth RIDGER's \(R\) exceeds theirs by 60.2 to 70.9 points.
\end{details}

\section{Reader study}
\label{S-sec:supp-I_readers}

This section gives the design, the results and the analysis plan of the blinded readings of Sec.~\ref{sec:readers}.

\subsection{Design}
\label{S-sec:supp-I-design}
The pulmonary reading asks whether the pieces that RIDGER and the control add away from the reference are vessels on CT. This question lies behind the pulmonary Dice and clDice loss of Sec.~\ref{sec:readers}. It uses the 49 pulmonary CT test patients scored in the main text (nnU-Net map, RIDGER at the final setting). Items are skeleton pieces of at least 3\,mm. The pieces of the three attached arms are connected to the root in the mask they come from; the distant fragments are not. Four arms are sampled, each by a length-weighted draw (random seed 20261029) from the full census of eligible pieces:
\begin{itemize}
\item \textbf{RIDGER additions} (32 items): pieces of RIDGER's mask whose every point lies more than 1\,mm from the pulmonary reference and from the default mask.
\item \textbf{Control additions} (32 items): the same rule applied to the float16 control at RIDGER's off-reference volume, the comparator of the main text.
\item \textbf{On-reference additions} (6 items): RIDGER pieces whose every point lies within 1\,mm of the reference; these anchor what a vessel looks like.
\item \textbf{Distant fragments} (6 items): raw fragments of the probability map at least 5\,mm from the reference, the default mask and both vessel labels, not connected to the root; these anchor noise.
\end{itemize}

Readers are blind to the arms. They are told the task and the interface, but not that two methods exist or which piece comes from which. Each reader first works two practice items. These have no prescribed answer and are not analysed. The screened readers judge one thing: \emph{Is the whole marked structure a vessel on CT?} The answers are vessel (a pulmonary artery or vein), not a vessel (for example an airway wall, a fissure or a scar), or indeterminate. The reader then gives a confidence rating from 1 to 3 and may add a note. Indeterminate answers stay in the denominator and never count as a success. The results are in Sec.~\ref{S-sec:supp-I-results}: the readers judged 97.9\% of the additions of both methods to be vessels and 16.7\% of the distant fragments.

\begin{details}
At most two pieces per patient enter the RIDGER and control arms, and the 76 items come from 39 patients. Arms are interleaved, and each reader sees an independent random order. For each item the reading tool shows linked CT planes, a curved-planar view of the marked piece, a 3D view with a local and a global pane, and a grey tree. The grey tree is the root-connected part of the default mask. The marked piece is outlined in amber. A dashed line shows where an attached piece would join the grey tree. The line serves only for orientation, and the 6 distant fragments are shown without it. The same tool, text and defaults apply to every item.

\paragraph{Admission.}
Before the study, each candidate received a standard training package. It held an overview of the full artery and vein labels of two public training patients, and six worked items with feedback. The candidate then read a 10-item screening quiz in the same tool. The quiz had five vessels and five non-vessels, taken from public reference labels and from patients that appear nowhere in the study. A candidate passed with at least 8 of 10 correct. Pass or fail was the only feedback, and the flat fee did not depend on any answer. Six candidates took the quiz, and all six passed. These six readers read the full study. All six are clinically trained and have completed a clinical rotation in respiratory medicine. None of them is a practising physician.
\end{details}

\subsection{Results}
\label{S-sec:supp-I-results}
\begin{figure*}[t]
\centering
\includegraphics[width=\textwidth]{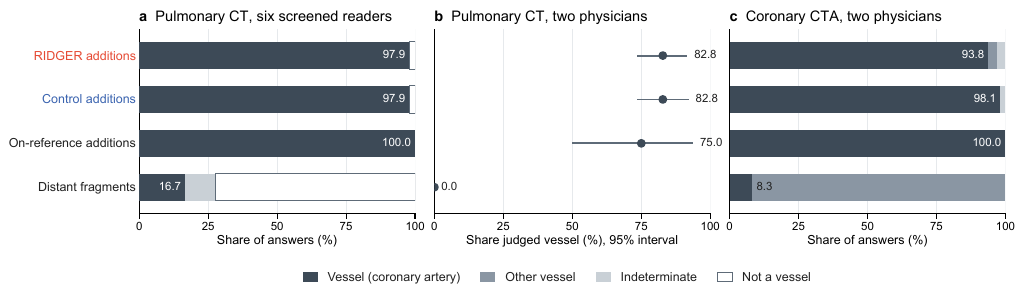}
\caption{\textbf{Answers of the readers for each arm.} (a) Six screened readers, pulmonary CT: share of answers in each category. (b) Two physicians, pulmonary CT: share judged vessel with its 95\% interval. (c) Two physicians, coronary CTA: share of answers in each category. The number at each bar or dot is the share judged vessel (coronary artery in c).}
\label{S-fig:supp-reader-answers}
\end{figure*}
The readers judged the additions of both methods to be vessels and rejected most of the distant fragments. On each of the two addition arms, 188 of 192 answers are vessel (Tab.~\ref{S-tab:readers-main}; Fig.~\ref{S-fig:supp-reader-answers}). The share is 97.9\% for RIDGER and for the control, and the interval of their difference includes zero. Every row of the table is the mean over the six readers with a 95\% patient-cluster bootstrap interval. All 4 rejections on each addition arm come from one reader (S06), whose other answers on these arms agree with the group.

All answers on the on-reference additions are vessel. The distant fragments are called vessel in 16.7\% of answers and not a vessel in 72.2\%, so the readers did not accept everything they were shown. Both addition arms are thus judged like the on-reference additions and unlike noise. Attached items (RIDGER, control and on-reference together) are judged vessel in 98.1\% [96.6, 99.4] of answers. Indeterminate answers make up 0.0\% on the three attached arms and 11.1\% on distant fragments (0.9\% over all 76 items), so excluding them leaves every estimate for the addition arms unchanged.

\begin{table*}[t]
\caption{\textbf{Share of answers judging the marked structure to be a vessel, by arm.} Screened non-physician readers, $n=6$; intervals resample patients.}
\label{S-tab:readers-main}
\centering
\scriptsize
\begin{tabular}{@{}lcccc@{}}
\toprule
Arm & Items & Answers judged vessel & Share (\%) & 95\% CI \\
\midrule
RIDGER additions & 32 & 188/192 & 97.9 & [95.5, 100.0] \\
Control additions (float16) & 32 & 188/192 & 97.9 & [95.6, 100.0] \\
RIDGER $-$ control (points) & & & $+$0.0 & [$-$3.4, $+$3.3] \\
On-reference additions & 6 & 36/36 & 100.0 & [100.0, 100.0] \\
Distant fragments & 6 & 6/36 & 16.7 & [0.0, 37.5] \\
\bottomrule
\end{tabular}
\end{table*}

\begin{details}
\paragraph{Per reader.}
Five of six readers judged every RIDGER, control and on-reference addition to be a vessel (Tab.~\ref{S-tab:readers-per}). Reader S06 judged 87.5\% of each addition arm to be vessels. No reader called more than 33.3\% of the distant fragments vessels.

\begin{table*}[t]
\caption{\textbf{Per-reader share (\%) judged vessel.} Intervals for S01 to S05 on the addition arms are degenerate at 100.0.}
\label{S-tab:readers-per}
\centering
\scriptsize
\begin{tabular}{@{}lcccc@{}}
\toprule
Reader & RIDGER additions & Control additions & On-reference & Distant fragments \\
\midrule
S01 & 100.0 & 100.0 & 100.0 & 0.0 \\
S02 & 100.0 & 100.0 & 100.0 & 16.7 \\
S03 & 100.0 & 100.0 & 100.0 & 16.7 \\
S04 & 100.0 & 100.0 & 100.0 & 16.7 \\
S05 & 100.0 & 100.0 & 100.0 & 33.3 \\
S06 & 87.5 [72.7, 100.0] & 87.5 [73.3, 100.0] & 100.0 & 16.7 \\
\midrule
Mean & 97.9 [95.5, 100.0] & 97.9 [95.6, 100.0] & 100.0 & 16.7 [0.0, 37.5] \\
\bottomrule
\end{tabular}
\end{table*}

\paragraph{Agreement.}
Readers agree closely. Because 98\% of the answers on the two addition arms are ``vessel'', Cohen's $\kappa$ is undefined or zero there, and Gwet's AC1 \cite{gwet2008computing} is reported instead. On the 64 addition items AC1 is 0.957 [0.921, 0.987], raw pairwise agreement is 95.8\%, and readers are unanimous on 56 of 64 items. All eight non-unanimous items are S06's rejections. On all 76 items AC1 is 0.942 [0.889, 0.981], mean pairwise $\kappa$ is 0.746 (range 0.386 to 1.000), and raw agreement is 95.1\%. Dropping any one reader leaves RIDGER's share at 97.5\% or higher and its difference from the control's share at $+$0.0.
\end{details}

\subsection{Physician reading}
\label{S-sec:supp-readers-physicians}
Two physicians (P01 and P02) read all 76 items in the same tool. Both are associate chief physicians in pulmonary vascular disease at a tertiary cardiovascular hospital. The physicians judge more than the screened readers. The screened readers judge whether the marked structure is a vessel (Sec.~\ref{S-sec:supp-I-design}). The physicians name the structure: pulmonary artery, pulmonary vein, airway, other non-vessel, part of the grey-tree vessel, or indeterminate. They also rate its continuity with the grey tree and their confidence. To set the two groups side by side, a physician's answer counts as vessel when it names a pulmonary artery, a pulmonary vein or part of the grey-tree vessel, and as not a vessel when it names an airway or another non-vessel structure. This rule was fixed before the key, the table that links each piece to its arm, was linked to the physicians' files. The physicians form a secondary group and are never pooled with the six screened readers.

\paragraph{Results.}
Neither physician called any RIDGER, control or on-reference addition a non-vessel. Under this rule the physicians judge 82.8\% of RIDGER's additions and 82.8\% of the control's additions to be vessels, so the two methods again receive the same share (Tab.~\ref{S-tab:readers-physicians}). The remaining answers are indeterminate: 17.2\% on each addition arm and 25.0\% on the on-reference additions. No distant fragment was called a vessel. A stricter rule, fixed at the same time, counts the answer ``part of the grey-tree vessel'' as indeterminate. It gives 60.9\% [51.5, 70.6] for RIDGER's additions and 56.2\% [45.8, 67.2] for the control's. The difference is $+$4.7 [$-$11.9, $+$20.2], an interval that includes zero. Under both rules the direction and the contrast agree with the screened readers.

\begin{table*}[t]
\caption{\textbf{Physician reading (secondary group, $n=2$).} Share (\%) of answers judging the marked structure to be a vessel, with the physicians' structure classes counted as in the text. Every other answer on the three attached arms is indeterminate; none is ``not a vessel''.}
\label{S-tab:readers-physicians}
\centering
\scriptsize
\begin{tabular}{@{}lccccc@{}}
\toprule
Reader & RIDGER additions & Control additions & RIDGER $-$ control (points) & On-reference & Distant fragments \\
\midrule
P01 & 93.8 [84.8, 100.0] & 87.5 [76.5, 97.0] & $+$6.2 [$-$7.9, $+$20.4] & 66.7 & 0.0 \\
P02 & 71.9 [54.8, 87.5] & 78.1 [60.6, 93.8] & $-$6.2 [$-$31.5, $+$19.8] & 83.3 & 0.0 \\
\midrule
Mean & 82.8 [73.5, 91.7] & 82.8 [73.5, 92.2] & $+$0.0 [$-$14.6, $+$14.1] & 75.0 [50.0, 93.8] & 0.0 \\
\bottomrule
\end{tabular}
\end{table*}

\begin{details}
\paragraph{Questionnaire.}
The physicians read with the structure-class questionnaire of the first reading package. Addendum 8 replaced it with the single vessel question for the screened readers, and Addendum 12 fixed the counting rule above and the analysis of the two physicians' files (Tab.~\ref{S-tab:readers-addenda} lists the addenda to the analysis plan). The analysis code was committed before the key was first linked to the files.

\paragraph{Analysis.}
An indeterminate answer stays indeterminate, and the continuity rating is not used. Estimators, presence rule and bootstrap are those of the primary analysis. Under the presence rule, a sampled piece counts as present if at least half of its skeleton lies in the final mask.
\end{details}

\subsection{Coronary reading}
\label{S-sec:supp-readers-coronary}
The coronary reading applies the pulmonary design to coronary CTA. It uses the same four arms, drawn in the same way from the 160 test patients: 32 RIDGER additions, 32 control additions at RIDGER's volume, 6 on-reference pieces and 6 distant fragments. It also uses the same tool, blinding, order and presence rule. It asks one question with four answers: coronary artery, other vessel such as a cardiac vein, not a vessel, or indeterminate. The primary endpoint is the share called coronary artery. The analysis code was tested and committed before the key was opened, as in Sec.~\ref{S-sec:supp-readers-plan}. Two physicians returned complete files by the cut-off (Addenda 13 and 14), so the physician group is primary under Addendum 5.

The two physicians call most additions of both methods coronary arteries, and no addition of either method a non-vessel. The share called coronary artery is 93.8\% [85.5, 100] for RIDGER's additions and 98.1\% [94.0, 100] for the control's (Tab.~\ref{S-tab:readers-coronary}). The difference of $-$4.4 [$-$13.5, 3.2] points favours the control, and its interval includes zero. All 32 RIDGER pieces and 27 of the 32 control pieces are present in the final masks. The RIDGER answers that are not coronary artery are other vessel or indeterminate. Every on-reference piece is called coronary artery, and 8.3\% of the distant fragments are; the rest are called other vessels. The two physicians agree on 93.4\% of all 76 items (Cohen's $\kappa$ 0.67 for coronary artery against the rest).

\begin{details}
\textbf{Exploratory analyses.} The pulmonary analyses of Sec.~\ref{S-sec:supp-readers-exploratory} were repeated after the key was opened and change no pre-registered number.
RIDGER's additions contain nearly all of what the control adds and extend beyond it. Beyond 1\,mm of the reference, RIDGER's coronary additions cover, to within 1\,mm, 99.5\% [99.1, 99.7] of the control's off-reference volume and 99.8\% of its off-reference centreline. RIDGER adds 2.1 times the control's volume and 5.0 times its centreline, pooled over the 160 test patients.

Crediting the additions judged coronary artery enlarges the Dice gain and recovers most of the skeleton-precision loss. Without any credit, the pooled Dice change of RIDGER against the default mask is $+$0.41, whereas it is negative on pulmonary CT. Crediting the physicians' share $s=0.938$ [0.855, 1.000] by length share raises it to $+$1.79 [$+$1.67, $+$1.88], which is positive in all 160 patients. The same credit raises the clDice change from $-$1.1 to $+$2.3. Skeleton precision is 94.9 for the default mask and 87.1 for RIDGER, and 93.3 for RIDGER with the credit. Off-reference additions are 6.6\% of RIDGER's skeleton length.

The vein measure of the pulmonary reading has no coronary counterpart, because ImageCAS-X has no cardiac-vein label. The only other-vessel answers on the addition arms fall on two RIDGER pieces, one per physician.

\end{details}

\begin{table}[t]
\caption{\textbf{Coronary reading, physician group (primary; two readers).} Reader-mean shares (\%) of present pieces, with 95\% patient-cluster intervals (10,000 draws) for the primary endpoint.}
\label{S-tab:readers-coronary}
\centering\scriptsize
\setlength{\tabcolsep}{1.5pt}
\begin{tabular*}{\columnwidth}{@{\extracolsep{\fill}}lcccc@{}}
\toprule
Arm & Coronary artery & Other vessel & Not a vessel & Indeterminate \\
\midrule
RIDGER additions (32) & 93.8 [85.5, 100] & 3.1 & 0.0 & 3.1 \\
Control additions (27) & 98.1 [94.0, 100] & 0.0 & 0.0 & 1.9 \\
On-reference pieces (6) & 100.0 & 0.0 & 0.0 & 0.0 \\
Distant fragments (6) & 8.3 & 91.7 & 0.0 & 0.0 \\
\bottomrule
\end{tabular*}
\end{table}

\subsection{Exploratory analyses}
\label{S-sec:supp-readers-exploratory}
These analyses ask what the pulmonary readings imply for Dice and skeleton precision, where the additions lie, and how the additions of the two methods differ. They were specified and run after the key was opened. Intervals are patient-cluster bootstraps.

\paragraph{Reader-completed Dice and precision.}
Counting the additions judged vessel as correct turns the pulmonary Dice loss into a gain. The estimated pooled Dice change of RIDGER against the default mask moves from $-$1.13 to $+$2.14 [$+$2.06, $+$2.21] points. The estimate credits the judged-vessel share $s=0.979$ [0.955, 1.000] to the additions by length share. This raises the reference precision of RIDGER's additions from 0.20 to 0.89. On the patient mean the change is $+$2.26, positive in 49 of 49 patients. The pooled change crosses zero at a share of 0.334, about a third of the readers' share. It crosses zero at 0.198 if the widening next to the default mask is credited as well. The descriptive value fixed in the pre-registration, 0.339, gives $+$0.02.

The same credit removes most of the skeleton-precision loss. Skeleton precision is 90.8 for the default mask and 57.7 for RIDGER, and 87.4 [86.6, 88.0] for RIDGER with the credit, a loss of 3.5 points [2.8, 4.2] in place of 33.2. This interval reflects only the uncertainty in $s$.

\begin{table}[t]
\caption{\textbf{Reader-completed pulmonary Dice and skeleton precision at each reader group's vessel share.} RIDGER on the pulmonary nnU-Net, 49 scored test patients; length-share assumption as above. $\Delta$Dice: pooled change against the default mask (points); positive: patients with a positive change; loss: skeleton precision of the default mask (90.8) minus RIDGER's.}
\label{S-tab:readers-scompletion}
\centering\scriptsize
\setlength{\tabcolsep}{3pt}
\begin{tabular}{@{}lcccc@{}}
\toprule
Reader group & $s$ & $\Delta$Dice & Positive & Loss \\
\midrule
Screened readers & 0.979 & $+$2.14 & 49/49 & 3.5 \\
Physicians & 0.828 & $+$1.64 & 49/49 & 8.1 \\
Physicians, stricter rule & 0.609 & $+$0.92 & 49/49 & 14.7 \\
Dice-neutral share & 0.334 & 0.00 & -- & 23.0 \\
None & 0 & $-$1.13 & 5/49 & 33.2 \\
\bottomrule
\end{tabular}
\end{table}

\paragraph{Vein label.}
RIDGER's additions lie near the vein label less often than the control's. Over the full census of eligible pieces on the 49 patients, the share of added length within 1\,mm of the vein label is 2.0\% for RIDGER and 5.0\% for the control. Among the sampled items, no RIDGER piece lies within 1\,mm of the vein label, against 23.9\% of the control's sampled centreline (4 of 32 control items touch it). This measure uses the label only; readers were not asked to separate arteries from veins.

\begin{details}
\paragraph{The same estimate for each reader group.}
The pooled Dice change is positive at the vessel share of every reader group, and positive in all 49 patients (Tab.~\ref{S-tab:readers-scompletion}). The groups are the screened readers, the two physicians (Sec.~\ref{S-sec:supp-readers-physicians}) and the physicians under the stricter rule. The gain lies between $+$0.92 and $+$2.14 points, and the skeleton-precision loss between 3.5 and 14.7 points, against 33.2 without credit.

\paragraph{Overlap with the control.}
RIDGER's additions contain most of what the control adds. Beyond 1\,mm of the reference, they cover, to within 1\,mm, 84\% of the control's off-reference volume, and RIDGER adds 1.5 times that volume, pooled over the 49 patients.

\paragraph{Where the loss in clDice lies, from the CT alone.}\label{S-sec:supp-readers-ctcost}
Almost all of RIDGER's loss in clDice from added skeleton falls on pieces that look like vessels on CT. A label-free check classes every piece of added skeleton outside the reference from the CT. A piece is vessel-like when its median shell contrast is at least 50\,HU and its median Hessian vesselness is at least 0.10; the thresholds were set earlier on validation patients. Of RIDGER's loss in clDice from added skeleton, 99.6 [99.5, 99.7]\% on the nnU-Net map and 99.5 [99.4, 99.6]\% on mask-prompted SR fall on vessel-like pieces, and none on pieces disconnected from the tree. On the nnU-Net map and on mask-prompted SR, 1.8 and 2.9\% of this added skeleton outside the reference lies on the vein label itself. The shares in the vein-label paragraph above differ because they count the eligible pieces within 1\,mm of that label. The float16 control at RIDGER's volume has the same profile (98.5 and 98.2\% vessel-like) at about a third of the loss.
\end{details}

\subsection{Analysis plan and amendments}
\label{S-sec:supp-readers-plan}
The analysis plan was committed with the sampling before any reader file had been returned, and it fixed everything that decides the result. The analysis code was tested on a synthetic key and committed before the real key was opened. It ran once on the unmodified files, and no analysis step changed after unblinding. The plan names the question (whether the pulmonary loss in Dice reflects unannotated vessels or false positives), the arms, the endpoints and the bootstrap. The primary endpoint is the share of RIDGER's additions judged vessel, averaged over readers, with indeterminate answers kept in the denominator. A second pre-registered estimate is the contrast between RIDGER's and the control's additions, the difference of the two reader-mean shares. Intervals are 95\% patient-cluster bootstraps with 10,000 draws and seed 20261029. Each draw resamples patients and carries all of a patient's items in both arms. Reliability is reported as agreement between readers, and the confidence distribution and the indeterminate rate are secondary endpoints.

\begin{details}
The plan also fixed how readings convert to Dice. Adding voxels raises Dice exactly when their precision against the reference exceeds half the Dice of the default mask. A reader-completed precision credits the judged-vessel share of the additions, by length share, as reference. This conversion is reported as an estimate and not as a re-annotated Dice.

Fourteen addenda were committed to the plan (Tab.~\ref{S-tab:readers-addenda}), each before the key was linked to the reader files that it concerns.

\begin{table*}[t]
\caption{\textbf{Addenda to the reader analysis plan.} Dates are in 2026. Every addendum was committed before the key, which links each piece to its arm, was linked to the reader files that it concerns.}
\label{S-tab:readers-addenda}
\centering\scriptsize
\setlength{\tabcolsep}{3pt}
\begin{tabular}{@{}rlp{0.84\linewidth}@{}}
\toprule
No. & Date & Change \\
\midrule
1 & 10-30 & Off-reference volume judged in three parts; control arm added so both methods are judged alike \\
2 & 10-30 & Five coronary answer options (replaced by four in Addendum 8) \\
3 & 10-30 & Option ``part of the grey-tree vessel'' added (removed in Addendum 8) \\
4 & 10-31 & RIDGER no longer links components that are detached from the root; a sampled piece counts as present if at least half of its skeleton lies in the final mask \\
5 & 10-31 & Two reader groups, physicians and screened non-physician readers, never pooled; physicians primary only if two return full files by the cut-off; quiz and pass rule fixed \\
6 & 11-01 & Five-minute standard walkthrough of the practice items, naming no method \\
7 & 11-01 & Standard artery and vein training package before the quiz \\
8 & 11-01 & One vessel question; primary endpoint becomes the vessel share; Dice break-even only descriptive; quiz and training rebuilt; items, sampling, blinding and order unchanged \\
9 & 11-01 & Readers grouped by who read; coronary admission rules \\
10 & 11-01 & Pulmonary cut-off set; recruitment of screened readers closed at six \\
11 & 11-02 & All six files complete, cut-off moved to the last return; one earlier partial export superseded \\
12 & 11-05 & Two physicians' files from the structure-class questionnaire analysed as a secondary group, with the counting rule \\
13 & 11-09 & Coronary cut-off set and analysis set fixed; coronary analysis code and wording committed \\
14 & 11-10 & Coronary cut-off moved to the commit time; reading closed and analysed on the two physicians who returned \\
\bottomrule
\end{tabular}
\end{table*}

The control arm was sampled at RIDGER's volume before Addendum 4 and keeps that sampling. Every sampled RIDGER, control and on-reference pulmonary piece is present in the final masks. No distant fragment lies within 5\,mm of the final RIDGER mask. No physician of the pulmonary reading had returned a file when the six screened readers' files were analysed. As Addendum 5 provides, the screened non-physician readers are therefore the primary group. The two physicians' files that arrived later form a secondary group.
\end{details}

\subsection{Ethics}
All images come from public, de-identified datasets released by their creators for research; no patient data were collected for this work. The reading asked six screened readers and two physicians to judge structures in these images. It records their judgements and, for admission, their stage of clinical training or professional title, and no other personal or health information. It involves no intervention. For these reasons we did not seek a formal ethics review. Participation was voluntary. Before taking part, the readers knew only that the work was to judge vessels on CT, and the screened readers knew the payment; no written consent form was used. The rules of the task were given with the training package. The screened readers were paid a flat fee that did not depend on their answers, and the physicians were not paid. To keep the reading blind, the purpose of the study and the comparison of two methods were not disclosed. Each screened reader declared that they completed the reading independently. Readers' identities are held only by the organisers and appear nowhere in the paper.

\section{Ablations}
\label{S-sec:supp-F_ablation}

This section gives the ablations of Sec.~\ref{sec:explain} on all four network pairs.

The rules of the lower part of Tab.~\ref{tab:ablation} are defined at the end of Sec.~\ref{S-sec:supp-F-contrasts}. The comparators are the same two as in the main text, the float16 control and the best 8-bit threshold. At equal volume the comparator uses at most RIDGER's off-reference volume \(V\) on each patient, and at equal length at most RIDGER's off-reference centreline length. Gains are at equal volume unless stated, in points of the reaching rate \(R\). Values are test-patient means over 160, 160, 49 and 50 patients (coronary Dice+CE, released coronary nnU-Net, pulmonary nnU-Net, mask-prompted SR). Intervals are 95\% patient bootstraps with 20,000 paired draws and are given for the gains only.

\subsection{Components, other selections and single changes}

Tables~\ref{S-tab:supp-ablation-cor} and~\ref{S-tab:supp-ablation-lung} repeat Tab.~\ref{tab:ablation} for every pair and add the gains over the best 8-bit threshold. Each block of rows has three groups: the components of RIDGER added one at a time, four other selections of the same segments, and single changes to RIDGER.

\textbf{Components.} The rooted subtree produces the gain. Alone it gains 3.12 to 7.49 points over the float16 control on the four pairs. The loop veto removes the spurious loops that the subtree creates. It returns the loop error $E_{\beta_1}$ to the value of the default mask on every pair, for example from 0.51 to 0.06 on coronary Dice+CE. The closure adds a small amount of volume.

\textbf{Other selections and single changes.} Four alternatives choose among RIDGER's own segments and are drawn by its renderer and veto. Table~\ref{S-tab:supp-selection-contrasts} pairs each of them with RIDGER. The single changes alter one setting of RIDGER at a time, and none removes the gain.

\begin{details}
\paragraph{Single changes.} The four single changes are thin rendering, a candidate threshold of $10^{-4}$, $\lambda\times0.3$ and $\lambda\times3$. None pulls the coronary gain over the float16 control below 5.51 points. The lowest coronary values come from raising RIDGER's candidate threshold to $10^{-4}$. The control already uses this threshold. The gains are then 5.51 [4.17, 7.01] and 5.63 [4.53, 6.77] on the two coronary pairs. RIDGER therefore still leads the control when both stop at the same threshold. A threefold change of $\lambda$ in either direction keeps every gain on all four pairs at 2.41 points or more.

\paragraph{Other selections.} \emph{Keep-all} keeps every segment. \emph{Per-segment gate} keeps a segment when $q_b>\theta$ and its parent is kept. The gate $\theta$ is chosen per anatomy on validation patients by the rule that chose $\lambda$ (0.99 coronary, 0.95 pulmonary). \emph{Budget-greedy} grows from the root segments and each time takes the frontier segment with the largest $q_bL_b/\mathrm{Vol}_b$. It keeps that segment while the summed volume stays within RIDGER's own, and drops a segment that does not fit together with its subtree. The row \emph{Control, RIDGER's width} redraws $S$ and each of the 12 float16 members of the control with RIDGER's renderer and veto. The additions of a member are skeletonised, linked to $S$ by shortest paths and drawn at their measured width. The best member at RIDGER's volume is taken, so the control and RIDGER use the same width rule.
\end{details}

\begin{table*}[t]
\caption{\textbf{Ablation on coronary CTA.} Test-patient means (160 patients). Gains are in points at RIDGER's off-reference volume, over the float16 control and over the best 8-bit threshold, with 95\% intervals. Selections share RIDGER's segments and renderer. The redrawn control is read at RIDGER's volume, so only $R$ and $V$ are defined for it. The single changes alter RIDGER one at a time.}
\label{S-tab:supp-ablation-cor}
\centering\scriptsize
\setlength{\tabcolsep}{3pt}
\begin{tabular*}{\textwidth}{@{\extracolsep{\fill}}lrrrrr@{}}
\toprule
 & $R\!\uparrow$ & $V$ & $E_{\beta_1}\!\downarrow$ & Gain over float16 control$\uparrow$ & Gain over best 8-bit threshold$\uparrow$ \\
\midrule
\multicolumn{6}{@{}l}{\textbf{Coronary, nnU-Net Dice+CE}} \\
Default threshold (0.5) & 77.31 & \textcolor{gray}{5.00} & 0.06 & $-$0.49 [$-$0.76, $-$0.27] & 0.00 [0.00, 0.00] \\
\rowcolor{subhead}\multicolumn{6}{@{}l}{\textit{Components, cumulative}} \\
\hspace{0.5em}Rooted subtree & 90.93 & \textcolor{gray}{6.56} & 0.51 & 6.51 [4.99, 8.20] & 12.04 [10.28, 13.87] \\
\hspace{0.5em}+ loop veto & 90.93 & \textcolor{gray}{6.56} & 0.06 & 6.51 [4.99, 8.20] & 12.04 [10.28, 13.87] \\
\rowcolor{ridger}\hspace{0.5em}+ closure = \textbf{RIDGER} & 91.03 & \textcolor{gray}{6.67} & 0.09 & 6.38 [4.85, 8.07] & 12.04 [10.27, 13.87] \\
\rowcolor{subhead}\multicolumn{6}{@{}l}{\textit{Other selections of the same segments}} \\
\hspace{0.5em}Keep all & 91.64 & \textcolor{gray}{7.51} & 0.19 & 5.40 [4.00, 6.94] & 11.68 [9.98, 13.46] \\
\hspace{0.5em}Per-segment gate & 90.08 & \textcolor{gray}{6.61} & 0.06 & 5.61 [4.22, 7.12] & 10.98 [9.36, 12.65] \\
\hspace{0.5em}Budget-greedy & 90.11 & \textcolor{gray}{6.81} & 0.14 & 4.86 [3.49, 6.35] & 10.83 [9.21, 12.55] \\
\hspace{0.5em}Control, RIDGER's width & 88.20 & \textcolor{gray}{5.73} & -- & -- & -- \\
\rowcolor{subhead}\multicolumn{6}{@{}l}{\textit{Single changes to RIDGER}} \\
\hspace{0.5em}Thin rendering & 91.13 & \textcolor{gray}{6.98} & 0.08 & 5.76 [4.29, 7.44] & 11.83 [10.07, 13.68] \\
\hspace{0.5em}Candidate floor $10^{-4}$ & 88.06 & \textcolor{gray}{6.04} & 0.07 & 5.51 [4.17, 7.01] & 9.51 [8.02, 11.09] \\
\hspace{0.5em}$\lambda\times0.3$ & 91.41 & \textcolor{gray}{7.05} & 0.12 & 5.64 [4.30, 7.15] & 11.98 [10.21, 13.80] \\
\hspace{0.5em}$\lambda\times3$ & 90.49 & \textcolor{gray}{6.32} & 0.08 & 6.92 [5.38, 8.59] & 11.73 [10.02, 13.50] \\
\midrule
\multicolumn{6}{@{}l}{\textbf{Coronary, released nnU-Net}} \\
Default threshold (0.5) & 75.13 & \textcolor{gray}{5.27} & 0.10 & $-$0.79 [$-$1.12, $-$0.51] & 0.00 [0.00, 0.00] \\
\rowcolor{subhead}\multicolumn{6}{@{}l}{\textit{Components, cumulative}} \\
\hspace{0.5em}Rooted subtree & 91.33 & \textcolor{gray}{7.13} & 0.50 & 7.49 [6.12, 8.93] & 13.47 [11.76, 15.24] \\
\hspace{0.5em}+ loop veto & 91.17 & \textcolor{gray}{7.13} & 0.10 & 7.33 [5.98, 8.75] & 13.30 [11.63, 15.05] \\
\rowcolor{ridger}\hspace{0.5em}+ closure = \textbf{RIDGER} & 91.25 & \textcolor{gray}{7.24} & 0.17 & 7.01 [5.69, 8.41] & 13.36 [11.67, 15.12] \\
\rowcolor{subhead}\multicolumn{6}{@{}l}{\textit{Other selections of the same segments}} \\
\hspace{0.5em}Keep all & 92.94 & \textcolor{gray}{11.51} & 0.53 & 4.82 [3.58, 6.19] & 12.16 [10.43, 13.95] \\
\hspace{0.5em}Per-segment gate & 88.75 & \textcolor{gray}{7.19} & 0.12 & 4.65 [3.57, 5.80] & 10.96 [9.46, 12.52] \\
\hspace{0.5em}Budget-greedy & 88.64 & \textcolor{gray}{7.96} & 0.26 & 2.37 [0.92, 3.87] & 9.64 [8.14, 11.21] \\
\hspace{0.5em}Control, RIDGER's width & 88.10 & \textcolor{gray}{6.26} & -- & -- & -- \\
\rowcolor{subhead}\multicolumn{6}{@{}l}{\textit{Single changes to RIDGER}} \\
\hspace{0.5em}Thin rendering & 91.35 & \textcolor{gray}{7.61} & 0.14 & 6.31 [4.99, 7.70] & 13.08 [11.41, 14.81] \\
\hspace{0.5em}Candidate floor $10^{-4}$ & 87.87 & \textcolor{gray}{6.65} & 0.17 & 5.63 [4.53, 6.77] & 10.63 [9.18, 12.13] \\
\hspace{0.5em}$\lambda\times0.3$ & 92.39 & \textcolor{gray}{8.23} & 0.21 & 6.17 [4.82, 7.60] & 13.45 [11.74, 15.22] \\
\hspace{0.5em}$\lambda\times3$ & 89.71 & \textcolor{gray}{6.75} & 0.16 & 6.78 [5.41, 8.15] & 12.48 [10.96, 14.03] \\
\bottomrule
\end{tabular*}
\end{table*}

\begin{table*}[t]
\caption{\textbf{Ablation on pulmonary CT.} Test-patient means (49 and 50 patients). Columns as in Tab.~\ref{S-tab:supp-ablation-cor}. Keep-all did not finish on pulmonary patients within the runtime cap and is not reported (--).}
\label{S-tab:supp-ablation-lung}
\centering\scriptsize
\setlength{\tabcolsep}{3pt}
\begin{tabular*}{\textwidth}{@{\extracolsep{\fill}}lrrrrr@{}}
\toprule
 & $R\!\uparrow$ & $V$ & $E_{\beta_1}\!\downarrow$ & Gain over float16 control$\uparrow$ & Gain over best 8-bit threshold$\uparrow$ \\
\midrule
\multicolumn{6}{@{}l}{\textbf{Pulmonary, nnU-Net}} \\
Default threshold (0.5) & 79.86 & \textcolor{gray}{8.75} & 4.51 & 0.00 [$-$0.01, 0.00] & 0.00 [0.00, 0.00] \\
\rowcolor{subhead}\multicolumn{6}{@{}l}{\textit{Components, cumulative}} \\
\hspace{0.5em}Rooted subtree & 96.17 & \textcolor{gray}{12.20} & 4.80 & 6.47 [4.74, 8.32] & 13.13 [10.39, 15.96] \\
\hspace{0.5em}+ loop veto & 96.17 & \textcolor{gray}{12.19} & 4.51 & 6.47 [4.74, 8.32] & 13.12 [10.38, 15.96] \\
\rowcolor{ridger}\hspace{0.5em}+ closure = \textbf{RIDGER} & 96.30 & \textcolor{gray}{12.87} & 4.37 & 5.79 [4.18, 7.54] & 12.86 [10.14, 15.66] \\
\rowcolor{subhead}\multicolumn{6}{@{}l}{\textit{Other selections of the same segments}} \\
\hspace{0.5em}Keep all & -- & -- & -- & -- & -- \\
\hspace{0.5em}Per-segment gate & 95.72 & \textcolor{gray}{16.87} & 4.14 & 2.55 [1.27, 4.01] & 9.88 [7.57, 12.29] \\
\hspace{0.5em}Budget-greedy & 91.39 & \textcolor{gray}{13.79} & 4.31 & $-$0.10 [$-$1.22, 1.15] & 7.21 [5.35, 9.15] \\
\hspace{0.5em}Control, RIDGER's width & 95.16 & \textcolor{gray}{12.04} & -- & -- & -- \\
\rowcolor{subhead}\multicolumn{6}{@{}l}{\textit{Single changes to RIDGER}} \\
\hspace{0.5em}Thin rendering & 96.28 & \textcolor{gray}{12.67} & 4.41 & 6.19 [4.48, 8.04] & 12.95 [10.22, 15.76] \\
\hspace{0.5em}Candidate floor $10^{-4}$ & 96.07 & \textcolor{gray}{12.04} & 4.49 & 6.62 [4.89, 8.46] & 13.13 [10.42, 15.91] \\
\hspace{0.5em}$\lambda\times0.3$ & 97.67 & \textcolor{gray}{17.82} & 4.14 & 4.14 [2.71, 5.75] & 11.33 [8.76, 14.03] \\
\hspace{0.5em}$\lambda\times3$ & 94.00 & \textcolor{gray}{10.91} & 4.49 & 6.05 [4.52, 7.67] & 12.14 [9.77, 14.56] \\
\midrule
\multicolumn{6}{@{}l}{\textbf{Pulmonary, mask-prompted SR}} \\
Default threshold (0.5) & 90.41 & \textcolor{gray}{15.36} & 4.24 & 0.00 [0.00, 0.00] & 0.00 [0.00, 0.00] \\
\rowcolor{subhead}\multicolumn{6}{@{}l}{\textit{Components, cumulative}} \\
\hspace{0.5em}Rooted subtree & 96.27 & \textcolor{gray}{17.52} & 4.38 & 3.12 [2.37, 3.93] & 5.46 [4.21, 6.82] \\
\hspace{0.5em}+ loop veto & 96.26 & \textcolor{gray}{17.52} & 4.24 & 3.11 [2.37, 3.92] & 5.45 [4.21, 6.81] \\
\rowcolor{ridger}\hspace{0.5em}+ closure = \textbf{RIDGER} & 96.35 & \textcolor{gray}{18.14} & 4.32 & 2.86 [2.16, 3.63] & 5.42 [4.18, 6.78] \\
\rowcolor{subhead}\multicolumn{6}{@{}l}{\textit{Other selections of the same segments}} \\
\hspace{0.5em}Keep all & -- & -- & -- & -- & -- \\
\hspace{0.5em}Per-segment gate & 96.01 & \textcolor{gray}{18.75} & 4.20 & 2.30 [1.68, 2.96] & 5.01 [3.87, 6.24] \\
\hspace{0.5em}Budget-greedy & 95.10 & \textcolor{gray}{18.55} & 4.22 & 1.40 [0.77, 2.10] & 4.09 [3.05, 5.22] \\
\hspace{0.5em}Control, RIDGER's width & 94.96 & \textcolor{gray}{17.36} & -- & -- & -- \\
\rowcolor{subhead}\multicolumn{6}{@{}l}{\textit{Single changes to RIDGER}} \\
\hspace{0.5em}Thin rendering & 96.32 & \textcolor{gray}{18.05} & 4.26 & 2.86 [2.16, 3.62] & 5.41 [4.18, 6.76] \\
\hspace{0.5em}Candidate floor $10^{-4}$ & 95.10 & \textcolor{gray}{16.86} & 4.22 & 2.46 [1.88, 3.08] & 4.42 [3.41, 5.50] \\
\hspace{0.5em}$\lambda\times0.3$ & 97.75 & \textcolor{gray}{21.89} & 4.70 & 2.83 [1.78, 3.90] & 6.20 [4.76, 7.75] \\
\hspace{0.5em}$\lambda\times3$ & 94.93 & \textcolor{gray}{16.70} & 4.30 & 2.41 [1.85, 3.02] & 4.28 [3.29, 5.34] \\
\bottomrule
\end{tabular*}
\end{table*}

\subsection{Equal-volume and equal-length selection contrasts}\label{S-sec:supp-F-contrasts}

These contrasts test whether the exact selection, rather than the segments or the width rule, carries RIDGER's lead (Secs.~\ref{sec:explain} and~\ref{sec:robust}). Table~\ref{S-tab:supp-selection-contrasts} pairs each alternative with RIDGER on the same patients. Each entry is RIDGER's gain over the float16 control minus the alternative's gain over the same control. Each entry is given once at equal volume and once at equal length.

\textbf{Equal volume.} The exact selection gains over every alternative on every pair, by 0.56 to 5.89 points. Every interval lies above zero, and the smallest lower bound is 0.21. The redrawn control shares RIDGER's width rule and still falls 1.14 to 3.15 points short, so the width rule alone does not produce the lead.

\textbf{Equal length.} Most contrasts stay above zero at equal length, and five do not. Keep-all has the larger gain on both coronary pairs. Three other intervals include zero. Thirteen of the 16 coronary intervals, at equal volume and at equal length, lie above zero.

\textbf{The lower part of Tab.~\ref{tab:ablation}.} The contrasts above fix each alternative at one setting. The lower part of the main-text table instead gives each alternative its best setting for each patient. Each rule defines a family of selections with one setting per member. Nearest segments first keeps the first $k$ segments in order of their path distance from $S$, and keep-all is its last member. The per-segment gate varies $\theta$. Budget-greedy varies its volume limit. A threshold on the tree keeps a segment when every node on its path to $S$ has a probability of at least $k$, and it varies $k$. \emph{Union greedy} uses RIDGER's segment value with $\mathrm{Vol}_b$ replaced by the volume that the segment adds to the tubes already kept. It repeatedly keeps the frontier segment with the largest positive value, and its setting is $\lambda$. Every member passes through the same renderer, loop veto and closure. For each patient, we consider the sampled members of the family whose off-reference volume does not exceed RIDGER's. The entry is RIDGER's \(R\) minus the highest \(R\) among these members. The APP2-style entry uses the settings chosen on validation patients (Sec.~\ref{S-sec:supp-F-app2}). These best-setting values appear only in Tab.~\ref{tab:ablation}; the tables of this section fix one setting for each rule.

\begin{details}
\paragraph{The five equal-length exceptions.} Keep-all has the larger gain on both coronary pairs, by 0.55 and 1.40 points, with both intervals below zero. It uses its extra volume for width, which the length measure does not count. The interval includes zero in three contrasts: budget-greedy on coronary Dice+CE, the redrawn control on pulmonary nnU-Net (0.65 [$-$0.15, 1.37]) and the per-segment gate on mask-prompted SR (0.08 [$-$0.11, 0.29]).
\end{details}

\begin{table*}[t]
\caption{\textbf{Paired selection contrasts.} RIDGER minus each alternative in the gain over the float16 control (points), with 95\% patient-bootstrap intervals, at equal volume and at equal length. $n$: patients. In the Equal volume column every interval lies above zero, and the rows ``Control, RIDGER's width'' range from 1.14 to 3.15. $^\dagger$ marks the coronary intervals that do not lie above zero. Keep-all did not finish on pulmonary patients within the runtime cap (--).}
\label{S-tab:supp-selection-contrasts}
\centering\scriptsize
\setlength{\tabcolsep}{3pt}
\begin{tabular*}{\textwidth}{@{\extracolsep{\fill}}llrcc@{}}
\toprule
Network & Alternative & $n$ & Equal volume & Equal length \\
\midrule
Coronary, nnU-Net Dice+CE & Keep all & 160 & 0.99 [0.35, 1.66] & $-$0.55 [$-$0.92, $-$0.22]$^\dagger$ \\
 & Per-segment gate & 160 & 0.78 [0.21, 1.39] & 0.93 [0.42, 1.50] \\
 & Budget-greedy & 160 & 1.52 [0.68, 2.47] & 0.52 [$-$0.15, 1.21]$^\dagger$ \\
 & Control, RIDGER's width & 160 & 2.83 [2.03, 3.72] & 2.84 [2.03, 3.72] \\
\midrule
Coronary, released nnU-Net & Keep all & 160 & 2.19 [1.27, 3.13] & $-$1.40 [$-$2.02, $-$0.83]$^\dagger$ \\
 & Per-segment gate & 160 & 2.35 [1.41, 3.37] & 1.98 [1.16, 2.93] \\
 & Budget-greedy & 160 & 4.64 [3.35, 5.99] & 2.48 [1.44, 3.53] \\
 & Control, RIDGER's width & 160 & 3.15 [2.14, 4.28] & 3.00 [2.01, 4.10] \\
\midrule
Pulmonary, nnU-Net & Keep all & 0 & -- & -- \\
 & Per-segment gate & 49 & 3.25 [2.44, 4.10] & 0.79 [0.48, 1.14] \\
 & Budget-greedy & 49 & 5.89 [4.67, 7.20] & 5.57 [4.48, 6.69] \\
 & Control, RIDGER's width & 49 & 1.14 [0.23, 2.02] & 0.65 [$-$0.15, 1.37] \\
\midrule
Pulmonary, mask-prompted SR & Keep all & 0 & -- & -- \\
 & Per-segment gate & 50 & 0.56 [0.33, 0.83] & 0.08 [$-$0.11, 0.29] \\
 & Budget-greedy & 50 & 1.46 [1.05, 1.91] & 1.49 [1.08, 1.94] \\
 & Control, RIDGER's width & 50 & 1.39 [0.93, 1.90] & 1.07 [0.68, 1.53] \\
\bottomrule
\end{tabular*}
\end{table*}

\subsection{APP2-style pruning set on validation patients}
\label{S-sec:supp-F-app2}
The APP2-style entry of Tab.~\ref{tab:ablation} compares RIDGER with APP2-style pruning \cite{xiao2013app2} run on RIDGER's own segments. The rule has two settings, a background threshold \(p_\mathrm{bg}\) and a length cut \(l_{\min}\). One pair \((p_\mathrm{bg}, l_{\min})\) per anatomy is chosen on the validation patients used for RIDGER's \(\lambda\). On the test patients, the comparison is at RIDGER's per-patient volume.

RIDGER reaches further than this rule on every pair. Its \(R\) is higher by 1.73 [1.05, 2.54] and 2.15 [1.25, 3.12] points on the Dice+CE and released coronary maps, where many patients tie (better/tied/worse 76/74/10 and 73/63/24). It is higher by 7.79 [5.94, 9.80] and 5.04 [3.88, 6.31] points on the pulmonary nnU-Net and mask-prompted SR maps, where RIDGER reaches further on all patients but one per map (48/1/0 and 49/1/0). APP2-style pruning draws level when both settings are chosen per patient from the reference.

\begin{details}
\paragraph{Settings.} The background threshold \(p_\mathrm{bg}\) drops the segments whose path minimum falls below it, and the long-segment decomposition is then recomputed. The rule also applies APP2's length cut \(l_{\min}\) and coverage test. The coverage test drops a long segment when kept tubes already cover more than two thirds of its probability-weighted centreline. The pair chosen on the validation patients is the one with the highest mean \(R\) whose mean off-reference volume does not exceed RIDGER's. This gives \(p_\mathrm{bg}=10^{-4.5}\), \(l_{\min}=2\)\,mm on coronary CTA and \(p_\mathrm{bg}=10^{-2}\), \(l_{\min}=2\)\,mm on pulmonary CT. Where the pair exceeds RIDGER's volume on a test patient, \(l_{\min}\) is raised at the same \(p_\mathrm{bg}\) until it fits.

\paragraph{Where the gain lies.} Where RIDGER gains over this rule, it crosses faint stretches. An explanatory analysis covers the 31 coronary test patients whose APP2-style outputs were available. We use the setting \(p_\mathrm{bg}=10^{-4.5}\), \(l_{\min}=4\)\,mm. We measure the reference centreline that RIDGER keeps beyond a stretch whose path minimum falls below \(p_\mathrm{bg}\). It averages 7.5\,mm on the patients where RIDGER reaches further and 1.5\,mm where the two tie (Spearman correlation 0.46, p-value 0.02). On one patient the map falls to \(2\times10^{-5}\) and then holds near \(10^{-3}\) for 21\,mm along the reference. RIDGER keeps that branch and the background threshold \(p_\mathrm{bg}\) drops it (21.2 points of \(R\)). The faded-map test of Sec.~\ref{S-sec:supp-repair-fade} isolates this case.
\end{details}

\subsection{Sensitivity to $\lambda$}

The main text fixes one $\lambda$ per anatomy on validation patients (Sec.~\ref{sec:impl}) and states that the lead does not depend on it (Sec.~\ref{sec:reach}). On the test patients, the rows $\lambda\times0.3$ and $\lambda\times3$ of Tabs.~\ref{S-tab:supp-ablation-cor} and~\ref{S-tab:supp-ablation-lung} give a threefold change in either direction. Table~\ref{S-tab:supp-lambda} adds a much wider sweep of the coronary value, from 0.03 to 3000, for the final readout on 40 coronary validation patients. These are every second of the 80 validation patients by case number, and $\lambda$ itself was chosen on all 80. Across these five decades of $\lambda$, the gain over the float16 control stays between 4.32 and 4.90 points, with every lower bound at least 2.65. A change of \(\lambda\) thus moves RIDGER along the volume axis, from a \(V\) of 8.83 at \(\lambda=0.03\) to 7.18 at \(\lambda=3000\), while its lead over the control barely changes.

\begin{table}[t]
\caption{\textbf{Gain across five decades of $\lambda$.} Coronary nnU-Net Dice+CE, every second validation patient by case number ($n=40$ of 80). Gain over the float16 control at equal volume (points) with 95\% intervals; $R$ and $V$ are patient means.}
\label{S-tab:supp-lambda}
\centering\scriptsize
\setlength{\tabcolsep}{3pt}
\begin{tabular*}{\columnwidth}{@{\extracolsep{\fill}}rrrrr@{}}
\toprule
$\lambda$ & $R\!\uparrow$ & $V$ & Gain over float16 control & Gain over best 8-bit threshold \\
\midrule
0.03 & 89.85 & \textcolor{gray}{8.83} & 4.32 [2.65, 6.17] & 10.17 [7.50, 13.06] \\
0.1 & 89.85 & \textcolor{gray}{8.83} & 4.32 [2.65, 6.17] & 10.17 [7.50, 13.06] \\
0.3 & 89.85 & \textcolor{gray}{8.83} & 4.32 [2.65, 6.17] & 10.17 [7.50, 13.06] \\
1 & 89.85 & \textcolor{gray}{8.82} & 4.32 [2.65, 6.17] & 10.20 [7.53, 13.08] \\
3 & 89.85 & \textcolor{gray}{8.79} & 4.32 [2.65, 6.17] & 10.20 [7.53, 13.08] \\
10 & 89.85 & \textcolor{gray}{8.72} & 4.63 [2.89, 6.53] & 10.41 [7.68, 13.34] \\
30 & 89.85 & \textcolor{gray}{8.59} & 4.64 [2.91, 6.54] & 10.51 [7.76, 13.44] \\
100 & 89.75 & \textcolor{gray}{8.27} & 4.90 [3.11, 6.86] & 10.77 [7.89, 13.88] \\
300 & 89.23 & \textcolor{gray}{7.86} & 4.56 [2.84, 6.47] & 10.69 [7.87, 13.75] \\
1000 & 88.95 & \textcolor{gray}{7.43} & 4.67 [2.90, 6.60] & 10.67 [7.83, 13.76] \\
3000 & 87.87 & \textcolor{gray}{7.18} & 4.55 [2.80, 6.51] & 9.82 [7.22, 12.65] \\
\bottomrule
\end{tabular*}
\end{table}

\subsection{Segment confidence}
\label{S-sec:supp-F-confidence}

Sec.~\ref{sec:prune} argues that probability is a poor confidence, because the flank of a thick vessel can have a higher probability than the ridge of a thin one. RIDGER uses the ridge contrast as the segment confidence \(q_b\). Replacing it with a constant or with the raw probability loses up to 2.1 points of gain over the float16 control. The two variants change only \(q_b\) in the selection objective~(\ref{eq:objective}) and keep the segments, the exact selection, the renderer, the loop veto and the closure. \emph{Constant} sets \(q_b=0.5\), so the selection sees only length and volume. \emph{Raw probability} sets \(q_b\) to the median map probability over the segment's nodes. Each variant has its own \(\lambda\), chosen on validation patients by the rule that chose RIDGER's (Sec.~\ref{S-sec:supp-A-constants}), so each variant is compared at its own best setting.

\textbf{Gain at equal volume.} Table~\ref{S-tab:supp-confidence} gives the three weights on every pair, with \(R\), \(V\), Dice and clDice. Each weight is compared with the float16 control at that weight's own off-reference volume \(V\). Equal volume and equal length are defined in Sec.~\ref{S-sec:supp-F_ablation}. On pulmonary CT, RIDGER leads both variants in this gain on both networks, by 0.72 to 2.08 points, with every interval above zero. On coronary maps, the constant weight trails RIDGER by 0.93 and 1.58 points, with both intervals above zero, and raw probability is level with RIDGER.

\begin{details}
\paragraph{Paired differences.} Each value is RIDGER's gain minus the variant's, in points. On coronary maps (Dice+CE, released nnU-Net), the difference is 0.93 [0.24, 1.77] and 1.58 [0.69, 2.56] for the constant weight, and \(-0.16\) [\(-0.59\), 0.25] and 0.60 [\(-0.09\), 1.37] for raw probability, where both intervals include zero. On pulmonary CT (nnU-Net, mask-prompted SR), it is 2.08 [1.59, 2.60] and 0.72 [0.07, 1.53] for the constant weight, and 1.90 [1.16, 2.76] and 1.68 [1.18, 2.24] for raw probability.

\paragraph{Choice of \(\lambda\).} The grid of each variant extends below RIDGER's because both variants need a smaller \(\lambda\). The chosen values are \(\lambda=1\) for the constant weight on both anatomies, and \(3\times10^{-4}\) (coronary) and \(3\times10^{-3}\) (pulmonary) for raw probability.

\paragraph{The pulmonary variants.} On pulmonary CT the constant weight matches or exceeds RIDGER in the reaching rate \(R\), but it adds 2.4 and 7.2 more points of off-reference volume and loses 10 and 19 clDice points. Raw probability has an \(R\) that is 1.8 and 2.6 points lower. It scores 6.8 and 10.4 clDice points higher than RIDGER, and it adds 2{,}300 and 3{,}100\,mm less centreline outside the reference per patient. clDice counts that centreline against a method, and readers judge 98\% of RIDGER's pulmonary additions to be vessels (Sec.~\ref{S-sec:supp-I_readers}).
\end{details}

\begin{table*}[t]
\caption{\textbf{Segment confidence.} Test-patient means (160, 160, 49 and 50 patients). Only \(q_b\) changes; each weight uses its own validation-chosen \(\lambda\). Gains are in points with 95\% patient-bootstrap intervals. Gains over the float16 control and the best 8-bit threshold are at the variant's own off-reference volume. Gains over the float16 control are also given at the variant's own off-reference length.}
\label{S-tab:supp-confidence}
\centering\scriptsize
\setlength{\tabcolsep}{3pt}
\begin{tabular*}{\textwidth}{@{\extracolsep{\fill}}lrrrrrrrr@{}}
\toprule
& & & & & & \multicolumn{2}{c}{Gain at equal \(V\) over} & \multicolumn{1}{c}{Gain at equal length over} \\
\cmidrule(lr){7-8}\cmidrule(l){9-9}
Weight \(q_b\) & \(\lambda\) & \(R\!\uparrow\) & \(V\) & Dice\(\!\uparrow\) & clDice\(\!\uparrow\) & float16 control & best 8-bit threshold & float16 control \\
\midrule
\multicolumn{9}{@{}l}{\textbf{Coronary, nnU-Net Dice+CE}} \\
\rowcolor{ridger}Ridge contrast (\textbf{RIDGER}) & 300 & 91.03 & \textcolor{gray}{6.67} & 90.06 & 90.94 & 6.38 [4.85, 8.07] & 12.04 [10.27, 13.87] & 2.48 [1.67, 3.38] \\
Constant 0.5 & 1 & 90.60 & \textcolor{gray}{6.80} & 89.27 & 89.73 & 5.45 [4.09, 6.92] & 11.37 [9.74, 13.04] & 2.76 [1.77, 3.81] \\
Raw probability & 0.0003 & 90.85 & \textcolor{gray}{6.51} & 90.13 & 91.34 & 6.54 [5.03, 8.18] & 11.91 [10.17, 13.76] & 2.39 [1.55, 3.36] \\
\midrule
\multicolumn{9}{@{}l}{\textbf{Coronary, released nnU-Net}} \\
\rowcolor{ridger}Ridge contrast (\textbf{RIDGER}) & 300 & 91.25 & \textcolor{gray}{7.24} & 89.89 & 90.87 & 7.01 [5.69, 8.41] & 13.36 [11.67, 15.12] & 2.58 [1.57, 3.68] \\
Constant 0.5 & 1 & 92.25 & \textcolor{gray}{8.74} & 88.82 & 87.58 & 5.42 [3.97, 6.96] & 12.69 [11.02, 14.44] & 3.28 [2.06, 4.56] \\
Raw probability & 0.0003 & 91.91 & \textcolor{gray}{7.85} & 89.67 & 90.12 & 6.41 [5.17, 7.74] & 13.24 [11.63, 14.89] & 3.07 [2.10, 4.11] \\
\midrule
\multicolumn{9}{@{}l}{\textbf{Pulmonary, nnU-Net}} \\
\rowcolor{ridger}Ridge contrast (\textbf{RIDGER}) & 30 & 96.30 & \textcolor{gray}{12.87} & 86.45 & 70.38 & 5.79 [4.18, 7.54] & 12.86 [10.14, 15.66] & $-$0.40 [$-$1.18, 0.17] \\
Constant 0.5 & 1 & 96.06 & \textcolor{gray}{15.26} & 85.28 & 60.16 & 3.71 [2.33, 5.20] & 11.07 [8.50, 13.72] & $-$2.06 [$-$2.85, $-$1.43] \\
Raw probability & 0.003 & 94.46 & \textcolor{gray}{12.82} & 86.57 & 77.20 & 3.89 [2.81, 4.98] & 11.01 [8.80, 13.28] & $-$0.31 [$-$1.02, 0.22] \\
\midrule
\multicolumn{9}{@{}l}{\textbf{Pulmonary, mask-prompted SR}} \\
\rowcolor{ridger}Ridge contrast (\textbf{RIDGER}) & 30 & 96.35 & \textcolor{gray}{18.14} & 85.87 & 70.78 & 2.86 [2.16, 3.63] & 5.42 [4.18, 6.78] & 0.10 [$-$0.63, 0.65] \\
Constant 0.5 & 1 & 97.67 & \textcolor{gray}{25.34} & 83.07 & 51.55 & 2.14 [1.15, 3.14] & 5.64 [4.30, 7.11] & 0.81 [$-$0.02, 1.57] \\
Raw probability & 0.003 & 93.76 & \textcolor{gray}{16.81} & 86.41 & 81.19 & 1.18 [0.89, 1.49] & 3.09 [2.41, 3.82] & $-$0.15 [$-$0.31, 0.00] \\
\bottomrule
\end{tabular*}
\end{table*}

\section{Robustness}
\label{S-sec:supp-R_robust}

This section repeats the comparison of Sec.~\ref{sec:robust} under changed conditions: equal off-reference length, other scoring windows, a boundary tolerance, changed probability maps, another connectivity, and the external ASOCA cohort. Each change is made one at a time. Units, intervals and counts are as in Sec.~\ref{S-sec:supp-E_results}.

\subsection{Equal-length comparison}
\label{S-sec:supp-E-length}

The off-reference volume counts width, so thin additions add little to it. The off-reference length counts every added millimetre of branch and no width (Sec.~\ref{sec:measurement}), so it tests whether RIDGER's lead comes from drawing thin branches (Sec.~\ref{sec:robust}). Each comparator is held to RIDGER's off-reference length on each patient. Table~\ref{S-tab:supp-E-length} gives the paired gains behind the equal-length rows of Tab.~\ref{tab:currencies}.

\textbf{Best 8-bit threshold.} At RIDGER's length, RIDGER beats the best 8-bit threshold by 1.48 to 7.42 points. Every interval excludes zero. This lead does not depend on the width at which RIDGER draws its branches.

\textbf{Float16 control.} RIDGER leads the control by 2.48 and 2.58 points on coronary CTA and by 1.62 on the airway (Sec.~\ref{S-sec:supp-H_domains}), with intervals above zero. On pulmonary CT the two are level, and both intervals include zero.

\begin{details}
The best 8-bit threshold at RIDGER's length is the member with the highest \(R\) among those whose off-reference length does not exceed RIDGER's (Sec.~\ref{S-sec:supp-D-length}). Against the float16 control, the coronary medians of the gain are 0.00. The pulmonary gains over the control are $-$0.40 [$-$1.18, 0.17] for the nnU-Net and 0.10 [$-$0.63, 0.65] for mask-prompted SR.
\end{details}

\begin{table*}[t]
\caption{\textbf{Equal off-reference length.} Gain of RIDGER in $R$ (points) over the float16 control and over the best 8-bit threshold of the same map, both at each patient's RIDGER length: mean [95\% CI], patients better/tied/worse and median.}
\label{S-tab:supp-E-length}
\centering\scriptsize
\setlength{\tabcolsep}{3pt}
\begin{tabular*}{\textwidth}{@{\extracolsep{\fill}}lrlrrlrr@{}}
\toprule
 & & \multicolumn{3}{c}{Float16 control} & \multicolumn{3}{c}{Best 8-bit threshold} \\
\cmidrule(lr){3-5}\cmidrule(l){6-8}
Network & $n$ & Mean [CI] & B/T/W & Median & Mean [CI] & B/T/W & Median \\
\midrule
\rowcolor{subhead}\multicolumn{8}{@{}l}{\textit{Coronary CTA}} \\
\hspace{0.5em}nnU-Net Dice+CE & 160 & 2.48 [1.67, 3.38] & 70/56/34 & 0.00 & 6.39 [5.05, 7.88] & 120/30/10 & 2.86 \\
\hspace{0.5em}released nnU-Net & 160 & 2.58 [1.57, 3.68] & 63/57/40 & 0.00 & 7.00 [5.69, 8.43] & 121/28/11 & 3.78 \\
\rowcolor{subhead}\multicolumn{8}{@{}l}{\textit{Pulmonary CT}} \\
\hspace{0.5em}nnU-Net & 49 & $-$0.40 [$-$1.18, 0.17] & 16/1/32 & $-$0.27 & 1.48 [0.52, 2.56] & 34/1/14 & 0.33 \\
\hspace{0.5em}mask-prompted SR & 50 & 0.10 [$-$0.63, 0.65] & 25/1/24 & 0.01 & 2.64 [1.92, 3.44] & 47/1/2 & 1.48 \\
\rowcolor{subhead}\multicolumn{8}{@{}l}{\textit{Airway CT}} \\
\hspace{0.5em}nnU-Net & 18 & 1.62 [0.68, 2.73] & 13/1/4 & 0.86 & 7.42 [5.13, 10.05] & 17/1/0 & 6.37 \\
\bottomrule
\end{tabular*}
\end{table*}

\subsection{Distal window and coverage tolerance}
\label{S-sec:supp-F-window}

\textbf{What is varied.} \(R\) counts the reference centreline that lies within the coverage tolerance \(d_\mathrm{on}\) (1\,mm) of the prediction, inside a distal window (20\,mm) upstream of each terminal. Neither setting enters RIDGER's objective. We rescore the final masks of the four primary pairs with windows of 10, 20 and 40\,mm and the whole tree, and with tolerances of 0.5, 1 and 2\,mm. Four pairs, four windows and three tolerances give 48 cells. Tab.~\ref{S-tab:supp-window} gives the gains at \(d_\mathrm{on}=1\)\,mm.

\textbf{Equal off-reference volume.} Every gain over the best 8-bit threshold and over the float16 control stays above zero in all 48 cells. The smallest lower bound of an interval is 2.28 points against the threshold and 1.09 points against the control. The lead at equal volume therefore holds for every window and tolerance tested. The gains shrink as the window widens, because a wider window adds proximal centreline that every readout already reaches.

\textbf{Equal off-reference length.} RIDGER stays above the best 8-bit threshold in every cell for both coronary networks and for mask-prompted SR. The same holds for the pulmonary nnU-Net at tolerances of 1 and 2\,mm. The exception is the pulmonary nnU-Net at 0.5\,mm, which ties the threshold from 20\,mm on. Against the float16 control at equal length, the coronary gains stay above zero. The pulmonary gains lie on both sides of zero, as at 20\,mm (Tab.~\ref{S-tab:supp-E-length}).

\begin{details}
\paragraph{Comparators in each cell.} The best 8-bit threshold is the exact maximum of \(R\) over every admissible 8-bit level. A level is admissible when its off-reference volume does not exceed RIDGER's. At equal length, a level is admissible when its off-reference length does not exceed RIDGER's. The float16 control is selected the same way from its stored members. At 20\,mm and 1\,mm the per-patient gains equal those of the main text exactly. In that cell RIDGER is better than the best 8-bit threshold on 144 and 149 of 160 coronary patients (worse on 2 and 1) and on 48 of 49 and 49 of 50 pulmonary patients (worse on none).

\paragraph{Coverage tolerance.} The tolerance matters little: 0.5 and 2\,mm move no mean by more than 0.95 points.

\paragraph{Exceptions at equal length.} For the pulmonary nnU-Net at 0.5\,mm, the gain over the best 8-bit threshold is 0.84 [$-$0.18, 1.95] at 20\,mm. The pulmonary gains over the float16 control lie between $-$1.09 and 0.43 points.
\end{details}

\begin{table*}[t]
\caption{\textbf{Gains in \(R\) across distal windows}, \(d_\mathrm{on}=1\)\,mm, test patients. RIDGER minus the comparator at RIDGER's per-patient off-reference volume or length, in points, with 95\% patient-bootstrap intervals (20,000 draws, seed 20261027).}
\label{S-tab:supp-window}
\centering\scriptsize
\setlength{\tabcolsep}{3pt}
\begin{tabular}{@{}lcccc@{}}
\toprule
Comparator & 10\,mm & 20\,mm & 40\,mm & Whole tree \\
\midrule
\rowcolor{subhead}\multicolumn{5}{@{}l}{\textit{Coronary nnU-Net Dice+CE} (\(n=160\))} \\
\quad Equal \(V\), best 8-bit & 15.68 [13.68, 17.73] & 12.04 [10.27, 13.87] & 8.25 [6.77, 9.82] & 5.12 [3.90, 6.50] \\
\quad Equal \(V\), float16 control & 7.86 [6.20, 9.67] & 6.38 [4.85, 8.07] & 4.71 [3.39, 6.18] & 3.23 [2.08, 4.58] \\
\quad Equal length, best 8-bit & 8.22 [6.67, 9.87] & 6.39 [5.05, 7.88] & 4.38 [3.30, 5.61] & 2.70 [1.85, 3.71] \\
\quad Equal length, float16 control & 3.24 [2.23, 4.32] & 2.48 [1.67, 3.38] & 1.75 [1.10, 2.50] & 1.06 [0.59, 1.66] \\
\midrule
\rowcolor{subhead}\multicolumn{5}{@{}l}{\textit{Released coronary nnU-Net} (\(n=160\))} \\
\quad Equal \(V\), best 8-bit & 16.73 [14.76, 18.76] & 13.36 [11.67, 15.12] & 9.41 [8.03, 10.89] & 5.52 [4.49, 6.66] \\
\quad Equal \(V\), float16 control & 8.66 [7.12, 10.31] & 7.01 [5.69, 8.41] & 4.96 [3.91, 6.10] & 3.04 [2.25, 3.94] \\
\quad Equal length, best 8-bit & 8.77 [7.25, 10.39] & 7.00 [5.69, 8.43] & 4.87 [3.84, 6.02] & 2.92 [2.15, 3.82] \\
\quad Equal length, float16 control & 2.96 [1.78, 4.22] & 2.58 [1.57, 3.68] & 2.08 [1.24, 3.04] & 1.43 [0.75, 2.26] \\
\midrule
\rowcolor{subhead}\multicolumn{5}{@{}l}{\textit{Pulmonary nnU-Net} (\(n=49\))} \\
\quad Equal \(V\), best 8-bit & 19.09 [15.26, 23.03] & 12.86 [10.14, 15.66] & 8.74 [6.84, 10.72] & 7.56 [5.90, 9.29] \\
\quad Equal \(V\), float16 control & 9.61 [7.18, 12.27] & 5.79 [4.18, 7.54] & 3.77 [2.61, 5.02] & 3.24 [2.23, 4.34] \\
\quad Equal length, best 8-bit & 2.98 [1.59, 4.63] & 1.48 [0.52, 2.56] & 0.82 [0.06, 1.61] & 0.68 [0.03, 1.36] \\
\quad Equal length, float16 control & $-$0.17 [$-$1.03, 0.59] & $-$0.40 [$-$1.18, 0.17] & $-$0.44 [$-$1.10, 0.03] & $-$0.40 [$-$0.94, $-$0.00] \\
\midrule
\rowcolor{subhead}\multicolumn{5}{@{}l}{\textit{Pulmonary mask-prompted SR} (\(n=50\))} \\
\quad Equal \(V\), best 8-bit & 8.66 [6.72, 10.76] & 5.42 [4.18, 6.78] & 3.65 [2.81, 4.57] & 3.14 [2.41, 3.93] \\
\quad Equal \(V\), float16 control & 4.75 [3.62, 5.98] & 2.86 [2.16, 3.63] & 1.90 [1.42, 2.42] & 1.63 [1.22, 2.08] \\
\quad Equal length, best 8-bit & 4.34 [3.18, 5.64] & 2.64 [1.92, 3.44] & 1.76 [1.27, 2.31] & 1.51 [1.09, 1.98] \\
\quad Equal length, float16 control & 0.40 [$-$0.41, 1.10] & 0.10 [$-$0.63, 0.65] & $-$0.03 [$-$0.66, 0.41] & $-$0.03 [$-$0.55, 0.34] \\
\bottomrule
\end{tabular}
\end{table*}

\subsection{Boundary tolerance}
\label{S-sec:supp-F-tolerance}

\textbf{Tolerant volume.} Annotated boundaries are uncertain by about a voxel, and the off-reference volume \(V\) counts that disagreement like any other error. We therefore let \(V\) forgive one voxel of boundary disagreement. This boundary tolerance \(\tau\) applies to \(V\) and is distinct from the coverage tolerance \(d_\mathrm{on}\) of the reaching rate \(R\) (Sec.~\ref{S-sec:supp-F-window}). The \emph{off-reference volume beyond tolerance} keeps only the predicted voxels farther than \(\tau\) from the reference,
\[
V_\tau(P)=100\,\bigl|\{x\in P:\ \operatorname{dist}(x,G)>\tau\}\bigr|/|G|,
\]
with \(\operatorname{dist}\) the exact Euclidean distance and \(\tau\) one native in-plane voxel, in the spirit of tolerance-based boundary scores \cite{nikolov2021clinically,maierhein2024metrics}. With this definition \(V=V_0\), and the volume that the tolerance forgives is the band \(V-V_\tau\). We repeat every equal-volume comparison (defined in Sec.~\ref{S-sec:supp-F_ablation}) with \(V_\tau\) in place of \(V\) and change nothing else (Tab.~\ref{S-tab:supp-vtau}).

\begin{table}[t]
\caption{\textbf{Gains at equal off-reference volume beyond tolerance.} RIDGER's gain in \(R\) (points) at equal \(V\) and at equal \(V_\tau\) (\(\tau\) = one in-plane voxel), test patients; mean [95\% CI].}
\label{S-tab:supp-vtau}
\centering\scriptsize
\setlength{\tabcolsep}{2pt}
\begin{tabular*}{\columnwidth}{@{\extracolsep{\fill}}lrrrr@{}}
\toprule
 & \multicolumn{2}{c}{Best 8-bit threshold} & \multicolumn{2}{c}{Float16 control} \\
\cmidrule(lr){2-3}\cmidrule(l){4-5}
Network & equal \(V\) & equal \(V_\tau\) & equal \(V\) & equal \(V_\tau\) \\
\midrule
Coronary, Dice+CE & 12.04 & 9.35 [7.74, 11.06] & 6.38 & 5.64 [4.24, 7.19] \\
Coronary, released & 13.36 & 10.10 [8.59, 11.67] & 7.01 & 6.28 [4.98, 7.69] \\
Pulmonary, nnU-Net & 12.86 & 10.27 [7.96, 12.67] & 5.79 & 5.19 [3.72, 6.80] \\
Pulmonary, m.-p.\ SR & 5.42 & 4.93 [3.80, 6.15] & 2.86 & 2.59 [1.96, 3.28] \\
\bottomrule
\end{tabular*}
\end{table}

Every gain is smaller at equal \(V_\tau\), and every interval stays above zero. The gains over the float16 control are 2.59 to 6.28 points on the four pairs, against 2.86 to 7.01 at equal \(V\). The lead at equal volume therefore does not depend on how \(V\) scores voxels at the reference boundary.

\begin{details}
\paragraph{Protocol.} The repeat keeps the same families, tie rules and selector as at equal \(V\) and rebuilds all members exactly.

\paragraph{The band.} The band holds a larger share of the comparators' volume than of RIDGER's, so the boundary tolerance forgives the comparators more. The band holds 38 to 45\% of RIDGER's \(V\). Almost all of this lies on the default mask's own boundary. RIDGER's additions put 0.08 to 0.25\% of the reference volume in the band. For the float16 member the band holds 42 to 52\% of \(V\), and for the 8-bit member chosen at RIDGER's volume 46 to 62\%. Thresholds place more of their volume along the boundary, and RIDGER places its volume on branches away from it.

\paragraph{Tolerance in millimetres.} A fixed tolerance in millimetres keeps the gains at a similar size. With \(\tau\) set to 0.5 and \(1.0\,\)mm, the gains over the float16 control on coronary CTA are 5.41 and 5.21 for Dice+CE and 6.24 and 6.32 for the released nnU-Net. On pulmonary CT they are 5.79 and 4.71 for nnU-Net and 2.86 and 2.50 for mask-prompted SR. No pulmonary voxel is finer than 0.5\,mm, so there \(V_{0.5}=V\).
\end{details}

\subsection{Changes to the probability map}
\label{S-sec:supp-F-mapchanges}

RIDGER reads the map of one network, so its gain could depend on how sharp, smooth or noisy that map is.

\textbf{Design.} We change the coronary nnU-Net Dice+CE map (seed 1) in three ways, each applied to the log-odds map \(\ell\). Temperature gives \(\ell/T\) and tests a map that is sharper (\(T<1\)) or softer (\(T>1\)) than the network's own. Blur applies a Gaussian of standard deviation \(w\) mm to \(\ell\) and tests a map with less fine detail. Noise gives \(\ell+s\,\eta\), with \(\eta\) a smoothed Gaussian noise field, and tests a map with spatially correlated errors. The same RIDGER, with all constants unchanged, reads every changed map. The gain is RIDGER's reaching rate \(R\) minus the \(R\) of the best 8-bit threshold of the changed map. The comparison is at equal volume, as defined in Sec.~\ref{S-sec:supp-F_ablation} (Tab.~\ref{S-tab:supp-mapchanges}). On the unchanged map the gain is 12.04 points on all 160 test patients.

\textbf{Noise.} Noise up to \(s=2\) logits leaves the gain where it was, at 12.04 to 12.33 points, with 145 of 160 patients better. At \(s=4\) the gain is smaller and its interval stays above zero, at 8.91 [7.49, 10.39] with 138 patients better. The noise then opens new candidate paths. RIDGER's \(V\) rises from 6.67 to 18.04, and its clDice falls from 90.94 to 72.81.

\textbf{Temperature.} Sharpening shrinks the gain and keeps it positive. The gain falls from 12.04 to 8.89 for \(T=0.75\) and to 5.33 for \(T=0.5\). Softening removes the gain. For \(T\geq1.5\), RIDGER returns the default mask on every patient of a seeded 20-patient subset.

\textbf{Blur.} RIDGER recovers the thin vessels that light blur pulls below 0.5. It does not recover those that heavier blur erases from the map. At \(w=0.5\)\,mm the \(R\) of the default mask falls from 77.31 to 22.74, and RIDGER brings it back to 81.99, a gain of 44.47 over the best 8-bit threshold of the blurred map. From \(w=1\)\,mm on, the distal vessels are gone from the map, and neither readout recovers the tree. The gain therefore holds as long as the map keeps the distal ridges and a low background, which noise, sharpening and light blur leave in place.

\begin{details}
\paragraph{Settings.} The noise field \(\eta\) is unit-variance Gaussian noise smoothed with a Gaussian of 1\,mm and renormalised. Each patient has one noise field for every \(s\). RIDGER reads every changed map with \(\lambda=300\). The default mask, its anchor and the 8-bit thresholds are recomputed on each changed map.

\paragraph{Sharpening against the float16 control.} Over the float16 control of the unchanged map, the gains at \(T=0.75\) and \(T=0.5\) are 5.44 and 3.16, against 6.38 without sharpening.

\paragraph{Softening.} A temperature of \(T\geq1.5\) lifts every voxel above the candidate threshold \(10^{-5}\), because the stored log-odds are clipped at \(-16\). These maps are read without the map normalisation, so the candidates fill the volume, as on the unnormalised maps of CAS-Net (Sec.~\ref{S-sec:supp-J-casnet}).

\paragraph{Blur.} At \(w=0.5\)\,mm, when RIDGER starts from the default mask of the unchanged map, the gain over the best 8-bit threshold is 6.62 [5.45, 7.86]. From \(w=1\)\,mm on, the default mask reaches 1.36 or less and RIDGER 6.07 or less.
\end{details}

\begin{table*}[t]
\caption{\textbf{RIDGER on changed coronary maps.} nnU-Net Dice+CE (seed 1), ImageCAS-X test patients. \(R_{0.5}\): \(R\) of the changed map's default mask. RIDGER's \(R\), \(V\), Dice and clDice in \%. Gain: RIDGER minus the best 8-bit threshold of the changed map at RIDGER's per-patient off-reference volume, in points of \(R\), with 95\% patient-bootstrap intervals (10,000 draws) and patients better/tied/worse. Last column: gain over the float16 control of the unchanged map. Temperatures 1.5 to 4 were run on a seeded 20-patient subset and give identical rows.}
\label{S-tab:supp-mapchanges}
\centering\scriptsize
\setlength{\tabcolsep}{3pt}
\begin{tabular}{@{}llccccccrc@{}}
\toprule
Change & \(n\) & \(R_{0.5}\) & \(R\!\uparrow\) & \(V\) & Dice$\!\uparrow$ & clDice$\!\uparrow$ & Gain over best 8-bit threshold & B/T/W & Gain over float16 control \\
\midrule
None & 160 & 77.31 & 91.03 & \textcolor{gray}{6.67} & 90.06 & 90.94 & 12.04 [10.34, 13.91] & 144/14/2 & 6.38 [4.89, 8.05] \\
\midrule
Noise \(s=0.5\) & 160 & 76.96 & 91.34 & \textcolor{gray}{6.71} & 90.05 & 90.96 & 12.29 [10.53, 14.19] & 145/13/2 & \\
Noise \(s=1\) & 160 & 76.81 & 91.07 & \textcolor{gray}{6.72} & 89.98 & 91.02 & 12.33 [10.54, 14.23] & 145/14/1 & \\
Noise \(s=2\) & 160 & 76.33 & 90.60 & \textcolor{gray}{6.90} & 89.74 & 90.91 & 12.04 [10.38, 13.84] & 145/12/3 & \\
Noise \(s=4\) & 160 & 73.16 & 89.26 & \textcolor{gray}{18.04} & 84.37 & 72.81 & 8.91 [7.49, 10.39] & 138/17/5 & \\
\midrule
\(T=0.75\) & 160 & 77.31 & 87.17 & \textcolor{gray}{5.85} & 90.11 & 92.34 & 8.89 [7.51, 10.31] & 140/18/2 & 5.44 [4.23, 6.74] \\
\(T=0.5\) & 160 & 77.31 & 83.07 & \textcolor{gray}{5.41} & 89.81 & 92.44 & 5.33 [4.34, 6.39] & 130/27/3 & 3.16 [2.36, 4.06] \\
\(T=1.5\) to 4 & 20 & 86.13 & 86.13 & \textcolor{gray}{4.95} & 91.30 & 95.47 & 0.00 & 0/20/0 & \\
\midrule
Blur \(w=0.5\)\,mm & 160 & 22.74 & 81.99 & \textcolor{gray}{1.16} & 69.82 & 89.11 & 44.47 [41.73, 47.25] & 160/0/0 & \\
Blur \(w=1\)\,mm & 160 & 1.36 & 6.07 & \textcolor{gray}{0.05} & 26.08 & 51.99 & 4.55 [3.63, 5.56] & 86/70/4 & \\
Blur \(w=1.5\)\,mm & 160 & 0.17 & 0.17 & \textcolor{gray}{0.00} & 1.27 & 2.10 & $-$0.26 [$-$0.49, $-$0.08] & 0/143/17 & \\
Blur \(w=2\)\,mm & 160 & 0.00 & 0.00 & \textcolor{gray}{0.00} & 0.00 & 0.01 & $-$0.00 [$-$0.01, 0.00] & 0/159/1 & \\
\bottomrule
\end{tabular}
\end{table*}

\subsection{Connectivity and grid}
\label{S-sec:supp-E-conn}

The reaching rate counts a branch only when it connects to the root, and the main text reads connection with 26-connectivity on the native grid (Sec.~\ref{sec:measurement}). A 6-connected lumen tests whether RIDGER's thin drawn branches reach the root only through edge or corner contacts.

\textbf{Conventions.} Table~\ref{S-tab:supp-E-conn} scores the same final masks with a 6-connected lumen instead of the 26-connected one of the main text. It does so on the native grid and on a 0.5\,mm grid onto which each mask is resampled. Reference, root rule, tolerance and distal window are unchanged.

\textbf{Result.} Under both 6-connected readings, the gain on the test patients is 5.04 to 6.44 points on coronary CTA and 2.50 to 5.10 on pulmonary CT, with every interval above zero. On the test patients, RIDGER's gain therefore holds under a 6-connected lumen. On the 40 ASOCA cases, there are six intervals, one for each of the two networks under each of the three conventions. Four of them include zero and two are above zero.

\begin{details}
\paragraph{Resampling.} On the 0.5\,mm grid, a grid voxel is foreground when any native voxel of the mask overlaps it. The comparators are selected at RIDGER's own off-reference volume under each convention.
\end{details}

\begin{table*}[t]
\caption{\textbf{Gain over the float16 control under three connectivity and grid conventions.} Mean [95\% CI] and patients better/tied/worse, in points of $R$ at RIDGER's off-reference volume. The shaded column is RIDGER's $R$ under that convention. The 26-connected rows are the numbers of Tabs.~\ref{S-tab:supp-E-all} and~\ref{S-tab:supp-E-asoca}.}
\label{S-tab:supp-E-conn}
\centering\scriptsize
\setlength{\tabcolsep}{3pt}
\begin{tabular*}{\textwidth}{@{\extracolsep{\fill}}llrrlr@{}}
\toprule
Network & Convention & $n$ & RIDGER $R\!\uparrow$ & Gain [CI] & B/T/W \\
\midrule
\multicolumn{6}{@{}l}{\textbf{Test patients}} \\
\hspace{0.5em}nnU-Net Dice+CE & Native 26-connected (main text) & 160 & \cellcolor{ridger}91.03 & 6.38 [4.85, 8.07] & 111/37/12 \\
\hspace{0.5em}nnU-Net Dice+CE & Native 6-connected & 160 & \cellcolor{ridger}89.08 & 5.04 [3.29, 6.89] & 99/26/35 \\
\hspace{0.5em}nnU-Net Dice+CE & 0.5\,mm grid, 6-connected & 160 & \cellcolor{ridger}90.89 & 5.93 [4.08, 7.80] & 104/48/8 \\
\hspace{0.5em}released nnU-Net & Native 26-connected (main text) & 160 & \cellcolor{ridger}91.25 & 7.01 [5.69, 8.41] & 123/29/8 \\
\hspace{0.5em}released nnU-Net & Native 6-connected & 160 & \cellcolor{ridger}89.77 & 6.44 [4.92, 8.01] & 115/20/25 \\
\hspace{0.5em}released nnU-Net & 0.5\,mm grid, 6-connected & 160 & \cellcolor{ridger}91.52 & 6.39 [4.63, 8.02] & 121/32/7 \\
\hspace{0.5em}nnU-Net & Native 26-connected (main text) & 49 & \cellcolor{ridger}96.30 & 5.79 [4.18, 7.54] & 47/1/1 \\
\hspace{0.5em}nnU-Net & Native 6-connected & 49 & \cellcolor{ridger}93.92 & 3.88 [2.47, 5.41] & 43/1/5 \\
\hspace{0.5em}nnU-Net & 0.5\,mm grid, 6-connected & 49 & \cellcolor{ridger}96.82 & 5.10 [3.61, 6.72] & 47/1/1 \\
\hspace{0.5em}mask-prompted SR & Native 26-connected (main text) & 50 & \cellcolor{ridger}96.35 & 2.86 [2.16, 3.63] & 49/1/0 \\
\hspace{0.5em}mask-prompted SR & Native 6-connected & 50 & \cellcolor{ridger}96.12 & 2.74 [2.03, 3.53] & 49/1/0 \\
\hspace{0.5em}mask-prompted SR & 0.5\,mm grid, 6-connected & 50 & \cellcolor{ridger}96.86 & 2.50 [1.86, 3.20] & 48/2/0 \\
\addlinespace[2pt]
\multicolumn{6}{@{}l}{\textbf{ASOCA, all 40 cases}} \\
\hspace{0.5em}nnU-Net Dice+CE & Native 26-connected (main text) & 40 & \cellcolor{ridger}86.17 & 4.08 [2.20, 6.30] & 21/14/5 \\
\hspace{0.5em}nnU-Net Dice+CE & Native 6-connected & 40 & \cellcolor{ridger}83.00 & 1.24 [$-$1.73, 4.13] & 16/12/12 \\
\hspace{0.5em}nnU-Net Dice+CE & 0.5\,mm grid, 6-connected & 40 & \cellcolor{ridger}86.42 & 4.02 [2.08, 6.36] & 18/19/3 \\
\hspace{0.5em}released nnU-Net & Native 26-connected (main text) & 40 & \cellcolor{ridger}85.40 & 3.69 [$-$0.82, 7.56] & 25/12/3 \\
\hspace{0.5em}released nnU-Net & Native 6-connected & 40 & \cellcolor{ridger}82.73 & 1.46 [$-$3.46, 5.83] & 22/9/9 \\
\hspace{0.5em}released nnU-Net & 0.5\,mm grid, 6-connected & 40 & \cellcolor{ridger}85.84 & 3.94 [$-$0.48, 7.78] & 22/15/3 \\
\bottomrule
\end{tabular*}
\end{table*}

\subsection{ASOCA}
\label{S-sec:supp-E-asoca}

The external ASOCA cohort of Sec.~\ref{sec:robust} tests whether RIDGER's settings carry over to another dataset without retuning. Inference and readout are applied to it untouched. On all 40 labelled cases, RIDGER gains 6.15 and 6.48 points of \(R\) over the best 8-bit threshold (nnU-Net Dice+CE, released nnU-Net). It gains 4.08 and 3.69 points over the float16 control (Tab.~\ref{S-tab:supp-E-asoca}). Of these four intervals, only the one for the released nnU-Net over the float16 control includes zero. Dice rises by 0.34 and 0.82 points against the default mask, and clDice falls by 5.01 and 6.32. RIDGER's settings thus carry over to the external cohort, with both gains over the best 8-bit threshold above zero.

\begin{table*}[t]
\caption{\textbf{ASOCA, 40 cases.} The upper block gives case means (\%; $E_{\beta_1}$ absolute) at 0.5 and as read by RIDGER. The lower block gives gains in points of $R$ at RIDGER's off-reference volume, with 95\% patient-bootstrap intervals. B/T/W counts the cases better, tied and worse than the float16 control, and the median is that of the gain over the control.}
\label{S-tab:supp-E-asoca}
\centering\scriptsize
\setlength{\tabcolsep}{3pt}
\begin{tabular*}{\textwidth}{@{\extracolsep{\fill}}lr*{5}{r}@{\hspace{6pt}}*{5}{r}@{}}
\toprule
 & & \multicolumn{5}{c}{0.5} & \multicolumn{5}{c}{RIDGER} \\
\cmidrule(lr){3-7}\cmidrule(l){8-12}
Network & $n$ & $R\!\uparrow$ & $V$ & Dice$\!\uparrow$ & clDice$\!\uparrow$ & $E_{\beta_1}\!\downarrow$ & $R\!\uparrow$ & $V$ & Dice$\!\uparrow$ & clDice$\!\uparrow$ & $E_{\beta_1}\!\downarrow$ \\
\midrule
\hspace{0.5em}nnU-Net Dice+CE & 40 & 78.61 & \textcolor{gray}{13.03} & 80.86 & 80.34 & 1.32 & \cellcolor{ridger}86.17 & \cellcolor{ridger}\textcolor{gray}{16.23} & \cellcolor{ridger}81.20 & \cellcolor{ridger}75.33 & \cellcolor{ridger}1.32 \\
\hspace{0.5em}released nnU-Net & 40 & 73.56 & \textcolor{gray}{11.09} & 79.52 & 79.42 & 1.43 & \cellcolor{ridger}85.40 & \cellcolor{ridger}\textcolor{gray}{16.57} & \cellcolor{ridger}80.34 & \cellcolor{ridger}73.10 & \cellcolor{ridger}1.38 \\
\bottomrule
\end{tabular*}

\vspace{4pt}
\begin{tabular*}{\textwidth}{@{\extracolsep{\fill}}lrllrr@{}}
\toprule
Network & $n$ & Over best 8-bit threshold & Over float16 control & B/T/W & Median \\
\midrule
\hspace{0.5em}nnU-Net Dice+CE & 40 & 6.15 [3.85, 8.72] & 4.08 [2.20, 6.30] & 21/14/5 & 0.61 \\
\hspace{0.5em}released nnU-Net & 40 & 6.48 [3.04, 10.04] & 3.69 [$-$0.82, 7.56] & 25/12/3 & 0.65 \\
\bottomrule
\end{tabular*}
\end{table*}

\section{Airways, retina and roads}
\label{S-sec:supp-H_domains}

This section gives every endpoint on the three other tree-like structures of Tab.~\ref{tab:domains}.

\textbf{Design.} We repeat the comparison of the main text on each domain. Each domain's map comes from its own network, and RIDGER reads it with the same code. Only $\lambda$ changes, chosen on each domain's validation images by the rule of Sec.~\ref{S-sec:supp-A-anatomy}. Every comparator is held to RIDGER's off-reference volume \(V\) per image, which is an area in 2D. The comparators are the best 8-bit threshold and, on airways, the float16 control. The 2D maps are stored in float32, so the second comparator there is hysteresis over every attained float32 level. $R_{\mathrm{glob}}$ is the connected recall over the whole reference centreline. \(R\) scores only the distal part of that centreline.

\subsection{AeroPath airway CT}

Airway CT is the one further domain whose map is stored in float16, so it repeats both comparators of the main text. On the 18 AeroPath \cite{stoverud2024aeropath} test patients, RIDGER reaches 19.58 [15.19, 23.93] points further than the best 8-bit threshold and 6.96 [5.21, 8.85] points further than the float16 control. Both intervals exclude zero. Table~\ref{S-tab:supp-airway} gives all endpoints.

\textbf{Branches detected and clDice.} RIDGER detects fewer branches than the float16 control and has the lowest clDice of the four readouts. RIDGER detects 3.94 points fewer branches than the float16 control of the table. This member of the control is the one selected for \(R\). Against the best 8-bit threshold, RIDGER detects more branches (75.66 against 69.10) and has a lower clDice (78.36 against 85.53).

\begin{table*}[t]
\caption{\textbf{AeroPath airway CT (18 test patients).} All endpoints at patient means. $V$ is the off-reference volume in \% of the reference. TD is tree length detected, BD branches detected, T\_prec and T\_sens skeleton precision and sensitivity, $E_{\beta}$ Betti errors. Comparators are the best 8-bit threshold and the float16 control at RIDGER's volume.}
\label{S-tab:supp-airway}
\centering\scriptsize
\setlength{\tabcolsep}{3pt}
\begin{tabular}{@{}lrrrrrrrrr@{}}
\toprule
Readout & $V$ & $R\!\uparrow$ & TD$\!\uparrow$ & BD$\!\uparrow$ & Dice$\!\uparrow$ & clDice$\!\uparrow$ & $T_\text{prec}\!\uparrow$ & $T_\text{sens}\!\uparrow$ & $E_{\beta_0}$\,/\,$E_{\beta_1}\!\downarrow$ \\
\midrule
Default mask (0.5) & \textcolor{gray}{6.94} & 70.74 & 76.29 & 63.22 & 91.18 & 84.10 & 95.06 & 76.29 & 2.67\,/\,4.11 \\
Best 8-bit threshold & \textcolor{gray}{8.45} & 75.66 & 80.27 & 69.10 & 92.13 & 85.53 & 92.94 & 80.27 & 2.67\,/\,3.50 \\
Float16 control & \textcolor{gray}{8.29} & 88.28 & 87.82 & 79.60 & 91.11 & 86.91 & 86.64 & 87.82 & 2.67\,/\,3.56 \\
\rowcolor{ridger}\textbf{RIDGER} & \textcolor{gray}{8.54} & 95.25 & 88.33 & 75.66 & 90.83 & 78.36 & 71.43 & 88.33 & 2.67\,/\,3.44 \\
\bottomrule
\end{tabular}
\end{table*}

\begin{details}
\textbf{Setup.} $\lambda$ is $30\,\text{mm}^{-2}$, chosen on 9 airway validation cases. $S$ is the largest 0.5 component.

\textbf{Skeleton precision and Dice.} The clDice deficit lies in skeleton precision. RIDGER's skeleton precision is 71.43, against 92.94 for the best 8-bit threshold. RIDGER's skeleton sensitivity is the highest of the four readouts. RIDGER's Dice is 1.30 points below that of the best 8-bit threshold and 0.35 points below that of the default mask.
\end{details}

\subsection{FIVES retinal fundus images}

Retinal fundus images test RIDGER on a 2D vessel tree. On the 150 FIVES \cite{jin2022fives} test images, RIDGER reaches 2.39 [1.74, 3.16] points further than the best 8-bit threshold. It reaches 2.36 [1.70, 3.13] points further than full-precision hysteresis at RIDGER's area, with 129 images better, 1 tied and 20 worse. Table~\ref{S-tab:supp-fives} gives all endpoints.

\begin{table*}[t]
\caption{\textbf{FIVES retinal fundus images (150 test images).} All endpoints at image means. $V$ is the off-reference area in \% of the reference area. TD and BD are 2D tree and branch detection. The second comparator is full-precision hysteresis at RIDGER's area.}
\label{S-tab:supp-fives}
\centering\scriptsize
\setlength{\tabcolsep}{3pt}
\begin{tabular}{@{}lrrrrrrrrrr@{}}
\toprule
Readout & $V$ & $R\!\uparrow$ & $R_{\mathrm{glob}}\!\uparrow$ & TD$\!\uparrow$ & BD$\!\uparrow$ & Dice$\!\uparrow$ & clDice$\!\uparrow$ & $T_\text{prec}\!\uparrow$ & $T_\text{sens}\!\uparrow$ & $E_{\beta_0}$\,/\,$E_{\beta_1}\!\downarrow$ \\
\midrule
Default mask (0.5) & \textcolor{gray}{3.96} & 83.33 & 83.03 & 81.85 & 71.41 & 89.50 & 87.92 & 97.18 & 81.71 & 4.03\,/\,21.47 \\
Best 8-bit threshold & \textcolor{gray}{6.30} & 85.98 & 85.62 & 84.51 & 75.12 & 89.95 & 89.00 & 95.79 & 84.37 & 4.03\,/\,20.69 \\
Full-precision hysteresis & \textcolor{gray}{6.34} & 86.01 & 85.64 & 84.54 & 75.18 & 89.95 & 89.00 & 95.76 & 84.40 & 4.03\,/\,20.68 \\
\rowcolor{ridger}\textbf{RIDGER} & \textcolor{gray}{6.37} & 88.36 & 87.99 & 86.29 & 74.44 & 89.97 & 89.09 & 93.48 & 86.20 & 4.03\,/\,21.41 \\
\bottomrule
\end{tabular}
\end{table*}

\begin{details}
\textbf{Setup.} $\lambda$ is $10\,\text{px}^{-1}$, chosen on 170 retinal validation images. $S$ is the largest 0.5 component.

\textbf{Overlap scores.} Overlap scores do not move: RIDGER's Dice is within 0.3 points of the best 8-bit threshold, and its clDice is 0.09 higher.
\end{details}

\subsection{Massachusetts roads}

Road maps test RIDGER on a tree-like structure that is not anatomical and that has its own path score, APLS \cite{van2018spacenet}. On the 49 Massachusetts road \cite{mnih2013machine} test images, RIDGER reaches 4.07 [3.15, 5.05] points further than the best 8-bit threshold and 4.05 [3.13, 5.03] points further than full-precision hysteresis at RIDGER's area (45 images better, 2 tied, 2 worse). The road benchmark's path score moves the same way. RIDGER raises APLS by 4.51 points over the default mask, to 71.92. The best 8-bit threshold gives 69.38. Overlap scores fall. RIDGER's clDice is 2.14 points lower than the threshold's (86.71 against 88.85), and its Dice is 0.30 points lower. Table~\ref{S-tab:supp-roads} gives all endpoints.

\begin{table*}[t]
\caption{\textbf{Massachusetts roads (49 test images).} All endpoints at image means. $V$ is the off-reference area in \% of the reference area. APLS is the average path length similarity. The second comparator is full-precision hysteresis at RIDGER's area.}
\label{S-tab:supp-roads}
\centering\scriptsize
\setlength{\tabcolsep}{3pt}
\begin{tabular}{@{}lrrrrrrrrr@{}}
\toprule
Readout & $V$ & $R\!\uparrow$ & $R_{\mathrm{glob}}\!\uparrow$ & Dice$\!\uparrow$ & clDice$\!\uparrow$ & $T_\text{prec}\!\uparrow$ & $T_\text{sens}\!\uparrow$ & APLS$\!\uparrow$ & $E_{\beta_0}$\,/\,$E_{\beta_1}\!\downarrow$ \\
\midrule
Default mask (0.5) & \textcolor{gray}{17.28} & 80.43 & 86.68 & 79.31 & 88.80 & 93.81 & 84.73 & 67.41 & 3.76\,/\,15.61 \\
Best 8-bit threshold & \textcolor{gray}{22.33} & 81.15 & 87.09 & 78.97 & 88.85 & 92.92 & 85.53 & 69.38 & 2.59\,/\,14.53 \\
Full-precision hysteresis & \textcolor{gray}{22.39} & 81.17 & 87.11 & 78.96 & 88.86 & 92.90 & 85.55 & 69.40 & 2.59\,/\,14.55 \\
\rowcolor{ridger}\textbf{RIDGER} & \textcolor{gray}{22.42} & 85.21 & 89.76 & 78.67 & 86.71 & 86.43 & 87.41 & 71.92 & 2.69\,/\,14.63 \\
\bottomrule
\end{tabular}
\end{table*}

\begin{details}
\textbf{Setup.} $\lambda$ is $1\,\text{px}^{-1}$, chosen on 74 road validation images. $S$ is every 0.5 component large enough to hold a road, each with its own anchor.
\end{details}

\section{Training on extended labels}
\label{S-sec:supp-L_distillation}

This section gives the retraining experiments of Sec.~\ref{sec:labels}.

\textbf{Terms.} A student is a network trained on labels completed by a readout of a teacher map, and it is scored at the default threshold 0.5. RIDGER students train on labels completed by RIDGER. Control students train on labels completed by the volume-matched control. On each training case, this control is the float16 member of the same teacher map whose added volume best matches RIDGER's. Two quantities are reported. The gain \(\Delta R_{0.5}\) is the reaching rate \(R\) of the student's 0.5 mask minus that of the baseline network of the same seed. The amortised fraction $A=\Delta R_{0.5}/[R(\text{baseline},\text{RIDGER})-R(\text{baseline},0.5)]$ is the share of RIDGER's gain that the student reaches at 0.5.

\subsection{Pulmonary students trained on RIDGER-completed labels}
A pulmonary network trained on RIDGER-completed labels holds about 72\% of RIDGER's gain in \(R\) in its 0.5 mask.

\textbf{Design.} The teacher is the pulmonary nnU-Net seed 2. RIDGER reads its map on the 170 training cases, and each training label is completed with RIDGER's additions beyond the 0.5 mask. Two students (seeds 1 and 3) are trained like the baseline networks of the same seeds and differ from them only in these labels. Validation and test labels are not altered. Each student is scored once on the 49 readable test patients against the original HiPaS labels with the evaluator of the main text.

\begin{table}[h]
\caption{\textbf{Students trained on RIDGER-completed labels, scored at 0.5.} Pulmonary test patients ($n=49$); 95\% patient bootstraps, 20{,}000 draws.}
\label{S-tab:supp-distill-primary}
\centering\scriptsize
\setlength{\tabcolsep}{2pt}
\begin{tabular*}{\columnwidth}{@{\extracolsep{\fill}}lrr@{}}
\toprule
 & Seed 1 & Seed 3 \\
\midrule
$R$, student at 0.5 & 91.65 & 91.16 \\
$R$, baseline at 0.5 & 79.86 & 79.78 \\
$R$, baseline read by RIDGER & 96.30 & 95.34 \\
\midrule
$\Delta R_{0.5}$ & 11.79 [9.62, 13.99] & 11.38 [9.34, 13.44] \\
\quad better/tied/worse & 48/1/0 & 48/1/0 \\
Amortised fraction $A$ & 0.717 [0.682, 0.754] & 0.731 [0.701, 0.764] \\
\bottomrule
\end{tabular*}
\end{table}

\textbf{Result.} At 0.5 the students reach 11.8 and 11.4 points further than their baselines, on 48 of 49 patients each and with none worse (Sec.~\ref{sec:labels}; Tab.~\ref{S-tab:supp-distill-primary}). Measured against what RIDGER adds to the baselines' maps, the students' amortised fractions \(A\) are 0.717 and 0.731. The students' 0.5 masks now hold most of what RIDGER adds at readout.

\begin{details}
\textbf{Training.} The teacher's final checkpoint is used, and RIDGER reads its map with $\lambda=30\,\text{mm}^{-2}$. RIDGER's additions to a training label are a median of 13{,}533 voxels (3.57\% of the label volume) and 4.04\,m of centreline, against a median labelled centreline of 14.7\,m. The students use the same residual-encoder nnU-Net, Dice+CE loss, plans, split and 1{,}000 epochs as the baselines, with the final checkpoint.

\textbf{Off-reference volume.} The student's off-reference volume \(V\) lies about halfway between the baseline's 0.5 mask and the baseline read by RIDGER. For the two students (seeds 1 and 3), \(V\) is 2.02 [1.61, 2.51] and 1.77 [1.44, 2.11] points above the baseline's 0.5 mask. It is 2.10 and 1.82 points below the baseline read by RIDGER.

\textbf{Equal off-reference volume.} At equal \(V\), the student's 0.5 mask reaches 9.88 [7.87, 11.97] and 9.83 [7.91, 11.82] points further than the baseline's best 8-bit threshold. It reaches 3.86 [2.69, 5.09] and 4.21 [3.05, 5.44] points further than the baseline's float16 control ($n=47$). Two patients per student have no admissible 8-bit level. On these patients, every level of the baseline map exceeds the student's off-reference volume. Every interval excludes zero, so the student leads both comparators of the baseline map at that volume.

\textbf{Other endpoints.} Dice is nearly unchanged, at $-0.31$ and $-0.27$ points (87.41 and 87.50 against 87.72 and 87.77). clDice falls by 7.08 and 6.41 points (79.75 and 80.40 against 86.83 and 86.81), because the learned branches beyond the incomplete labels count as false centreline. Tree length detected rises by 6.49 and 6.32 points, and branches detected by 8.85 and 8.48, so the gain is not confined to the distal window of \(R\).

\textbf{Stacking.} RIDGER can also read the student's own map. It then reaches \(R\) of 98.49 [97.69, 99.03] and 98.37 [97.57, 98.92]. Its gain over the student's own float16 control is 0.98 [0.16, 1.61] and 0.91 [0.09, 1.55] points ($n=47$), and both intervals exclude zero.
\end{details}

\subsection{Volume-matched control}
This control separates the effect of RIDGER's particular additions from that of extending the labels at all. Two control students train on labels completed by the volume-matched control, with the same seeds, trainer and data pipeline as the RIDGER students. The endpoint is the difference in \(R\) at 0.5 between the RIDGER student and the control student of the same seed.

\begin{table}[h]
\caption{\textbf{Volume-matched control students.} RIDGER-label student minus control student of the same seed, pulmonary test patients ($n=49$; stacking $n=47$); 95\% patient bootstraps, 20{,}000 draws.}
\label{S-tab:supp-distill-control}
\centering\scriptsize
\setlength{\tabcolsep}{2pt}
\begin{tabular*}{\columnwidth}{@{\extracolsep{\fill}}lrr@{}}
\toprule
 & Seed 1 & Seed 3 \\
\midrule
$R$, control student at 0.5 & 91.46 & 91.50 \\
$\Delta R_{0.5}$ & 0.18 [$-$0.48, 0.75] & $-$0.34 [$-$0.98, 0.22] \\
\quad better/tied/worse & 26/1/22 & 24/1/24 \\
$\Delta V_{0.5}$ & $-$2.31 [$-$2.86, $-$1.84] & $-$2.42 [$-$2.92, $-$1.96] \\
$\Delta$Dice$_{0.5}$ & $-$0.03 [$-$0.26, 0.21] & 0.25 [0.02, 0.47] \\
$\Delta$clDice$_{0.5}$ & $-$4.20 [$-$4.70, $-$3.70] & $-$3.58 [$-$4.14, $-$3.01] \\
$\Delta R$, both read by RIDGER & 0.39 [$-$0.39, 0.93] & 0.57 [$-$0.23, 1.20] \\
\bottomrule
\end{tabular*}
\end{table}

\textbf{Result.} The RIDGER students and the control students reach the same \(R\) at 0.5: both intervals of the difference include zero (Tab.~\ref{S-tab:supp-distill-control}). The control students gain 11.60 [9.60, 13.61] and 11.72 [9.69, 13.73] points over their baselines, which is 98\% and 103\% of the RIDGER students' gain. The gain at 0.5 therefore comes from extending the labels below the threshold, not from the particular readout that extends them.

\begin{details}
\textbf{Matching.} The control labels use the 12-level members of the float16 control. On 52.4\% of the training cases the control's added volume lies within $\pm25\%$ of RIDGER's, and the median ratio of the control's added volume to RIDGER's is 1.23.

\textbf{What the labels add.} The two sets of labels add different content. The control's additions carry a median 0.94\,m of centreline against RIDGER's 4.04\,m. They overlap RIDGER's with a median Jaccard of 0.14 and mostly thicken branches the label already has.

\textbf{Other endpoints.} The RIDGER students reach the same \(R\) at 0.5 as the control students with 2.31 and 2.42 points less off-reference volume. They are 4.20 and 3.58 clDice points below the controls, and about 1.2 and 1.9 points below them in tree length and branches detected.

\textbf{Equal off-reference volume and stacking.} An equal-volume comparison between the two students is not defined, because the control students' 0.5 masks already exceed the RIDGER students' per-patient volume in 46 of 47 and 47 of 47 patients. RIDGER at readout adds more to a control student's map. It gains 2.90 [2.07, 3.81] and 2.87 [2.06, 3.76] points over the control students' float16 controls, against 0.98 and 0.91 points on the RIDGER students' maps.
\end{details}

\subsection{Coronary students}
Both arms are repeated on coronary CTA, the second anatomy of Sec.~\ref{sec:labels}, where the gain is smaller than on pulmonary CT.

\textbf{Design.} The coronary design follows the pulmonary one. The teacher is the nnU-Net Dice+CE of the main text (seed 1) with its 559 training cases. Students have seeds 1 and 3, with a RIDGER arm and a control arm per seed. Each student is scored once on the 160 test patients against the unaltered ImageCAS-X labels and is paired with the baseline network of the same seed.

\begin{table}[h]
\caption{\textbf{Coronary students.} Test patients ($n=160$; equal volume $n=157$ and 138 for seeds 1 and 3, where the baseline has an admissible 8-bit level); 95\% patient bootstraps, 20{,}000 draws. The two equal-volume rows compare the RIDGER student's 0.5 mask with the best 8-bit threshold and the float16 control of the baseline map. The RIDGER student is trained on RIDGER-completed labels and the control student on the volume-matched control labels; $\Delta R_{0.5}$ and $A$ are defined in the text. Stacking: RIDGER reads the map of the RIDGER student, and the gain is over the float16 control of that map.}
\label{S-tab:supp-distill-coronary}
\centering\scriptsize
\setlength{\tabcolsep}{0.5pt}
\begin{tabular*}{\columnwidth}{@{\extracolsep{\fill}}lrr@{}}
\toprule
 & Seed 1 & Seed 3 \\
\midrule
$R$: baseline, 0.5 / RIDGER & 77.31 / 91.03 & 77.70 / 91.17 \\
$R$ at 0.5: RIDGER / control student & 82.25 / 82.55 & 82.03 / 84.23 \\
\midrule
$\Delta R_{0.5}$ & 4.95 [3.82, 6.11] & 4.32 [3.38, 5.32] \\
Amortised fraction $A$ & 0.36 [0.29, 0.43] & 0.32 [0.26, 0.39] \\
$\Delta R_{0.5}$, RIDGER $-$ control & $-$0.29 [$-$1.18, 0.57] & $-$2.21 [$-$3.22, $-$1.29] \\
$\Delta V_{0.5}$, RIDGER $-$ control & $-$0.37 [$-$0.49, $-$0.26] & $-$0.43 [$-$0.60, $-$0.29] \\
\midrule
Eq.\ $V$ vs best 8-bit & 3.67 [2.68, 4.70] & 3.64 [2.78, 4.53] \\
Eq.\ $V$ vs float16 control & $-$1.59 [$-$2.56, $-$0.64] & $-$0.26 [$-$1.19, 0.67] \\
Stacking: RIDGER vs float16 & 4.71 [3.38, 6.21] & 5.43 [4.07, 6.92] \\
\bottomrule
\end{tabular*}
\end{table}

\textbf{Result.} Both RIDGER students reach further at 0.5 than their baselines, by 4.95 and 4.32 points, with 130 and 123 of 160 patients better (Tab.~\ref{S-tab:supp-distill-coronary}). These gains are 36\% and 32\% of RIDGER's gain, about half the share reached on pulmonary CT. The control students gain 5.24 [4.30, 6.23] and 6.53 [5.43, 7.69] points, at 0.37 and 0.43 points more off-reference volume. The RIDGER student is level with the control student for seed 1 (Tab.~\ref{S-tab:supp-distill-coronary} gives both seeds).

\begin{details}
\textbf{Labels.} The control labels use the 12-level members of the float16 control. On 78.2\% of the training cases the control's added volume lies within $\pm25\%$ of RIDGER's. The median ratio of the control's added volume to RIDGER's is 1.004. RIDGER's additions per coronary training case are a median of 421 voxels and about 51\,mm of centreline, against a median labelled centreline of 655\,mm. The control's additions carry a median 19\,mm and overlap RIDGER's with a median Jaccard of 0.23.

\textbf{Other endpoints.} At 0.5 the RIDGER students add 1.13 and 0.73 points of off-reference volume and 0.56 and 0.20 Dice points over the 0.5 masks of their baselines. Tree length detected rises by 2.16 and 1.73 points and branches detected by 2.81 and 2.78. clDice is unchanged (+0.12 and $-$0.06), whereas it fell on pulmonary CT.

\textbf{Equal off-reference volume and stacking.} At equal off-reference volume the RIDGER students reach 3.7 and 3.6 points further than the baseline's best 8-bit threshold. Their 0.5 masks reach 8.8 and 9.1 points less than the baseline read by RIDGER, so RIDGER at readout remains the larger step on coronary CTA. RIDGER on the maps of the four coronary students reaches \(R\) of 92.8 to 93.8 and gains 3.9 to 5.4 points over each student's own float16 control.
\end{details}

\section{Length penalty, width and input maps}
\label{S-sec:supp-M_limits}

This section gives the measurements behind the limitations of Sec.~\ref{sec:limits}.

\subsection{A length penalty in the selection objective}
\label{S-sec:supp-F-length-price}
A length penalty in the selection objective changes no test result. The selection objective \eqref{eq:objective} penalises an addition by its volume alone. A variant also penalises its length, \(u_b=q_bL_b-(1-q_b)(\lambda\,\mathrm{Vol}_b+\mu L_b)\), where \(\mu=0\) gives RIDGER and everything else is unchanged. The dynamic program stays exact. With \(\lambda\) fixed, we select \(\mu\in\{0,0.03,0.1,0.3,1,3\}\) per anatomy on the validation patients, and the test patients are scored once.

Validation selects \(\mu=1\) on coronary CTA and \(\mu=0\), which is RIDGER itself, on pulmonary CT. On the coronary test patients every gain over the float16 control is unchanged at \(\mu=1\): 6.38 and 7.01 at equal volume, 2.48 and 2.58 at equal length (Dice+CE and released nnU-Net).

\begin{details}
\textbf{Selection criterion.} \(\mu\) is selected by the balanced mean of the two validation gains over the float16 control, at equal volume and at equal length (both defined in Sec.~\ref{S-sec:supp-F_ablation}).

\textbf{Coronary CTA.} The validation score of \(\mu=1\) lies 0.005 points above that of \(\mu=0\), so the two are practically tied. Near the keep-or-drop boundary, the volume penalty per millimetre of advancing length has a median of about 150. A length penalty of \(\mu=1\) is small beside it and removes few twigs.

\textbf{Pulmonary CT.} A length penalty lowers both validation gains. The equal-length gain falls from 0.07 and 0.67 at \(\mu=0\) to $-$0.18 and 0.48 at \(\mu=3\) on nnU-Net and mask-prompted SR. The segments it removes contribute to the reaching rate \(R\).
\end{details}

\subsection{Width}
\label{S-sec:supp-F-width}

RIDGER's lead over the float16 control survives when the width is measured on the log odds. RIDGER reads width at the half height of the map's cross-section in probability (Sec.~\ref{sec:width}). On faint distal ridges this edge lies inside the lumen, so the rendered coronary additions are too thin (Sec.~\ref{sec:limits}). Thin additions add little off-reference volume, so the test also asks whether RIDGER's lead at equal volume comes from its thin rendering (Sec.~\ref{sec:robust}). We therefore test an edge halfway between background and peak in log odds, selected on validation patients and scored once on the test patients. Coronary pairs are given in the order Dice+CE, released nnU-Net.

\textbf{Result.} The log-odds edge halves the coronary width error, to 0.11\,mm, and the selected tree is unchanged. The wider lumen adds off-reference volume \(V\), and the control is then also allowed more volume (equal volume and equal length are defined in Sec.~\ref{S-sec:supp-F_ablation}). The equal-volume gains over the control therefore fall to 3.68 [2.53, 5.01] and 2.78 [1.76, 3.90] on coronary CTA, with both intervals above zero, and to 5.10 and 2.79 on pulmonary CT. The equal-length gains do not fall (2.60 and 3.06 on coronary CTA). The paper uses the half-height width throughout.

\begin{details}
\textbf{Width error.} The error is measured on the added coronary nodes that meet the reference. With the half-height edge, the rendered radius is 0.23 and 0.22\,mm too small on average, while the default mask's own radius errs by 0.06\,mm. The log-odds edge lowers the error to 0.11\,mm, with a lower error for every patient, and raises Dice near the matched nodes from 56.0 to 75.5 and from 60.1 to 76.7. On pulmonary CT the selected edge is close to the half-height edge, and the error falls by 0.02 to 0.04\,mm.

\textbf{Reaching rate and added volume.} With the selected tree unchanged, the reaching rate \(R\) rises from 91.03 to 91.32 and from 91.25 to 91.67 on coronary CTA. Of the voxels that the wider lumen adds, 86 to 92\% lie on branches the reference does not draw.
\end{details}

\subsection{Map normalisation on CAS-Net}
\label{S-sec:supp-J-casnet}

Without the map normalisation, RIDGER returns the default mask on the CAS-Net maps.

\textbf{Diffuse background.} The CAS-Net map keeps the ridges of the vessels but has a diffuse background. A reference centreline point counts as a ridge when the map value at the point is at least the map value 1\,mm away in the normal plane. It must also be above the map value 3\,mm away. In the CAS-Net map, 96.3\% of reference centreline points are such a ridge, against 89.3\% for nnU-Net Dice+CE and 93.5\% for the released nnU-Net (72 patients), so its ridges are no weaker than those of the two nnU-Net maps. The background differs. Of the voxels of the CAS-Net map with \(p\geq10^{-3}\), 4.1\% [3.5, 4.9] lie within 2\,mm of a vessel, against 92.8\% [89.2, 95.0] for the released nnU-Net (Tab.~\ref{S-tab:supp-casnet-haze}). Most voxels outside the default mask \(S\) therefore pass the candidate threshold \(10^{-5}\). When RIDGER reads this map at \(T=1\), the candidates merge into one mass with no tree to select, and RIDGER returns the default mask. Its gain in the reaching rate \(R\) is then 0.00 [$-$0.21, 0.22] points over the best 8-bit threshold and $-$0.49 [$-$1.14, 0.00] over the float16 control.

\textbf{The normalisation rule.} The map normalisation of Sec.~\ref{sec:robust} detects this background without labels. \(H_T\) is the share of voxels outside \(S\) whose sharpened probability \(\sigma(\ell/T)\) passes the candidate threshold \(10^{-5}\). RIDGER reads the map at the largest \(T\) in \(\{1,\tfrac12,\tfrac14,\tfrac18\}\) with \(H_T\leq h_0\), where the bound is \(h_0=28.88\%\). The rule leaves every map of the other networks unchanged. None of them exceeds \(h_0\), so all of these maps are read at \(T=1\), and every other result of this paper is read from the unchanged map. The CAS-Net maps lie far above the bound. On the 72 CAS-Net test patients \(H_1\) is at least 51.38\% and \(H_{1/2}\) is at most 2.24\% (Tab.~\ref{S-tab:supp-casnet-HT}), so the rule gives \(T=\tfrac12\) to all 72 patients.

\textbf{Test patients.} On the normalised map RIDGER extends the CAS-Net tree well beyond both comparators. At \(T=\tfrac12\), RIDGER lifts \(R\) from 73.62 to 93.35 at an off-reference volume \(V\) of 12.65. At that \(V\), RIDGER reaches 15.56 [12.96, 18.24] points further than the best 8-bit threshold of the map that RIDGER reads (64 patients better, 7 tied, 1 worse) and 5.75 [3.88, 7.70] points further than the float16 control (48/14/10). Both intervals exclude zero. The overlap scores fall: Dice moves from 89.11 at the default mask to 88.56, and clDice from 90.98 at \(T=1\) to 86.28. The gains hold against the original ImageCAS labels. On the 67 patients whose original label has a readable root, RIDGER lifts \(R\) from 54.24 to 66.83. It reaches 5.67 [4.04, 7.45] points further than the best 8-bit threshold and 4.54 [2.94, 6.26] points further than the float16 control. Read at \(T=1\), RIDGER returns the default mask on these labels as well.

\begin{details}
\textbf{The bound \(h_0\).} The bound \(h_0\) is twice the largest \(H_1\) among the 180 validation maps of the four primary networks. On these maps, \(H_1\) is 0.022 to 0.090\% for the two coronary networks and 5.62 to 14.44\% for the two pulmonary networks, so \(h_0=28.88\%\). No validation map and none of the 1375 test and ASOCA maps of the networks other than CAS-Net exceeds \(h_0\) (largest \(H_1\) 16.16\%). \(H_1\) stays at or below 0.134\% for the seven other coronary networks on the 160 test patients and at or below 0.093\% on ASOCA, and lies at 3.00 to 16.16\% for the three pulmonary networks. The temperatures of the CAS-Net maps do not depend on the exact value of \(h_0\). Any bound between 2.24\% and 51.38\% gives the same temperatures on the test patients, and any bound between 2.69\% and 51.38\% gives them on the test and the validation patients together (Tab.~\ref{S-tab:supp-casnet-HT}).

\textbf{Validation patients.} The rule gives the same temperature and a similar gain on the validation patients. The 80 official validation patients of ImageCAS-X were scored with the rule and \(h_0\) unchanged. Every patient has \(H_1\) above \(h_0\) and \(H_{1/2}\) below it (Tab.~\ref{S-tab:supp-casnet-HT}), so the rule gives \(T=\tfrac12\) to every patient. RIDGER lifts \(R\) from 74.11 to 94.07 at \(V=12.46\) and reaches 15.00 [12.43, 17.59] points further than the best 8-bit threshold of the read map (71/8/1), against 15.56 on the test patients. Dice moves from 90.93 to 89.74, and clDice is 86.47. The float16 control is built from members scored on test patients and has no counterpart on the validation patients. The normalisation was introduced for the CAS-Net test maps, on which RIDGER otherwise returns the default mask, and \(T=\tfrac12\) and \(\tfrac14\) were tried on those maps. The rule contains no label, and its bound comes from the validation maps of the other networks.
\end{details}

\begin{table}[t]
\caption{\textbf{Background share \(H_T\) of the CAS-Net maps (\%).} Median [minimum, maximum] over patients. The rule reads the map at the largest \(T\) with \(H_T\leq h_0=28.88\%\).}
\label{S-tab:supp-casnet-HT}
\centering\scriptsize
\setlength{\tabcolsep}{3pt}
\begin{tabular}{@{}lcc@{}}
\toprule
\(T\) & Test (72 patients) & Validation (80 patients) \\
\midrule
1 & 62.76\,[51.38, 76.37] & 63.16\,[54.37, 73.71] \\
1/2 & 1.07\,[0.62, 2.24] & 1.31\,[0.56, 2.69] \\
1/4 & 0.10\,[0.07, 0.15] & 0.10\,[0.07, 0.16] \\
1/8 & 0.05\,[0.03, 0.08] & 0.05\,[0.04, 0.09] \\
\bottomrule
\end{tabular}
\end{table}

\begin{table*}[t]
\caption{\textbf{Background of the CAS-Net map.} Median [IQR] over 72 ImageCAS-X test patients. Columns (a1) and (a2) give the share (\%) of the voxels with $p\geq10^{-3}$ and with $p\geq10^{-2}$ that lie within 2\,mm of the reference. Columns (b) and (b') give the share (\%) of reference voxels that have $p\geq10^{-2}$ and $p\geq10^{-3}$. Column (c) gives the median $p$ on the reference voxels that the 0.5 mask misses. The last column gives the share (\%) of reference voxels with $p\geq0.5$.}
\label{S-tab:supp-casnet-haze}
\centering\scriptsize
\setlength{\tabcolsep}{3pt}
\begin{tabular}{@{}lcccccc@{}}
\toprule
Model & (a1) & (a2) & (b) & (b') & (c) & Recall at 0.5 \\
\midrule
CAS-Net & 4.1\,[3.5, 4.9] & 43.3\,[40.3, 46.3] & 100.0\,[99.9, 100.0] & 100.0\,[100.0, 100.0] & 0.40\,[0.34, 0.41] & 89.7\,[87.1, 91.8] \\
released nnU-Net & 92.8\,[89.2, 95.0] & 94.5\,[91.4, 96.6] & 94.9\,[92.5, 96.1] & 96.6\,[94.9, 97.6] & 0.02\,[0.01, 0.03] & 88.0\,[85.3, 90.3] \\
\bottomrule
\end{tabular}
\end{table*}

\subsection{A split 0.5 mask on ASOCA}
\label{S-sec:supp-J-split}

RIDGER does not rejoin a part of the 0.5 mask that is split from the root. The default mask \(S\) of the released nnU-Net for ASOCA patient asoca\_diseased\_03 loses a branch that the reference keeps. A lower threshold of the same map joins the split component back to the root and gives \(R=68.98\). RIDGER gives \(R=38.18\), so the threshold reaches further than RIDGER on this patient. RIDGER grows only from \(S\), the set of 0.5 components touching the anchor, and its ridge tree does not bridge to a component outside \(S\). RIDGER gives \(R=100.00\) on the same patient read from the nnU-Net Dice+CE map, whose 0.5 mask is not split.

\section{Additional qualitative results}
\label{S-sec:supp-Z_gallery}

This section shows the whole trees of ten further test patients, chosen at fixed percentiles of RIDGER's gain over the float16 control.

\begin{figure*}[t]
\centering
\includegraphics[width=\textwidth]{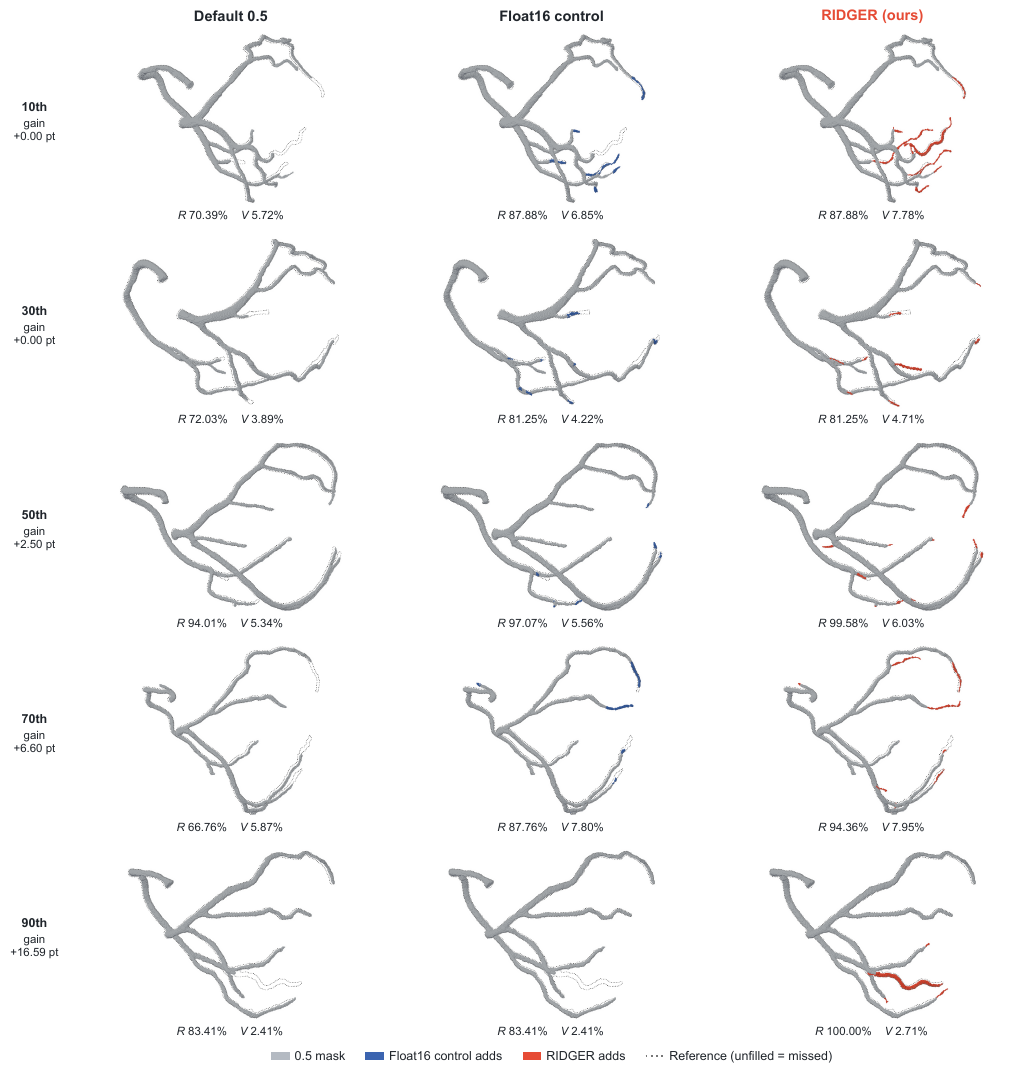}
\caption{\textbf{Five coronary test patients across the range of RIDGER's gain.} The map is that of the nnU-Net trained with Dice and cross-entropy. Rows are the patients at the 10th to 90th percentile of the gain in the reaching rate \(R\) over the float16 control, lowest gain at the top; the number at the left of each row is that gain in points. Columns show the default mask, the float16 control and RIDGER at RIDGER's per-patient off-reference volume \(V\). Grey is the default mask, blue the voxels the float16 control adds, vermilion the voxels RIDGER adds; the dotted outline is the reference, unfilled where missed. \(R\) and \(V\) under each panel are those of that patient.}
\label{S-fig:supp-gallery-coronary}
\end{figure*}

\begin{figure*}[t]
\centering
\includegraphics[width=\textwidth]{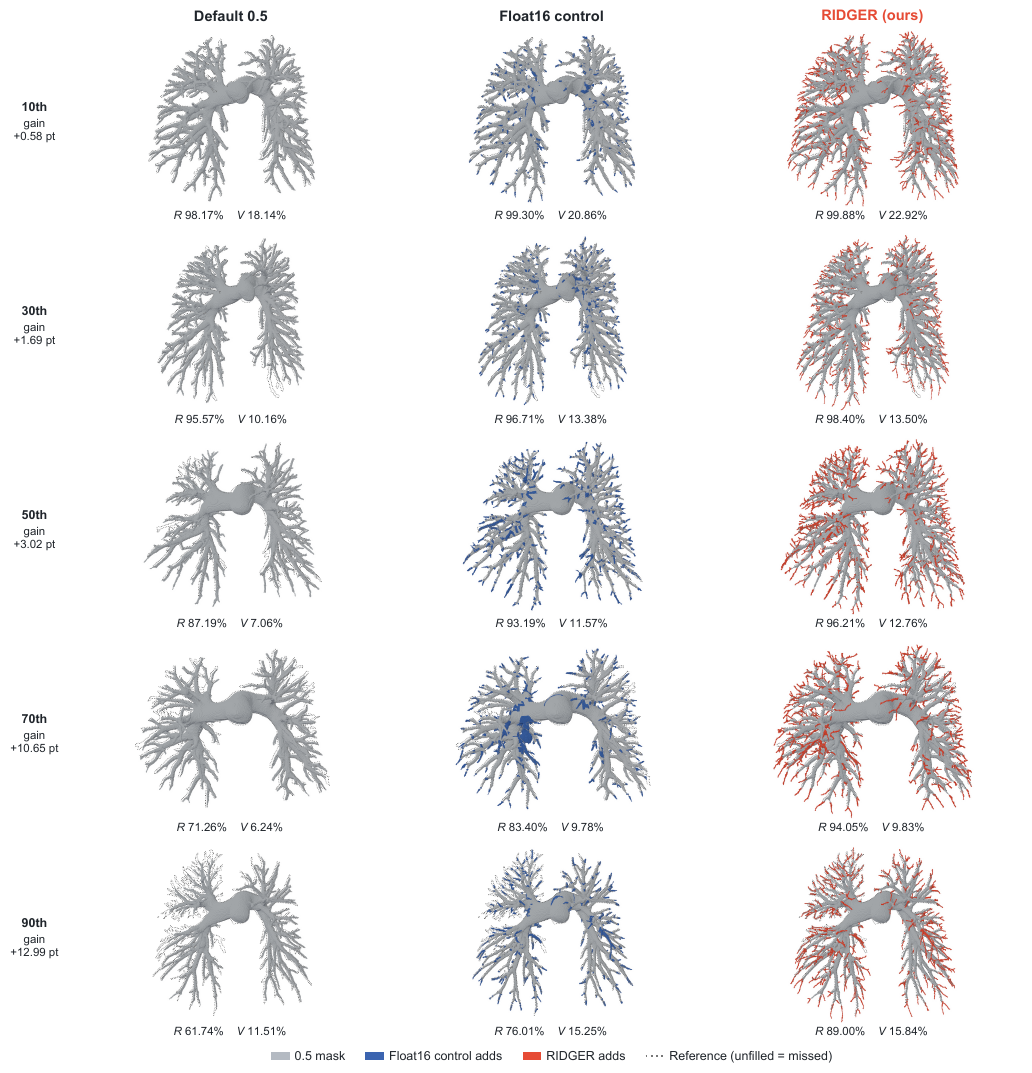}
\caption{\textbf{Five pulmonary CT test patients across the range of RIDGER's gain.} The map is that of the nnU-Net. Rows are the patients at the 10th to 90th percentile of the gain in the reaching rate \(R\) over the float16 control, lowest gain at the top; the number at the left of each row is that gain in points. Columns show the default mask, the float16 control and RIDGER at RIDGER's per-patient off-reference volume \(V\). Grey is the default mask, blue the voxels the float16 control adds, vermilion the voxels RIDGER adds; the dotted outline is the reference, unfilled where missed. \(R\) and \(V\) under each panel are those of that patient.}
\label{S-fig:supp-gallery-pulmonary}
\end{figure*}

Figs.~\ref{S-fig:supp-gallery-coronary} and~\ref{S-fig:supp-gallery-pulmonary} give five patients for each anatomy, read on the nnU-Net map. The patients are ranked by RIDGER's gain in the reaching rate \(R\) over the float16 control at RIDGER's off-reference volume \(V\), and those at the 10th, 30th, 50th, 70th and 90th percentile are shown, lowest gain first. Before ranking, we remove the patient of Fig.~\ref{fig:qual}. For pulmonary CT, we also remove the patients flagged by the slice-spacing screen and the patients whose scans also appear in PARSE. This leaves 159 coronary and 40 pulmonary patients. Every panel shows the default mask in grey and colours only the voxels that the float16 control (blue) or RIDGER (vermilion) adds to it; the dotted line is the reference outline, and reference left unfilled was missed.

\clearpage
\end{document}